\documentclass[12pt, onecolumn]{IEEEtran}
\usepackage{amsmath,amsfonts}
\usepackage{bm}
\usepackage{nicefrac}

\usepackage{siunitx}

\newcommand{\ve}{\mathbf}
\newcommand{\m}{\mathbf}

\usepackage{array}
\usepackage[caption=false,font=normalsize]{subfig}
\usepackage{textcomp}
\usepackage{stfloats}
\usepackage{url}
\usepackage{verbatim}

\usepackage{graphicx}
\usepackage{color}
\graphicspath{{./fig/}}
\usepackage[dvipsnames]{xcolor}	% load before tikz!!!
\usepackage{tikz}
\usetikzlibrary{calc}
\usetikzlibrary{positioning}
\usetikzlibrary{spy}
\tikzset{>=latex}
\usepackage{pgfplots}
\pgfplotsset{compat=newest}
\pgfplotsset{plot coordinates/math parser=false}
\newcommand*\circled[1]{\tikz[baseline=(char.base)]{
            \node[shape=circle,draw,inner sep=0.5pt] (char) {#1};}}

\definecolor{myred}{RGB}{161,23,23}
\definecolor{myblue}{RGB}{23,32,161}
\definecolor{mygreen}{RGB}{66, 172,21}
\definecolor{myorange}{RGB}{245,146,33}
\definecolor{mykaki}{RGB}{184,188,34}
\definecolor{myviolet}{RGB}{205,99,243}

\usepackage{algorithm, algpseudocode}
\usepackage{linegoal}
\makeatletter
\let\Z@E@linegoal\relax % undefine the property
\zref@newprop*{linegoal}[\linewidth]{%
   \the\dimexpr
   \ifdim\dimexpr
     \linewidth -\the\pdflastxpos sp
     +\ifodd\zref@extractdefault{linegoal/posx.\the\LNGL@unique}{page}\c@page
        \oddsidemargin
     \else\evensidemargin
     \fi
     +1in+\hoffset\relax<0pt
   \columnwidth+\columnsep+\fi
   \linewidth -\the\pdflastxpos sp
   +\ifodd\zref@extractdefault{linegoal/posx.\the\LNGL@unique}{page}\c@page
      \oddsidemargin
   \else\evensidemargin
   \fi
   +1in+\hoffset
   \relax
}% linegoal zref-property
\makeatother
\algnewcommand{\LongState}[1]{\State\parbox[t]{\linegoal}{\hangindent=0.3cm\hangafter=1 #1\strut}}
\algrenewcommand\algorithmiccomment[1]{{\color{gray}  \hfill \small $\triangleright$ #1}}

\usepackage{tabularx}
\usepackage{booktabs}
\usepackage{makecell}
\usepackage{multirow}
\usepackage{vcell}

\usepackage{cite}
\def\BibTeX{{\rm B\kern-.05em{\sc i\kern-.025em b}\kern-.08em
    T\kern-.1667em\lower.7ex\hbox{E}\kern-.125emX}}
\AtBeginDocument{\definecolor{tmlcncolor}{cmyk}{0.93,0.59,0.15,0.02}\definecolor{NavyBlue}{RGB}{0,86,125}}

\begin{document}

%\markboth{SICNN: Soft Interference Cancellation Inspired Neural Network Equalizers}{S. Baumgartner {et al.}}

\title{SICNN: Soft Interference Cancellation Inspired Neural Network Equalizers}

\author{Stefan~Baumgartner~\IEEEmembership{Graduate Student Member,~IEEE}, Oliver Lang~\IEEEmembership{Member,~IEEE}, and Mario~Huemer~\IEEEmembership{Senior Member,~IEEE}%
\thanks{This work has been supported by the ``University SAL Labs'' initiative of Silicon Austria Labs (SAL) and its Austrian partner universities for applied fundamental research for electronic based systems. This paper was presented in part at the 2022 IEEE 23rd International Workshop on Signal Processing Advances in Wireless Communication (SPAWC), Oulo, Finland, 2022~\cite{Baumgartner22_C1}.}
\thanks{Stefan Baumgartner and Mario Huemer are with the JKU LIT SAL eSPML Lab, Johannes Kepler University Linz, Austria, and with the Institute of Signal Processing, Johannes Kepler University Linz, Austria (e-mails: \{stefan.baumgartner, mario.huemer\}@jku.at).}%
\thanks{Oliver Lang is with the Institute of Signal Processing, Johannes Kepler University Linz, Austria (email: oliver.lang@jku.at).}%
}

% The paper headers
%\markboth{Journal of \LaTeX\ Class Files,~Vol.~14, No.~8, August~2021}%
%{Shell \MakeLowercase{\textit{et al.}}: A Sample Article Using IEEEtran.cls for IEEE Journals}

%\IEEEpubid{0000--0000/00\$00.00~\copyright~2021 IEEE}
% Remember, if you use this you must call \IEEEpubidadjcol in the second
% column for its text to clear the IEEEpubid mark.

\maketitle

\begin{abstract}
In recent years data-driven machine learning approaches have been extensively studied to replace or enhance traditionally model-based processing in digital communication systems. In this work, we focus on equalization and propose a novel neural network (NN-)based approach, referred to as SICNN. SICNN is designed by deep unfolding a model-based iterative soft interference cancellation (SIC) method. It eliminates the main disadvantages of its model-based counterpart, which suffers from high computational complexity and performance degradation due to required approximations. We present different variants of SICNN. SICNNv1 is specifically tailored to single carrier frequency domain equalization (SC-FDE) systems, the communication system mainly regarded in this work. SICNNv2 is more universal and is applicable as an equalizer in any communication system with a block-based data transmission scheme. Moreover, for both SICNNv1 and SICNNv2, we present versions with highly reduced numbers of learnable parameters. Another contribution of this work is a novel approach for generating training datasets for NN-based equalizers, which significantly improves their performance at high signal-to-noise ratios. We compare the bit error ratio performance of the proposed NN-based equalizers with state-of-the-art model-based and NN-based approaches, highlighting the superiority of SICNNv1 over all other methods for SC-FDE. Exemplarily, to emphasize its universality, SICNNv2 is additionally applied to a unique word orthogonal frequency division multiplexing (UW-OFDM) system, where it achieves state-of-the-art performance. Furthermore, we present a thorough complexity analysis of the proposed NN-based equalization approaches, and we investigate the influence of the training set size on the performance of NN-based equalizers. 
\end{abstract}


\maketitle

\section{Introduction}
\label{sec:Introduction}
\IEEEPARstart{D}{igital} communications at the physical layer level is traditionally a quite model-based discipline. That is, especially for the receiver processing blocks of digital communication systems, most algorithms have been developed based on physical and statistical models of the communication chain. With this established approach well interpretable methods can be obtained, their performance bounds can often be specified, and usually algorithms achieving optimal performance for the given models can be derived. Besides these advantageous properties, model-based approaches also have some downsides. Performance-optimal methods can in some cases exhibit an infeasible computational complexity, requiring the application of suboptimal algorithms in practice. Further, modeling errors, wrong (or oversimplified) assumptions, or insufficient model knowledge may lead to a considerable performance degradation. Since with data-driven machine learning methods many of the drawbacks of model-based approaches can be resolved, currently intensive research is conducted on machine learning approaches for several applications in communications engineering. This includes possible future scenarios like communications assisted by reconfigurable intelligent surfaces (RISs)~\cite{Sagir23}, molecular communications~\cite{Qian18}, or integrated sensing and communication~\cite{Mateos22}. However, also in traditional wireless communication systems promising results can be achieved by means of machine learning. This involves completely abandoning the block-based paradigm of current digital communication system design with the help of end-to-end learning~\cite{OShea17, Aoudia19}, or replacing / enhancing individual blocks of a standard communication chain~\cite{He19, Balatsoukas19, Hoydis21}. The latter includes machine learning approaches for channel estimation~\cite{Ye18,Hu21}, channel decoding~\cite{Nachmani18}, and self-interference cancellation~\cite{Ploder22, Balatsoukas18, Baumgartner23_C2}. In this work, we regard another important processing block at the receiver, namely equalization. Equalization, also referred to as data estimation, is the task of reconstructing transmitted data -- distorted during transmission over a channel -- at the receiver side of a communication system. Typically, equalization is conducted by model-based methods. However, also with machine learning methods auspicious results have already been demonstrated~\cite{Mohammadkarimi19, Pratik21, Samuel19, He20, Liao20, Khani20, Shlezinger20_1, Shlezinger21, Luong22, Wei20}.  In current publications on machine learning approaches for data estimation, mainly neural networks (NNs) are employed. NNs are known to be universal function approximators~\cite{Hornik91} and thus are expected to approximate the optimal data estimators. Many of the presented results are promising, but there are also some new challenges arising. More specifically, standard NNs like a fully-connected feedforward NN (FCNN) are black-box approaches, i.e., their inference is not interpretable, performance bounds can hardly be derived, and domain knowledge is not exploited. Especially due to the latter fact most NNs suffer from requiring large amounts of training data and a high inference complexity. Optimally, one can fuse model-based and data-driven approaches by, e.g., incorporating existing model knowledge into NNs, which is expected to lead to less complex and better performing NNs than the standard black-box NNs. One possibility of incorporating model knowledge into NNs is to design their layer structure accordingly, which leads to NNs we refer to as model-inspired NNs. Currently, one of the most promising and most popular approaches for obtaining NNs with a model-inspired layer structure is deep unfolding~\cite{Hershey14}. The idea of deep unfolding is to take a model-based iterative algorithm, which is conceived for finding the solution of an optimization problem, fix its number of iterations, and unfold every iteration to a layer of an NN. Depending on the aspired abstraction level of the NN (i.e., the similarity between the model-based algorithm and the NN), only a few parameters of the model-based iterative algorithm (e.g., its step size) or even whole parts are replaced by learnable parameters or modules, respectively. Those can then be optimized with tools known from NN optimization by utilizing available training data. A number of NN-based data estimators, e.g., the NNs in~\cite{Samuel19,Liao20, Shlezinger20_1, Shlezinger21, Luong22, Wei20} are designed by employing deep unfolding. In this work, we also apply deep unfolding for the design of our proposed NN-based equalizers.

Most NN-based data estimators are currently proposed for equalization in multiple-input multiple-output (MIMO) communication systems, often assuming data transmission over an uncorrelated Rayleigh fading channel. In this work, the developed NN-based equalizers are mainly evaluated for single carrier frequency domain equalization (SC-FDE) systems~\cite{Pancaldi08, Huemer03}. In an SC-FDE system, a single carrier transmission scheme is utilized, but the payload data is transmitted in a block-wise manner with guard intervals between successive blocks, as it is the case in orthogonal frequency division multiplexing (OFDM) systems. The received blocks are transformed to frequency domain before conducting matched filtering, downsampling, and equalization. This allows an efficient receiver implementation~\cite{Huemer03} and results in a system model similar to that of an OFDM system. In this work, we regard employing both a cyclic prefix (CP) and a so-called unique word (UW), which is a known deterministic sequence, as guard interval. The UW can advantageously be utilized, e.g., for synchronization purposes~\cite{Huemer03_1}, however, at the cost of equalization complexity. For a CP guard interval, the optimal linear equalizer is a low-complex single-tap equalizer, while for a UW guard interval, in turn, the optimal linear equalizer is more complex. In contrast to CP-OFDM systems, for SC-FDE optimal performance can only be obtained with computationally highly demanding nonlinear equalizers. This motivates developing NN-based data estimators for SC-FDE systems. Employing NN-based equalizers for SC-FDE systems necessitates, in contrast to MIMO systems over uncorrelated Rayleigh fading channels, an additional pre-processing step. As extensively described in~\cite{Baumgartner23_C1}, for the application of NN-based equalizers in SC-FDE systems a data normalization scheme is required for a well-behaved NN training and thus a satisfying performance of the NN equalizers. We also briefly review the necessary data normalization scheme in this work.

\subsection*{Contribution}

In this work, we propose the NN-based data estimators SICNNv1 and SICNNv2, which are designed by unfolding an iterative soft interference cancellation (SIC) method~\cite{Choi00}.  The main idea of iterative SIC is that in each iteration every single data symbol in the transmitted data vector is estimated on its own, by considering the influence of all other data symbols in the data vector as interference. This interference can be mitigated by incorporating estimates of the data symbols from the previous into the current iteration. By that, the data symbol estimates are refined from iteration to iteration. Although also DeepSIC~\cite{Shlezinger21} is inspired by the same iterative SIC method, SICNNv1 and SICNNv2 are fundamentally different from this NN. The idea of SIC is adopted by DeepSIC concerning its structure. That is, DeepSIC consists of multiple stages, where each stage is comprised of as many sub-FCNNs as there are data symbols in the transmitted data vector. Each of the sub-FCNNs is utilized to estimate one data symbol, whereby the input data of a sub-FCNN is made up of the received vector as well as of estimates provided by the sub-FCNNs for the remaining data symbols from the last stage. All estimates are refined stage by stage. That is, DeepSIC has similarities with the model-based SIC method only by refining the estimates of the posterior data symbol probabilities, but neither interference cancellation is conducted in a stage, nor model knowledge is utilized. In contrast, our proposed NN-based equalizers are far more similar to the underlying model-based method. More specifically, with SICNNv1 --~an adapted version of an NN-based equalizer called SICNN proposed in our previous work~\cite{Baumgartner22_C1}~-- we try to resemble the model-based iterative SIC method closely. However, we replace numerically demanding, computationally intensive operations, for which also approximations have to be made in the model-based approach, by low-complex NNs. This NN-based approach achieves significantly better performance than the corresponding model-based method, while exhibiting lower complexity. We tailor SICNNv1 for being employed as an NN-based equalizer in an SC-FDE system by exploiting some properties of this communication system in the NN architecture design. SICNNv2, in turn, is more abstracted from the model-based iterative SIC method, i.e., less model knowledge is utilized for the NN architecture design. However, it is more universal and can also be applied as an equalizer in any communication system with a block-based data transmission scheme. A further difference between DeepSIC and the proposed SICNNv1 and SICNNv2 is their generalization ability regarding different channels. As it is the case for, e.g., MMNet~\cite{Khani20} or ViterbiNet~\cite{Shlezinger20}, DeepSIC is trained for one specific channel. This generally allows lower complex NNs, but requires retraining as soon as the channel changes. SICNNv1 and SICNNv2 belong, like DetNet~\cite{Samuel19} or OAMP-Net~\cite{Liao20}, to the group of NN-based data estimators, which are trained with different channels sampled from a statistical channel model, and use the actual channel realization as an input. These NNs generally require an extensive offline training, and exhibit a higher computational inference complexity, but they do not have to be retrained as long as the specified statistical channel model is valid for the operating environment. 

Since in every stage\footnote{In order to avoid any misunderstandings, we refer to one unfolded iteration of the model-based iterative SIC method as \textit{stage} of SICNNv1/SICNNv2 instead of \textit{layer}.} of SICNNv1/SICNNv2 the same task has to be fulfilled, namely to refine estimated posterior data symbol probabilities, we additionally introduce two modified versions of SICNNv1 and SICNNv2. While in SICNNv1 and SICNNv2 for every stage different sub-NNs are utilized to estimate posterior data symbol probabilites, in SICNNv1Red and SICNNv2Red every stage uses the same sub-NNs, which drastically reduces the number of parameters to be trained.

We compare the proposed NN-based equalizers with state-of-the-art model-based and NN-based data estimators concerning achieved bit error ratio (BER) performance, and regarding their computational complexity during inference. The evaluation is conducted for SC-FDE systems, either employing a UW or a CP as a guard interval, for both quadrature phase shift keying (QPSK) and 16-QAM (quadrature amplitude modulation) alphabets, and with perfect and imperfect channel knowledge at the receiver. We investigate the required amount of training data for achieving satisfying performance of selected NN-based equalizers, pointing out the advantage of reducing the number of learnable parameters of an NN. Further, we demonstrate the universal applicability of SICNNv2 by presenting its achieved performance for a communication system employing the so-called UW-OFDM signaling scheme.

As another important contribution of this paper, we present a novel approach to generate training sets for NN-based equalizers. In this approach, only those sample data transmissions are included in the training set for which the number of data symbol estimation errors made by a baseline equalizer exceeds a specified quantity. This greatly enhances the performance of NN-based data estimators at high signal-to-noise ratios (SNRs).

The remainder of this paper is structured as follows. In Sec.~\ref{sec:Preliminaries}, we review the SC-FDE signaling scheme and data transmission model, and we discuss state-of-the-art model-based equalizers, including in particular the iterative SIC method our NN-based equalizers are inspired by. In Sec.~\ref{sec:Soft_Interference_Cancellation_Inspired_Neural_Network_Equalizers}, the novel NN-based equalizers are introduced and discussed in detail.  Further, we propose a novel approach for generating training datasets for NN-based data estimators in Sec.~\ref{sec:Training_Set_Generation_and_Data_Normalization}. We present BER performance results, and an in-depth analysis of the computational complexity of the regarded model-based and NN-based equalizers in Sec.~\ref{sec:Results}.

\subsection*{Notation}
Throughout this paper, we use lower-case bold face letters $\ve{x}$ for vectors and upper-case bold face letters $\m{X}$ for matrices, $x_k$ for the $k$th element of $\ve{x}$, $[\m{X}]_{kj}$ for the element of $\m{X}$ in row $k$ and column $j$, and $[\m{X}]_{k*}$ for the $k$th row of $\m{X}$. Further, $(.)^T$, $(.)^H$, and $(.)^*$ indicate transposition, conjugate transposition, and conjugation respectively, while $|\m{X}|$ is the determinant of the matrix $\m{X}$. We denote the probability density function (PDF) of a continuous random variable as $p(.)$, the probability mass function (PMF) of a discrete random variable as $p[.]$, a conditional PMF of a random variable $a$ given $b$ as $p[a|b]$, and a PMF evaluated at the value $\tilde{a}$ as $p[a=\tilde{a}]$. We describe the expectation operator averaging over the PDF/PMF of a random variable $a$ as $E_a[.]$, where the subscript of the expectation operator is omitted when the averaging PDF/PMF is clear from context. 

\section{Preliminaries}
\label{sec:Preliminaries}

In this section, we describe the system model for SC-FDE, and state-of-the-art model-based equalization approaches. Further, we discuss an iterative SIC approach for data estimation, and we highlight some properties of this method. 

\subsection{Single Carrier Frequency Domain Equalization}
\label{ssec:Single_Carrier_Frequency_Domain_Equalization}
In an SC-FDE communication system~\cite{Pancaldi08, Huemer03, Witschnig02, Huemer10}, a single carrier modulation scheme is employed for data transmission. At the transmitter, the data symbols to be transmitted, which are drawn from a modulation alphabet $\mathbb{S}$ (in this work, we mainly use QPSK as a modulation alphabet), are grouped into blocks of length $N_{\text{d}}$. These blocks of data symbols, which we refer to as data vectors $\ve{d}\in\mathbb{S}^{N_{\text{d}}}$, are strung together to generate a transmit data burst, whereby they are separated by guard intervals of length $N_{\text{g}}$. As a guard interval, in this work we consider using either a CP or a UW, which is a deterministic sequence known by the receiver. Depending on the employed guard interval, some processing steps in the receiver are different, which are described later in this section. The transmit data burst is upsampled and pulse shaped with a root-raised-cosine (RRC) filter, followed by transmitting the resulting signal over a multipath channel, which is additionally disturbed by additive white Gaussian noise (AWGN). At the receiver, the first processing step depends on the employed guard interval. While for a UW guard interval every received data vector including its succeeding guard interval is transformed individually to frequency domain, for a CP guard interval the CPs are removed first before transforming the remaining received data vectors individually to frequency domain. In frequency domain, the further processing steps matched filtering, downsampling, and equalization are conducted. Independent of the employed guard interval, the general model\footnote{Here we assume sufficiently long guard intervals such that each data block can be processed individually and independently of all other transmitted data blocks.} of the transmission of a data vector up to the input of the equalizer in the equivalent complex baseband can be written as~\cite{Reinhardt06}
\begin{gather}
\ve{y}_{\text{r}} = \widetilde{\m{H}}\m{F}_{N^\prime}\ve{x} + \ve{w}\,.\label{eq:SC-FDE_model_general}
\end{gather}
Here, $\ve{y}_{\text{r}}\in\mathbb{C}^{N^\prime}$ is the received vector after matched filtering and downsampling in frequency domain, where $N^\prime$ depends on the employed guard interval and is being specified later in this section. $\widetilde{\m{H}}\in\mathbb{R}^{N^\prime\times N^\prime}$ is a diagonal matrix containing the sampled frequency response of the cascade of upsampler, pulse shaping filter, multipath channel, matched filter, and downsampler on its main diagonal. Note, that $\widetilde{\m{H}}$ is a real-valued matrix since we conduct optimal matched filtering in frequency domain, i.e., the filter is matched to the channel distorted transmit pulse (for further details on optimal matched filtering in SC-FDE systems, we refer to~\cite{Huemer99, Yang07}). Furthermore, $\m{F}_{N^\prime}\in\mathbb{C}^{N^\prime\times N^\prime}$ is the $N^\prime$-point discrete Fourier transform (DFT) matrix and ${\ve{w}\sim\mathcal{C N}(\ve{0}, N^\prime\sigma_{\text{n}}^2\widetilde{\m{H}})}$ is circularly symmetric complex AWGN, with $\sigma_{\text{n}}^2$ being the variance of the AWGN in time domain. The structure of the transmitted vector ${\ve{x}\in\mathbb{C}^{N^\prime}}$ as well as the final system model differ for UW and CP guard intervals, which we further detail in the following.

\subsubsection{Unique Word Guard Interval}
\label{ssec:UW_Guard_interval}
As already mentioned, in case of a UW guard interval~\cite{Witschnig02, Huemer10}, at the receiver both a received data vector and its succeeding UW are transformed to frequency domain for the further processing steps. Hence, the vector $\ve{x}\in\mathbb{C}^{N^\prime}$ in~\eqref{eq:SC-FDE_model_general} has the form ${\ve{x} = [\ve{d}^T, \ve{u}^T]^T}$, where $\ve{d}\in\mathbb{S}^{N_{\text{d}}}$ is the transmitted data vector to be estimated, $\ve{u}\in\mathbb{C}^{N_{\text{g}}}$ is the UW, and ${N^\prime = N = N_{\text{d}}+N_{\text{g}}}$. Inserting into~\eqref{eq:SC-FDE_model_general} leads to
\begin{gather}
\ve{y}_{\text{r}} = \widetilde{\m{H}}\m{F}_N\begin{bmatrix}
\ve{d}\\ \ve{u}
\end{bmatrix}+ \ve{w}\,.\label{eq:system_model_UW_not_removed}
\end{gather}
By assuming perfect channel knowledge on receiver side, the influence of the known UW~$\ve{u}$ on the received vector $\ve{y}_{\text{r}}$ can be removed according to
\begin{gather}
\ve{y}_{\text{r}} - \widetilde{\m{H}}\m{M}^\prime\ve{u} = \widetilde{\m{H}}\m{M}_{\text{uw}}\ve{d} + \ve{w}\,,\label{eq:system_model_SC-FDE_UW}
\end{gather}
with $\m{F}_N = [\m{M}_{\text{uw}}\,\m{M}^\prime]$, where $\m{M}_{\text{uw}}\in\mathbb{C}^{N\times N_{\text{d}}}$ and $\m{M}^\prime\in\mathbb{C}^{N\times N_{\text{g}}}$ are built by the first $N_{\text{d}}$ columns and the remaining $N_{\text{g}}$ columns of $\m{F}_N$, respectively. 

\subsubsection{Cyclic Prefix Guard Interval}
\label{ssec:CP_Guard_interval}
In case of a CP guard interval~\cite{Pancaldi08, Huemer03}, the guard intervals are removed at the receiver before transforming the received blocks of data to frequency domain, which means that $\ve{x}$ in~\eqref{eq:SC-FDE_model_general} is realized as the transmitted data vector $\ve{d}\in\mathbb{S}^{N_{\text{d}}}$, and ${N^\prime = N_{\text{d}}}$. Consequently, for a CP guard interval, the data transmission is modeled as
\begin{gather}
\ve{y}_{\text{r}} = \widetilde{\m{H}}\m{F}_{N_{\text{d}}}\ve{d} + \ve{w}\,.\label{eq:system_model_SC-FDE_CP}
\end{gather}

\subsubsection{System Model for Single Carrier Frequency Domain Equalization}
\label{ssec:System_Model_for_Singel_Carrier_FDE}
As elaborated above, the model for data transmission in an SC-FDE system is given by~\eqref{eq:system_model_SC-FDE_UW} for a UW guard interval and by~\eqref{eq:system_model_SC-FDE_CP} for a CP. For the ease of notation, in the remainder of this work the SC-FDE system model is given for both guard intervals by
\begin{gather}
\ve{y} = \widetilde{\m{H}}\m{M}\ve{d} + \ve{w} = \m{H}\ve{d}+\ve{w}\,,\label{eq:system_model_SC-FDE}
\end{gather}
with $\ve{y}\in\mathbb{C}^{N^\prime}$, $\m{M}\in\mathbb{C}^{N^\prime\times N_{\text{d}}}$, and $\m{H} = \widetilde{\m{H}}\m{M}$, where we employ for a
\begin{itemize}
\item UW guard: $N^\prime=N=N_{\text{d}}+N_{\text{g}}$, $\ve{y} = \ve{y}_{\text{r}} - \widetilde{\m{H}}\m{M}^\prime\ve{u}$, $\m{M} = \m{M}_{\text{uw}}$.
\item CP guard: $N^\prime = N_{\text{d}}$, $\ve{y} = \ve{y}_{\text{r}}$, $\m{M} = \m{F}_{N_{\text{d}}}$.
\end{itemize}

\subsubsection{Model-Based Equalization}
Based on~\eqref{eq:system_model_SC-FDE}, data estimation can be conducted for a given received vector~$\ve{y}$ and a channel matrix~$\m{H}$. As thoroughly elucidated in~\cite{Baumgartner23_J1}, depending on the optimality criterion, there exist different optimal equalizers. The bit-wise maximum a-posteriori (MAP) estimator yields for every transmitted bit the bit value featuring the highest posterior probability. It is known to be the optimal estimator regarding the bit error probability. The vector MAP, in turn, is optimal regarding the error probability of the data vector estimate. However, the computational complexity of both of the aforementioned estimators grows exponentially with the data vector length $N_{\text{d}}$, which makes them in general prohibitive for practical applications. Hence, one usually has to resort to suboptimal linear or nonlinear estimation methods. 

The best linear estimator in the Bayesian sense is the linear minimum mean square error (LMMSE) estimator, which is given by~\cite{Kay93}
\begin{gather}
\hat{\ve{d}} = \Big(\m{M}^H\widetilde{\m{H}}\m{M} + \frac{N\sigma_{\text{n}}^2}{\sigma_{\text{d}}^2}\m{I}\Big)^{-1}\m{M}^H\ve{y} = \m{E}_{\text{LMMSE}}\ve{y}\,,\label{eq:LMMSE_v1}
\end{gather}
with $\m{I}$ and $\sigma_{\text{d}}^2$ being the identity matrix with appropriate dimensions and the variance of the symbol alphabet, respectively. In case of a CP guard interval ($\m{M} = \m{F}_{N_{\text{d}}}$), \eqref{eq:LMMSE_v1} can be simplified to~\cite{Huemer10} 
\begin{gather}
\hat{\ve{d}} = \frac{1}{N^\prime}\m{M}^H\Big(\widetilde{\m{H}} + \frac{\sigma_{\text{n}}^2}{\sigma_{\text{d}}^2}\m{I}\Big)^{-1}\ve{y} = \frac{1}{N^\prime}\m{M}^H\m{E}_{\text{LMMSE,dg}}\ve{y}\,,\label{eq:LMMSE_diag}
\end{gather}
where $\m{E}_{\text{LMMSE,dg}} = \big(\widetilde{\m{H}} + \frac{\sigma_{\text{n}}^2}{\sigma_{\text{d}}^2}\m{I}\big)^{-1}$ is a diagonal matrix, and thus also the inversion required to compute this estimator matrix can be realized efficiently. Typically, instead of multiplying by $\frac{1}{N^\prime}\m{M}^H$, an inverse DFT is conducted. Note, that this low-complex equalizer can also be employed for a UW guard interval when applying LMMSE estimation to~\eqref{eq:system_model_UW_not_removed} instead of~\eqref{eq:system_model_SC-FDE_UW}, i.e., the known UW $\ve{u}$ is not being removed before data estimation, but is estimated as well. Compared to the LMMSE estimator~\eqref{eq:LMMSE_v1}, this approximate\footnote{By neglecting the knowledge about the UW, only an approximate LMMSE estimator is obtained.} LMMSE estimator allows a lower-complex equalization, however, at the cost of performance degradation~\cite{Huemer10}.

A popular suboptimal nonlinear estimator is the decision feedback equalizer (DFE), which is an iterative method. There, in every iteration LMMSE estimation of the data symbol with the smallest error variance is conducted, followed by removing the influence of the hard decision data symbol estimate on the received vector. However, in case of wrong data symbol estimates, this method suffers from error propagation deteriorating the estimation performance. For more details, we refer to~\cite{Baumgartner23_J1}, where the DFE is elaborated for a so-called unique word orthogonal frequency division multiplexing (UW-OFDM) system. In the following, we address another suboptimal nonlinear method, namely iterative soft interference cancellation (SIC), in more detail, since the proposed NN-based equalizers are deduced from this model-based approach. 

\subsection{Iterative Soft Interference Cancellation}
\label{ssec:Iterative_Soft_Interference_Cancellation}

The idea of the iterative SIC method proposed in~\cite{Choi00} is to estimate each data symbol $d_k$, $k\in\{0, ..., N_{\text{d}}-1\}$, in the data vector $\ve{d}$ separately, and refine the estimates from iteration to iteration. For the estimation of the $k$th data symbol $d_k$, all other data symbols $d_l$, $l \neq k$, are treated as interference, and thus their influence on the received vector $\ve{y}$ is cancelled as far as possible. Since the data symbols $d_l$ are unknown, their currently available estimates $\hat{d}_l$ are utilized for interference cancellation. Interference cancellation reduces the unknown variable to be estimated from an $N_{\text{d}}$-dimensional data vector $\ve{d}$ to a single data symbol $d_k$, and thus nonlinear minimum mean square error (MMSE) estimation --~which is generally computationally infeasible for estimating $\ve{d}$~-- can easily be applied for estimating $d_k$. In order to prevent error propagation, instead of using hard decision data symbol estimates for interference cancellation, soft information of every data symbol estimate from the previous iteration is utilized in form of the MMSE estimate, which is the posterior mean, and the corresponding conditional mean square error (MSE). The soft estimates are refined iteratively. In~\cite{Choi00}, this approach is proposed for a MIMO system where all entries of the channel matrix $\m{H}$ are independent of each other (all entries of the channel matrix are modeled as independent random variables following a normal distribution). This allows for simplifications in the iterative SIC method that cannot be applied in general. In the following, we present the iterative SIC method following the approach proposed in~\cite{Choi00}. However, we adapt this method, which is also summarized in Algorithm~\ref{alg:Model-based_SIC_SC_FDE_Algorithm}, for an SC-FDE system, where the assumption of independent elements of $\m{H}$ is not fulfilled.

\begin{algorithm}[t]
\caption{Model-based iterative SIC for SC-FDE.}
\label{alg:Model-based_SIC_SC_FDE_Algorithm}
\begin{algorithmic}[1]
\Function{IterativeSoftIC\_SC-FDE}{$\m{H}$, $\ve{y}$, $\sigma_{\text{n}}^2$, $\sigma_{\text{d}}^2$}
\State $\hat{d}_{k}^{(-1)} \gets 0, e_k^{(-1)}\gets\sigma_{\text{d}}^2 \quad\forall k = 0, ..., N_{\text{d}}-1$
\For{$q = 0, ..., Q-1$}
\For{$k = 0, ..., N_{\text{d}}-1$}
\State Compute $\ve{y}_{\text{ic},k}^{(q)}$ according to~\eqref{eq:system_model_interference_cancellation}
\LongState{Compute $\m{C}_{\ve{v}\ve{v},k}^{(q)}$ following \eqref{eq:C_vv_SIC_SC-FDE_first_iteration} for $q=0$, or \eqref{eq:C_vv_SIC_SC-FDE_approx} for $q > 0$}\label{alg_line:Cvv_est}
\State Evaluate posterior PMF $p[d_k|\ve{y}_{\text{ic},k}^{(q)}]$\label{alg_line:posterior_pmf}
\LongState{Update soft information: $\hat{d}_k^{(q)}$ using \eqref{eq:MMSE_data_symbol_estimate} and $e_k^{(q)}$ via \eqref{eq:MMSE_data_symbol_MSE}}
\EndFor
\EndFor
\State\Return{$\hat{\ve{d}}^{(Q-1)}$}
\EndFunction
\end{algorithmic}
\end{algorithm}

Let us regard the $q$th iteration, $q = 0, ..., Q-1$, of $Q$ total iterations of the iterative SIC method. We assume that for every data symbol $d_k$, $k\in\{0, ..., N_{\text{d}}-1\}$, a soft estimate from the previous iteration $(q-1)$ is available, namely, the MMSE data symbol estimate, which is the posterior mean
\begin{gather}
\hat{d}_k^{(q-1)} = E_{d_k|\ve{y}_{\text{ic},k}^{(q-1)}}\Big[d_k\big|\ve{y}_{\text{ic},k}^{(q-1)}\Big]\,,
\end{gather}
and the corresponding MSE given $\ve{y}_{\text{ic},k}^{(q-1)}$
\begin{gather}
e_k^{(q-1)} = E_{d_k|\ve{y}_{\text{ic},k}^{(q-1)}}\Big[\big|d_k-\hat{d}_k^{(q-1)}\big|^2\big|\ve{y}_{\text{ic},k}^{(q-1)}\Big]\,.
\end{gather}
Here, $\ve{y}_{\text{ic},k}^{(q-1)}$ is the received vector without the interference of all but the $k$th data symbol estimates in iteration $(q-1)$. For the estimation of a data symbol $d_k$, one can reformulate the system model~\eqref{eq:system_model_SC-FDE} to 
\begin{gather}
\ve{y} = \ve{h}_k d_k + \bar{\m{H}}_k\bar{\ve{d}}_k + \ve{w}\,,\label{eq:system_model_reformulated}
\end{gather}
where $\bar{\m{H}}_k$ is $\m{H}$ after removing the $k$th column $\ve{h}_k$, and $\bar{\ve{d}}_k$ is the data vector without the $k$th data symbol $d_k$. The term $\bar{\m{H}}_k\bar{\ve{d}}_k$ denotes the interference caused by all but the $k$th data symbol in the data vector, which should ideally be removed from $\ve{y}$ for the estimation of $d_k$. SIC can be conducted by removing $\bar{\m{H}}_k\hat{\bar{\ve{d}}}_k^{(q-1)}$ from $\ve{y}$, leading to
\begin{gather}
\ve{y}_{\text{ic},k}^{(q)} = \ve{y} - \bar{\m{H}}_k\hat{\bar{\ve{d}}}_k^{(q-1)} = \ve{h}_kd_k \underbrace{- \bar{\m{H}}_k\bar{\bm{\delta}}_k^{(q-1)}+\ve{w}}_{\ve{v}_k^{(q)}}\,,\label{eq:system_model_interference_cancellation}
\end{gather}
where $\hat{\bar{\ve{d}}}_k^{(q-1)}$ consists of all but the $k$th data symbol estimates from iteration $(q-1)$, and $\bar{\bm{\delta}}_k^{(q-1)} = \hat{\bar{\ve{d}}}_k^{(q-1)} - \bar{\ve{d}}_k$ contains the (unknown) data symbol estimation errors from the previous iteration step. 
For estimating $d_k$ based on~\eqref{eq:system_model_interference_cancellation}, the statistics of the total noise vector $\ve{v}_k^{(q)}$, which is composed of the Gaussian noise vector $\ve{w}$ and the noise due to data symbol estimation errors, have to be specified. We start by considering the noise statistics for the first iteration ($q=0$). As we will elaborate later in this section, initializing the data symbol estimates with the mean of the symbol alphabet is a rational choice, i.e., $\hat{\bar{\ve{d}}}_k^{(-1)}=\ve{0}$, leading to $\ve{y}_{\text{ic},k}^{(0)}=\ve{y}$ and $\bar{\bm{\delta}}_k^{(0)} = -\bar{\ve{d}}_k$. Assuming independent and identically distributed (i.i.d.) data symbols with uniform prior probability and reasonably large $N_{\text{d}}$, following central limit theorem arguments, $\bar{\m{H}}_k\bar{\ve{d}}_k$ can be considered to follow a circularly symmetric complex Gaussian distribution with zero mean, and to be independent of $\ve{w}$. Hence, $\ve{v}_k^{(0)}$ approximately also follows a circularly symmetric complex Gaussian distribution with zero mean and a covariance matrix
\begin{gather}
\m{C}_{\ve{v}\ve{v},k}^{(0)} = E\big[\ve{v}_k^{(0)}\ve{v}_k^{(0)\,H}\big] = \sigma_{\text{d}}^2\widetilde{\m{H}}\bar{\m{M}}_k\bar{\m{M}}_k^H\widetilde{\m{H}} + N\sigma_{\text{n}}^2\widetilde{\m{H}}\,,\label{eq:C_vv_SIC_SC-FDE_first_iteration}
\end{gather}
where $\bar{\m{M}}_k$ is the matrix $\m{M}$ without the $k$th column. 
For all further iterations ($q > 0$), we start by specifying the type of the statistical distribution of the vector $\ve{r}_k^{(q-1)} = \bar{\m{H}}_k\bar{\bm{\delta}}_k^{(q-1)}$. Based on central limit theorem arguments and unbiased MMSE estimates (cf. Appendix~\ref{apx:noise_statistics_in_SIC_step}), $\ve{r}_k^{(q-1)}$ can be approximated to follow a circularly symmetric complex Gaussian distribution with zero mean, and thus the same assumption holds for $\ve{v}_{k}^{(q)}$. As shown in Appendix~\ref{apx:noise_statistics_in_SIC_step}, the noise covariance matrix $\m{C}_{\ve{v}\ve{v},k}^{(q)}$ is given by
\begin{gather}
\begin{aligned}
\m{C}_{\ve{v}\ve{v},k}^{(q)} &= \bar{\m{H}}_{k}\widetilde{\m{E}}_k^{(q-1)}\bar{\m{H}}_{k}^H + N\sigma_{\text{n}}^2\widetilde{\m{H}}\\ &\quad - \bar{\m{H}}_k E_{\left(\bar{\ve{d}}_k|\ve{y},\hat{\ve{d}}^{(q-2)}\right), \ve{w}}\big[\bar{\bm{\delta}}_{k}^{(q-1)}\ve{w}^H\big]\\&\quad - E_{\left(\bar{\ve{d}}_k|\ve{y},\hat{\ve{d}}^{(q-2)}\right), \ve{w}}\big[\ve{w}\bar{\bm{\delta}}_{k}^{(q-1)\,H}\big]\bar{\m{H}}_k^H\,,\label{eq:Cvv_SIC}
\end{aligned}
\end{gather}
where $\widetilde{\m{E}}_k^{(q-1)} = E_{\bar{\ve{d}}_k|\ve{y},\hat{\ve{d}}^{(q-2)}}\big[\bar{\bm{\delta}}_k^{(q-1)}\bar{\bm{\delta}}_k^{(q-1)\,H}\big|\ve{y},\hat{\ve{d}}^{(q-2)}\big]$ is the conditioned error covariance matrix. For the off-diagonal entries of $\widetilde{\m{E}}_k^{(q-1)}$ and for the third and the fourth term in~\eqref{eq:Cvv_SIC} no exact closed form solution is available. A possible workaround is to employ the approximation
\begin{gather}
\m{C}_{\ve{v}\ve{v},k}^{(q)}\approx\bar{\m{H}}_k\m{E}_k^{(q-1)}\bar{\m{H}}_k^H+N\sigma_{\text{n}}^2\widetilde{\m{H}}\,,\label{eq:C_vv_SIC_SC-FDE_approx}
\end{gather}
where $\m{E}_k^{(q-1)} = \text{diag}\big(\big[e_0^{(q-1)}, ..., e_{k-1}^{(q-1)}, e_{k+1}^{(q-1)}, ..., e_{N_{\text{d}}-1}^{(q-1)}\big]\big)$, i.e., correlations between the data symbol estimates as well as correlations between the estimation errors and the AWGN noise are neglected. However, this may lead to inaccuracies in the estimation process. Especially at high SNRs, for deep fading channels, and when $q$ is increasing (then the $e_k^{(q-1)}$ are usually becoming smaller), $\m{C}_{\ve{v}\ve{v},k}^{(q)}$ can become ill-conditioned, which is --~in combination with the occurring approximation errors~-- an issue for computing its inverse required for the next steps in the estimation process.

For computing a data symbol estimate $\hat{d}_k^{(q)}$, the posterior PMF $p\big[d_k|\ve{y}_{\text{ic},k}^{(q)}\big]$ is needed, which can be obtained via the Bayesian rule by utilizing the likelihood function $p\big(\ve{y}_{\text{ic},k}^{(q)}\big|d_k\big)$. Given the noise covariance matrix $\m{C}_{\ve{v}\ve{v},k}^{(q)}$, the likelihood function follows to
\begin{align}
p\big(\ve{y}_{\text{ic},k}^{(q)}\big|d_k\big) &= \frac{1}{\pi^N|\m{C}_{\ve{v}\ve{v},k}^{(q)}|}\exp\big(-\ve{x}_k^{(q)\,H}\m{C}_{\ve{v}\ve{v},k}^{(q)^{-1}}\ve{x}_k^{(q)}\big)\nonumber\\
&=\alpha_k^{(q)} f_k^{(q)}(d_k)\,,
\end{align}
with $\ve{x}_k^{(q)}=\ve{y}_{\text{ic},k}^{(q)}-\ve{h}_k d_k$, a scaling factor
\begin{gather}
\alpha_k^{(q)} = \frac{1}{\pi^N|\m{C}_{\ve{v}\ve{v},k}^{(q)}|}\exp\big(-\ve{y}_{\text{ic},k}^{(q)\,H}\m{C}_{\ve{v}\ve{v},k}^{(q)^{-1}}\ve{y}_{\text{ic},k}^{(q)}\big)\,,
\end{gather}
which is independent of $d_k$, and a function 
\begin{gather}
\begin{aligned}
f_k^{(q)}(d_k) &= \exp\big(d_k^*\ve{h}_k^H\m{C}_{\ve{v}\ve{v},k}^{(q)^{-1}}\ve{y}_{\text{ic},k}^{(q)}+\ve{y}_{\text{ic},k}^{(q)\,H}\m{C}_{\ve{v}\ve{v},k}^{(q)^{-1}}\ve{h}_k d_k\big)\\
&\quad\cdot\exp\big(-d_k^*\ve{h}_k^H\m{C}_{\ve{v}\ve{v},k}^{(q)^{-1}}\ve{h}_k d_k\big)
\end{aligned}\label{eq:posterior_prob_f_k}
\end{gather}
depending on $d_k$. With the above stated results at hand, the MMSE data symbol estimate $\hat{d}_k^{(q)}$ and its corresponding MSE $e_k^{(q)}$ can then be computed as
\begin{align}
\hat{d}_k^{(q)} &= E\big[d_k|\ve{y}_{\text{ic},k}^{(q)}\big] = \sum_{s^\prime\in\mathbb{S}}s^\prime p\big[d_k=s^\prime\big|\ve{y}_{\text{ic},k}^{(q)}\big]\nonumber\\
&= \frac{\sum_{s^\prime\in\mathbb{S}}s^\prime p\big(\ve{y}_{\text{ic},k}^{(q)}\big|d_k=s^\prime\big)}{\sum_{s^\prime\in\mathbb{S}}p\big(\ve{y}_{\text{ic},k}^{(q)}\big|d_k=s^\prime\big)} = \frac{\sum_{s^\prime\in\mathbb{S}}s^\prime f_k^{(q)}(s^\prime)}{\sum_{s^\prime\in\mathbb{S}}f_k^{(q)}(s^\prime)}\label{eq:MMSE_data_symbol_estimate}
\end{align}
and
\begin{gather}
e_k^{(q)} = E\big[\big|d_k-\hat{d}_k^{(q)}\big|^2|\ve{y}_{\text{ic},k}^{(q)}\big] = \frac{\sum_{s^\prime\in\mathbb{S}}\big|s^\prime -\hat{d}_k^{(q)}\big|^2 f_k^{(q)}(s^\prime)}{\sum_{s^\prime\in\mathbb{S}}f_k^{(q)}(s^\prime)}\,,\label{eq:MMSE_data_symbol_MSE}
\end{gather}
where a uniform prior PMF $p[d_k]$ is assumed. 

Updating the data symbol soft estimates concludes one iteration. The succeeding iteration starts with conducting SIC following~\eqref{eq:system_model_interference_cancellation}. 

A quite interesting and not obvious result is verified in Appendix~\ref{apx:estimates_of_soft_interference_cancellation_method_first_iteration}, namely, that when using a zero vector as initialization for the estimated data symbol vector, after one iteration the iterative SIC exhibits  the same hard decision bit error probability as the LMMSE estimator. Hence, we employ the zero vector, which is also the prior mean of the data vector (since we assume a symmetric modulation alphabet and uniformly distributed data symbol probabilities), as initialization of the iterative SIC method.

\section{Soft Interference Cancellation Inspired Neural Network Equalizers}
\label{sec:Soft_Interference_Cancellation_Inspired_Neural_Network_Equalizers}
As already mentioned in Sec.~\ref{ssec:Iterative_Soft_Interference_Cancellation}, the model-based SIC method suffers from the issue that the computation of the inverse noise covariance matrix $\m{C}_{\ve{v}\ve{v},k}^{(q)^{-1}}$, also known as precision matrix, is computationally and numerically demanding while approximations have to be made in addition. With our proposed NN equalizers SICNNv1 and SICNNv2, whose layer structures are inspired by the iterative SIC method, we aim to overcome the weaknesses of the model-based SIC method. However, the SIC operation from the model-based method is preserved in the developed NNs, which is expected to help SICNNv1 and SICNNv2 to provide reliable soft estimates (required, e.g., to compute log-likelihood ratios in case of channel coded transmission), and allows to obtain interpretable intermediate quantities / variables inside the NN-based equalizers.
While the structure of SICNNv1 is very similar to the model-based method and is specifically designed for SC-FDE communication systems, SICNNv2 is more general and can be applied for any communication system, where the system model can be formulated as in~\eqref{eq:system_model_SC-FDE} with any matrix $\m{H}$.
For both SICNNv1 and SICNNv2, we additionally present a version with a reduced number of learnable parameters, since we exploit the knowledge that every stage of SICNNv1/SICNNv2 has to provide estimates of the posterior data symbol probabilities, given the estimates of the previous stage and the received vector, and thus should work with the same set of learnable parameters. The parameter-reduced versions are referred to as SICNNv1Red and SICNNv2Red.
Besides, in this section we also briefly review a normalization scheme of the input data, which is required for a satisfying performance of NN-based data estimators in SC-FDE systems. 

\subsection{SICNNv1}
\begin{figure*}[t]
\centering
\includegraphics[width=\textwidth]{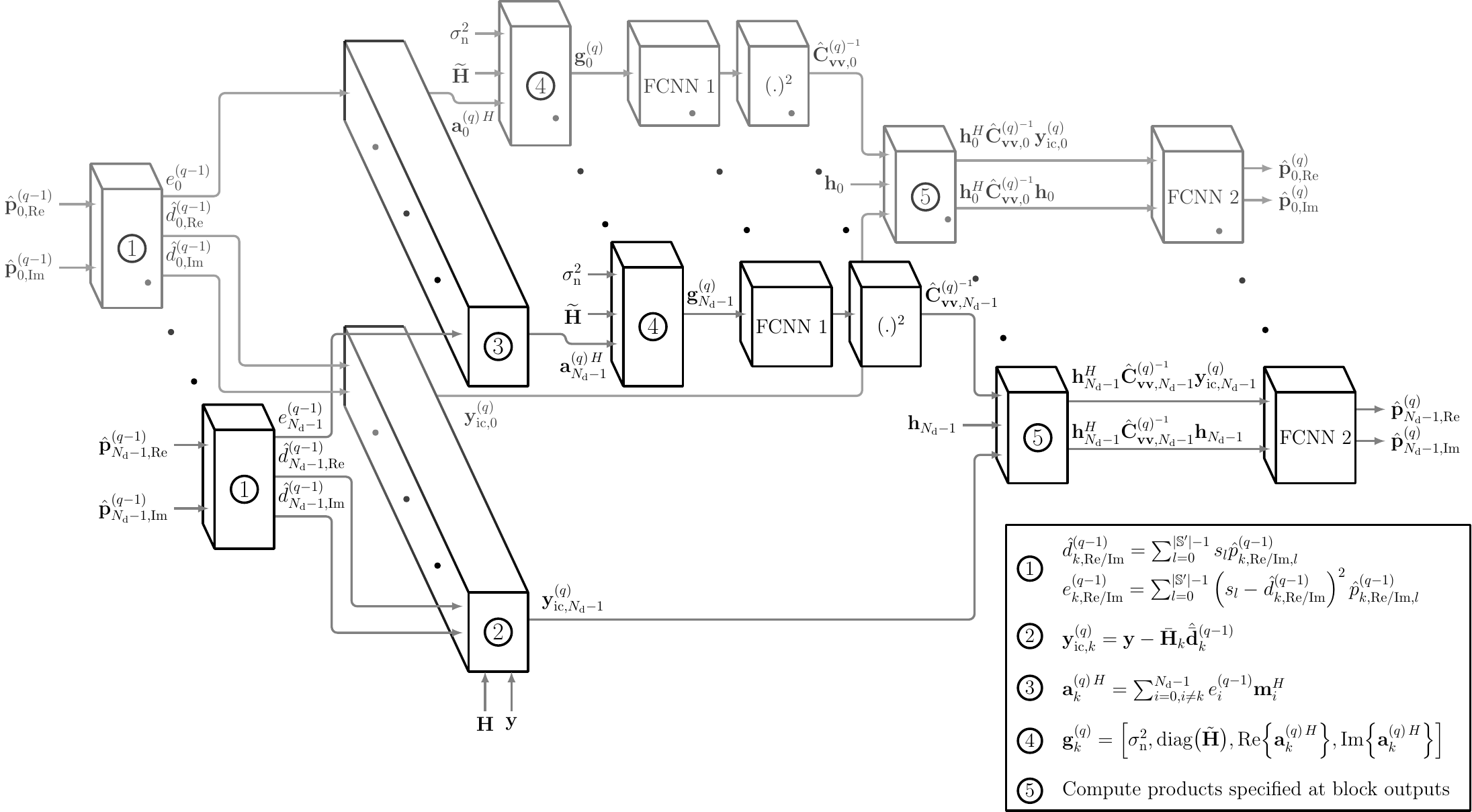}
\caption{Schematic structure of one stage of SICNNv1.}
\vspace{-0.3cm}
\label{fig:Layer_structure_SICNNv1}
\end{figure*}

The NN architecture is deduced by deep unfolding~\cite{Hershey14} the iterative SIC method described in Sec.~\ref{ssec:Iterative_Soft_Interference_Cancellation} to $Q$ stages of SICNNv1. That is, every iteration of the model-based SIC method (the outer loop of Algorithm~\ref{alg:Model-based_SIC_SC_FDE_Algorithm}), corresponds to one stage of SICNNv1. The steps conducted in one stage of SICNNv1, which is schematically shown in Fig.~\ref{fig:Layer_structure_SICNNv1}, are very similar to those of the model-based method described in Algorithm~\ref{alg:Model-based_SIC_SC_FDE_Algorithm}, however, the model-based computations in line~\ref{alg_line:Cvv_est} and line~\ref{alg_line:posterior_pmf} of Algorithm~\ref{alg:Model-based_SIC_SC_FDE_Algorithm}, are accomplished using data-driven FCNNs. Let us describe the structure of stage $q$ of SICNNv1, ${q\in\{0, ..., Q-1\}}$, in more detail, starting at its input. The inputs of stage $q$ are the received vector $\ve{y}$, the sampled frequency response $\text{diag}(\widetilde{\m{H}})$, the noise variance $\sigma_{\text{n}}^2$, and the vectors $\hat{\ve{p}}_k^{(q-1)} = \big[\hat{\ve{p}}_{k,\text{Re}}^{(q-1)\,T},\,\hat{\ve{p}}_{k,\text{Im}}^{(q-1)\,T}\big]^T$, $\hat{\ve{p}}_k^{(q-1)}\in [0,1]^{2|\mathbb{S}^\prime|}$, $k\in\{0, ..., N_{\text{d}}-1\}$, where $\mathbb{S}^\prime = \text{Re}\{\mathbb{S}\} = \text{Im}\{\mathbb{S}\}$ (assuming a symmetric alphabet~$\mathbb{S}$). The elements $\hat{p}_{k,\text{Re},l}^{(q-1)}$ and $\hat{p}_{k,\text{Im},l}^{(q-1)}$ of the vectors $\hat{\ve{p}}_{k,\text{Re}}^{(q-1)}$ and $\hat{\ve{p}}_{k,\text{Im}}^{(q-1)}$, respectively, are the estimates of stage $(q-1)$ for the data symbol posterior probabilities $p\big[\text{Re}\{d_k\}= s_l|\ve{y}_{\text{ic},k}^{(q-1)}\big]$ and $p\big[\text{Im}\{d_k\}= s_l|\ve{y}_{\text{ic},k}^{(q-1)}\big]$, respectively, where $s_l\in\mathbb{S}^\prime$ are the uniquely numbered symbols of $\mathbb{S}^\prime$, $l\in\{0, ..., |\mathbb{S}^\prime|-1\}$. For the inputs of the first stage ($q=0$), $\hat{\ve{p}}_{k,\text{Re}}^{(-1)} = \hat{\ve{p}}_{k,\text{Im}}^{(-1)} = \frac{1}{|\mathbb{S}^\prime|}\ve{1}$ is chosen, i.e., the estimated posterior probabilities are initialized uniformly, which represents the prior data symbol probability distribution. The values of the elements of $\hat{\ve{p}}_{k,\text{Re}}^{(q)}$ and $\hat{\ve{p}}_{k,\text{Im}}^{(q)}$ are updated in every stage following the procedure described below.

Similar to the model-based method, in the first step ($\circled{1}$ in Fig.~\ref{fig:Layer_structure_SICNNv1}) of the $q$th stage
\begin{gather}
\hat{d}_{k,\text{Re}}^{(q-1)}=\sum_{l=0}^{|\mathbb{S}^\prime|-1}s_l\hat{p}_{k,\text{Re},l}^{(q-1)}\,,\qquad \hat{d}_{k,\text{Im}}^{(q-1)}=\sum_{l=0}^{|\mathbb{S}^\prime|-1}s_l\hat{p}_{k,\text{Im},l}^{(q-1)}\,,\label{eq:d_k_estimate_SICNN}
\end{gather} 
and
\begin{gather}
e_k^{(q-1)} = \sqrt{\big(e_{k,\text{Re}}^{(q-1)}\big)^2 + \big(e_{k,\text{Im}}^{(q-1)}\big)^2}\,,
\end{gather}
with the estimated MSEs
\begin{gather}
e_{k,\text{Re/Im}}^{(q-1)} = \sum_{l=0}^{|\mathbb{S}^\prime|-1}\big(s_l - \hat{d}_{k,\text{Re/Im}}^{(q-1)}\big)^2\hat{p}_{k,\text{Re/Im},l}^{(q-1)}\,,\label{eq:error_k_re_im_estimate_SICNN}
\end{gather}
are computed, where $e_k^{(q-1)}$ is utilized as a reliability measure of the corresponding data symbol estimate. Note that these quantities can be computed independently for every data symbol index~$k$, and thus the blocks~$\circled{1}$ are drawn isolated from each other in Fig.~\ref{fig:Layer_structure_SICNNv1}. 

In a second step ($\circled{2}$ in Fig.~\ref{fig:Layer_structure_SICNNv1}), model-based interference cancellation is carried out by computing
\begin{gather}
\ve{y}_{\text{ic},k}^{(q)} = \ve{y} - \bar{\m{H}}_k\hat{\bar{\ve{d}}}_k^{(q-1)}\,,\label{eq:interference_cancellation_SICNN}
\end{gather}
using the data symbol estimates $\hat{\bar{d}}_{k,l}^{(q-1)} = \hat{d}_{l,\text{Re}}^{(q-1)} + j\hat{d}_{l,\text{Im}}^{(q-1)}$, $l\in\{0, ..., k-1, k+1, ..., N_{\text{d}}-1\}$ from stage $(q-1)$ for~$\hat{\bar{\ve{d}}}_k^{(q-1)}$. 

Instead of computing (an approximate of) the precision matrix $\m{C}_{\ve{v}\ve{v},k}^{(q)^{-1}}$ in a model-based fashion, we estimate it by utilizing fully-connected feedforward layers, which we also refer to as sub-NNs. A straightforward approach is to estimate the real and the imaginary part of the $N^2$ elements of $\m{C}_{\ve{v}\ve{v},k}^{(q)^{-1}}$. We exploit two observations for both reducing the number of parameters to be estimated by a sub-NN and ensuring that the estimated precision matrix satisfies the properties implied by its definition. Firstly, the covariance matrix $\m{C}_{\ve{v}\ve{v},k}^{(q)}$, and thus also its inverse, has to be a Hermitian, positive definite matrix. That is, $\m{C}_{\ve{v}\ve{v},k}^{(q)^{-1}}$ can be decomposed into the matrix product $\m{C}_{\ve{v}\ve{v},k}^{(q)^{-1}} = \m{B}_k^{(q)\,H}\m{B}_k^{(q)}$, where $\m{B}_k^{(q)}$ is the matrix to be estimated by the sub-NNs. Secondly, our empirical investigations showed, that for the regarded SC-FDE communication system a precision matrix $\m{C}_{\ve{v}\ve{v},k}^{(q)^{-1}}$ exhibits significant non-zero values only on the major and the first few minor diagonals, and thus can be approximated as a band matrix. In the initial version of SICNNv1 described in~\cite{Baumgartner22_C1} (where it is simply referred to as SICNN), we specify $\m{B}_k^{(q)}$ to be a lower triangular matrix containing non-zero values only on the main diagonal and the first $n_{\text{md}}$ minor diagonals, where $n_{\text{md}}\in\mathbb{N}_0$ is a hyperparameter of SICNN, which tremendously reduces the number of non-zero elements of $\m{B}_{k}^{(q)}$ to be estimated. The non-zero elements of the complex-valued matrix $\m{B}_{k}^{(q)}$ are to be estimated by two separate sub-NNs, using one sub-NN for the real part and one for the imaginary part. As stated in~\cite{Baumgartner22_C1}, hyperparameter optimization turns out that the best equalization performance is achieved with $n_{\text{md}} = 0$, i.e., the precision matrix is assumed to be a diagonal matrix. This insight motivates an adaption of the architecture of SICNN for the structure of SICNNv1. As depicted in Fig.~\ref{fig:Layer_structure_SICNNv1}, a single sub-NN FCNN~1 is employed to estimate the major diagonal of $\m{C}_{\ve{v}\ve{v},k}^{(q)^{-1}}$. To ensure positive definiteness, the final estimates of the major diagonal of $\m{C}_{\ve{v}\ve{v},k}^{(q)^{-1}}$ are obtained by squaring the outputs of FCNN~1. FCNN~1 has $n_{\text{L,C}}$ hidden layers with $n_{\text{H,C}}$  neurons per hidden layer, ReLU activation functions, and a batch norm layer after the input layer. 
For determining the required inputs of FCNN~1, we reconsider the computation of $\m{C}_{\ve{v}\ve{v},k}^{(q)}$ in~\eqref{eq:Cvv_SIC} and the quantities involved there. Besides the terms describing correlations between $\ve{w}$ and $\hat{\bar{\ve{d}}}_k^{(q-1)}$, only terms consisting of $\sigma_{\text{n}}^2$, $\widetilde{\m{H}}$, and $\bar{\m{M}}_k\widetilde{\m{E}}_k^{(q-1)}\bar{\m{M}}_k^H$ occur in~\eqref{eq:Cvv_SIC}. When replacing $\widetilde{\m{E}}_k^{(q-1)}$ by its approximation $\m{E}_k^{(q-1)}$, the latter term can be expressed as $\bar{\m{M}}_k\m{E}_k^{(q-1)}\bar{\m{M}}_k^H = \sum_{i=0, i\neq k}^{N_{\text{d}}-1} e_i^{(q-1)}\ve{m}_i\ve{m}_i^H =: \m{A}_k^{(q)}$, with $\ve{m}_i$ as the $i$th column of $\m{M}$. $\m{A}_k^{(q)}$ is a Hermitian Toeplitz matrix, consequently it is already fully described by its first row $\ve{a}_k^{(q)\,H} = \big[\m{A}_k^{(q)}\big]_{0*} = \sum_{i=0,i\neq k}^{N_{\text{d}}-1}e_i^{(q-1)}\ve{m}_i^{H}$, where we exploit $m_{i,0} = 1$ for all~$i$ in the last step. The vectors $\ve{a}_k^{(q)\,H}$, which are computed in block~$\circled{3}$, are concatenated in block~$\circled{4}$ with $\sigma_{\text{n}}^2$ and $\text{diag}\big(\widetilde{\m{H}}\big)$ to the input vector of FCNN~1
\begin{gather}
\ve{g}_k^{(q)} = \big[\sigma_{\text{n}}^2,\,\text{diag}\big(\widetilde{\m{H}}\big),\,\text{Re}\big\{\ve{a}_k^{(q)\,H}\big\},\,\text{Im}\big\{\ve{a}_k^{(q)\,H}\big\}\big]\,.
\end{gather}

With a given estimated precision matrix $\hat{\m{C}}_{\ve{v}\ve{v},k}^{(q)^{-1}}$, the posterior PMF $p\big[d_k|\ve{y}_{\text{ic},k}^{(q)}\big]$ should be estimated. Experiments, where the posterior PMF is computed as described in Sec.~\ref{ssec:Iterative_Soft_Interference_Cancellation} in a model-based fashion did not lead to satisfying performance. We assume a major reason for this issue is that the estimate $\hat{\m{C}}_{\ve{v}\ve{v},k}^{(q)^{-1}}$ provided by FCNN1 is not precise enough to be treated as the exact precision matrix. Hence, we utilize another sub-NN (FCNN~2 in Fig.~\ref{fig:Layer_structure_SICNNv1}), which is trained (jointly with FCNN1) to estimate the posterior PMF $p\big[d_k|\ve{y}_{\text{ic},k}^{(q)}\big]$ and can cope with inaccuracies in the estimated precision matrix. More specifically, the output of FCNN2 is the vector $\hat{\ve{p}}_k^{(q)} = \big[\hat{\ve{p}}_{k,\text{Re}}^{(q)\,T},\,\hat{\ve{p}}_{k,\text{Im}}^{(q)\,T}\big]^T$ (which is also the output of the $q$th SICNNv1 stage), containing estimates for the data symbol posterior probabilities, as introduced earlier in this section. To specify the required input quantities of FCNN~2 for estimating the posterior PMF $p\big[d_k|\ve{y}_{\text{ic},k}^{(q)}\big]$, let us consider its representation
\begin{gather}
p\big[d_k|\ve{y}_{\text{ic},k}^{(q)}\big] = \frac{p(\ve{y}_{\text{ic},k}^{(q)}|d_k)}{\sum_{s^\prime\in\mathbb{S}}p(\ve{y}_{\text{ic},k}^{(q)}|s^\prime)} = \frac{f_k^{(q)}(d_k)}{\sum_{s^\prime\in\mathbb{S}}f_k^{(q)}(s^\prime)}\,,
\end{gather} 
by applying the Bayesian rule and assuming a uniform prior data symbol probability, where $f_k^{(q)}(.)$ is defined in~\eqref{eq:posterior_prob_f_k}. Due to the definition of $f_k^{(q)}(.)$, the posterior PMF $p\big[d_k|\ve{y}_{\text{ic},k}^{(q)}\big]$ depends only on $\ve{h}_k^{H}\m{C}_{\ve{v}\ve{v},k}^{(q)^{-1}}\ve{y}_{\text{ic},k}^{(q)}$, its complex conjugate, and $\ve{h}_k^H\m{C}_{\ve{v}\ve{v},k}^{(q)^{-1}}\ve{h}_k$. Therefore, the input vector of FCNN~2 is chosen to be
\begin{gather}
\begin{aligned}
\ve{s}_k^{(q)} = \big[&\text{Re}\big\{\ve{h}_k^H\hat{\m{C}}_{\ve{v}\ve{v},k}^{(q)^{-1}}\ve{y}_{\text{ic},k}^{(q)}\big\},\,\text{Im}\big\{\ve{h}_k^H\hat{\m{C}}_{\ve{v}\ve{v},k}^{(q)^{-1}}\ve{y}_{\text{ic},k}^{(q)}\big\},\,\\&\quad\ve{h}_k^H\hat{\m{C}}_{\ve{v}\ve{v},k}^{(q)^{-1}}\ve{h}_k\big]^T\,,
\end{aligned}
\label{eq:input_FCNN2}
\end{gather}
where the elements of $\ve{s}_k^{(q)}$ are computed in block~$\circled{5}$. FCNN~2 consists of $n_{\text{L,Pr}}$ fully-connected hidden layers with $n_{\text{H,Pr}}$  neurons per hidden layer, ReLU activation functions between each hidden layer, and a batch norm layer after the input layer. Further, two independent softmax functions are utilized as output activation functions to obtain $\hat{\ve{p}}_{k,\text{Re}}^{(q)}$ and $\hat{\ve{p}}_{k,\text{Im}}^{(q)}$, which are concatenated to the stage output $\hat{\ve{p}}_k^{(q)}$. 

An investigation of the inputs of sub-NN FCNN~2 as defined in~\eqref{eq:input_FCNN2} reveals a large variation of the values of $\ve{s}_k^{(q)}$, which depends on both the data symbol index~$k$ and the SICNNv1 stage index~$q$. Hence, in the $q$th SICNNv1 stage we suggest to multiply $\ve{y}_{\text{ic},k}^{(q)}$ and $\ve{h}_k$  by $||\ve{y}_{\text{ic},k}^{(q)}||_2^{-\nicefrac{1}{2}}$, which turns out to lead to a more robust training procedure. 

In one stage, we use the same sub-NNs for estimating the precision matrix and the posterior data symbol probabilities for all $N_{\text{d}}$ data symbols to be estimated. However, different sub-NNs are utilized from stage to stage, i.e. their learnable parameters are in general different to those of the sub-NNs of the remaining stages, whereby the hyperparameters of the sub-NNs are the same for all stages.

We optimize the parameters of SICNNv1 by employing a custom loss function based on the cross entropy loss, which can be computed as
\begin{gather*}
f_{\text{CE}}(\bm{\alpha},\bm{\beta}) := -\frac{1}{D}\sum_{d=0}^{D-1}\alpha_d\ln(\beta_d)\,,
\end{gather*}
with any vectors $\bm{\alpha}, \bm{\beta}\in [0, 1]^{D}$ and ${\sum_d \alpha_d = \sum_d \beta_d = 1}$. More specifically, instead of utilizing only the final output of SICNNv1 for computing the loss value, the custom loss is based on $Q$ partial cross entropy losses of all $Q$ stage outputs, which are weighted by the corresponding stage index $q$. That is, the employed custom loss function is given by
\begin{gather}
\begin{aligned}
\ell(\ve{d}_{\text{oh}},\hat{\mathcal{P}}) &= \frac{1}{Q N_{\text{d}}}\sum_{q=0}^{Q-1}\sum_{k=0}^{N_{\text{d}}-1}w_q\big(f_{\text{CE}}(\ve{d}_{\text{oh},k,\text{Re}},\hat{\ve{p}}_{k,\text{Re}}^{(q)})
\\&\qquad+f_{\text{CE}}(\ve{d}_{\text{oh},k,\text{Im}},\hat{\ve{p}}_{k,\text{Im}}^{(q)})\big)\,,
\end{aligned}
\label{eq:loss_function_SICNN}
\end{gather}
where $\hat{\mathcal{P}}:=\big\{\{\hat{\ve{p}}_{k,\text{Re/Im}}^{(q)}\}_{k=0}^{N_{\text{d}}-1}\big\}_{q=0}^{Q-1}$ is the collection of all stage outputs, $w_q=(q+1)/\sum_{q=1}^{Q^-1}q$ is the weighting factor of the partial losses, and $\ve{d}_{\text{oh},k,\text{Re}}$ and $\ve{d}_{\text{oh},k,\text{Im}}$ are one-hot vectors corresponding to the real and imaginary part of a data symbol $d_k$, respectively. This custom loss function is inspired by the loss function employed for training DetNet~\cite{Samuel19} and by the auxiliary classifiers of GoogLeNet~\cite{Szegedy15}, and should lead to a faster converging training.

\subsection{SICNNv2}
\label{ssec:SICNNv2}
\begin{figure*}[t]
\centering
\includegraphics[width=0.83\textwidth]{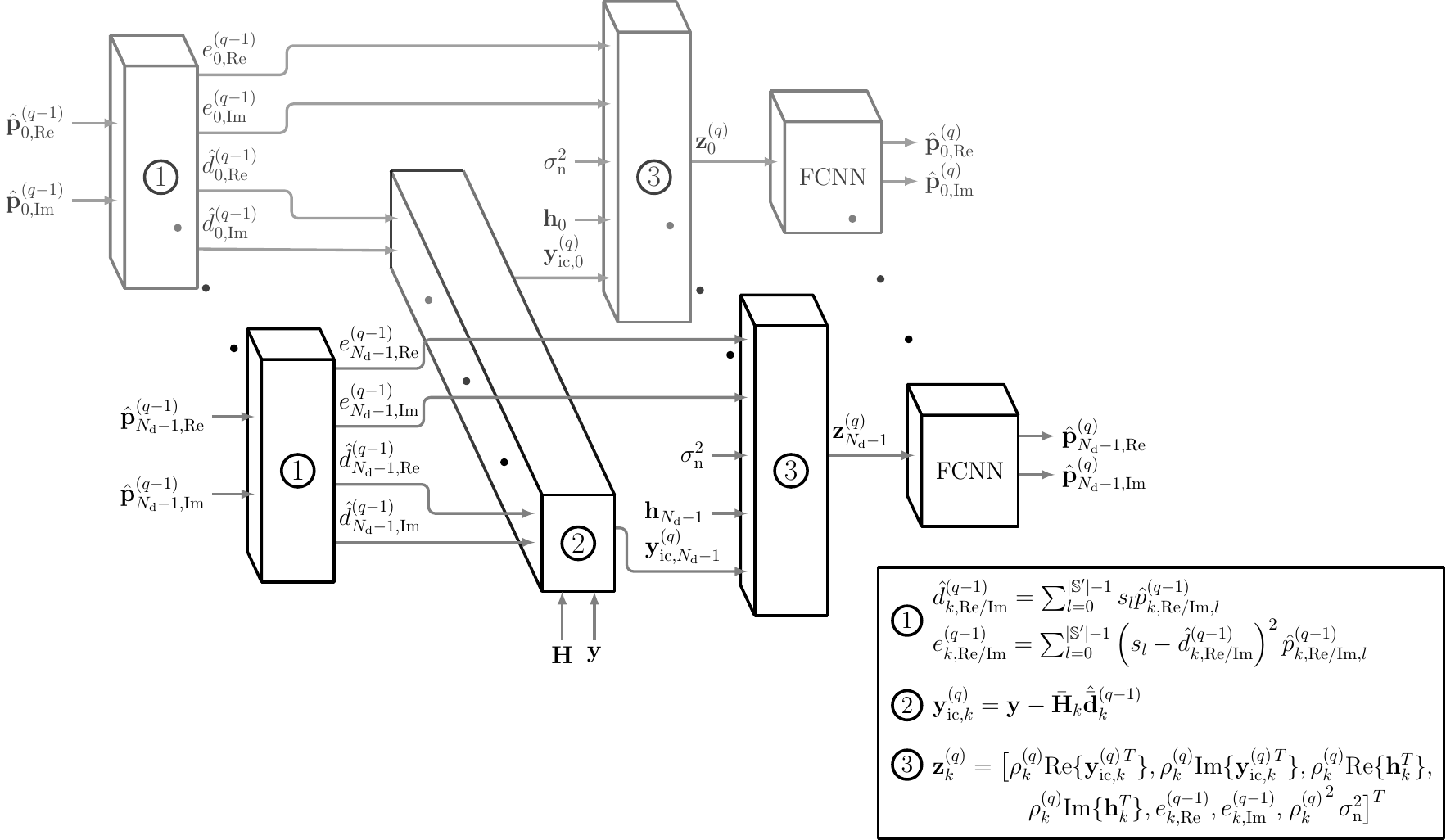}
\caption{Schematic structure of one stage of SICNNv2.}
\vspace{-0.3cm}
\label{fig:Layer_structure_SICNNv2}
\end{figure*}

The first operations conducted in a stage of SICNNv2 are equivalent to those in a stage of SICNNv1. That is, given the estimated posterior probabilities $\hat{\ve{p}}_k^{(q-1)}$ from the previous stage $(q-1)$, the corresponding data symbol estimates $\hat{d}_{k,\text{Re/Im}}^{(q-1)}$ and estimated MSEs $e_{k,\text{Re/Im}}^{(q-1)}$ are computed according to~\eqref{eq:d_k_estimate_SICNN} and \eqref{eq:error_k_re_im_estimate_SICNN}, respectively. With the data symbol estimates $\hat{d}_{k,\text{Re/Im}}^{(q-1)}$ for all data symbols in the data vector at hand, interference cancellation is conducted as a next step according to~\eqref{eq:interference_cancellation_SICNN} to obtain $\ve{y}_{\text{ic},k}^{(q)}$. However, as shown in Fig.~\ref{fig:Layer_structure_SICNNv2}, the remaining structure of a stage of SICNNv2 differs from the stage structure of SICNNv1. Specifically, while the further inference steps conducted in a stage of SICNNv1 to obtain the stage output are similar to the steps of the model-based algorithm, in an SICNNv2 stage an FCNN  is employed for directly estimating the posterior data symbol probabilities using the input vector
\begin{gather*}
\begin{aligned}
\ve{z}_k^{(q)} =& \big[\rho_k^{(q)}\text{Re}\{\ve{y}_{\text{ic},k}^{(q)\,T}\}, \rho_k^{(q)}\text{Im}\{\ve{y}_{\text{ic},k}^{(q)\,T}\}, \rho_k^{(q)}\text{Re}\{\ve{h}_k^T\}, \\&\quad \rho_k^{(q)}\text{Im}\{\ve{h}_k^T\}, e_{k,\text{Re}}^{(q-1)}, e_{k,\text{Im}}^{(q-1)}, \left.\rho_k^{(q)}\right.^2\sigma_{\text{n}}^2\big]^T\,,
\end{aligned}
\end{gather*}
with a normalization factor $\rho_k^{(q)} = ||\ve{y}_{\text{ic},k}^{(q)}||_2^{-\nicefrac{1}{2}}$. The estimated posterior data symbol probabilites are contained in the output vector $\hat{\ve{p}}_k^{(q)} = \big[\hat{\ve{p}}_{k,\text{Re}}^{(q)\,T}, \hat{\ve{p}}_{k,\text{Im}}^{(q)\,T}\big]^T$ of the FCNN, which is also the output of the $q$th stage of SICNNv2. The FCNN has $n_{\text{L}}$ hidden layers, $n_{\text{H}}$  neurons per hidden layer, a batch norm layer after the input layer and every third hidden layer, and ReLU activation. SICNNv2 is trained with the same loss function~\eqref{eq:loss_function_SICNN} as SICNNv1. 

The architecture of SICNNv2 is solely based on the idea of SIC, and is more universal than that of SICNNv1 since no properties of an SC-FDE system are utilized. Hence, SICNNv2 can be employed for equalization in other communication systems like, e.g., general MIMO systems or UW-OFDM systems. However, for SC-FDE we expect a higher equalization complexity and maybe a worse BER performance of SICNNv2 compared to SICNNv1 as less model knowledge is incorporated.

\subsection{Parameter Reduction: SICNNv1Red and SICNNv2Red}
For reducing the number of parameters to be trained, we exploit the fact that in every stage of SICNNv1 and SICNNv2 the same task is fulfilled, namely, the posterior data symbol probabilities are to be estimated given the estimated posterior data symbol probabilities from the previous stage, the received vector, the channel matrix, and the noise variance. Hence, we employ the same sub-NNs in every stage of the NN-based equalizers, leading to the corresponding parameter-reduced versions SICNNv1Red and SICNNv2Red. These parameter-reduced NNs can also be viewed as a single stage where its output is fed back $Q$ times. 

This NN architecture distinctly reduces the number of parameters to be optimized, which reduces the computational effort for training, and is also supposed to lead to a more robust training procedure and a smaller amount of training data to be required. However, it turns out that the employed loss function for training the parameter-reduced NNs has to be slightly altered for obtaining a good performance. More specifically, the outputs of the stages with higher stage index $q$ are given a higher importance by changing the weights $w_q$ of the loss function~\eqref{eq:loss_function_SICNN} for training SICNNv1Red and SICNNv2Red to
\begin{gather}
w_q = \frac{(q+1)^r}{\sum_{q=1}^{Q-1} q^r}\,,
\label{eq:weights_loss_function_SICNNRed}
\end{gather}
where $r$ is a hyperparameter, which we choose for SC-FDE systems to be $r=4$.

\section{Training Set Generation and Data Normalization}
\label{sec:Training_Set_Generation_and_Data_Normalization}

In this section, we describe a novel approach for generating training sets for NN-based equalizers. For the regarded SC-FDE systems, this approach considerably improves the performance of NN-based equalizers at high SNRs. Further, we briefly describe a data normalization scheme specifically tailored for SC-FDE, which was already presented in~\cite{Baumgartner23_C1}.

\subsection{Training Set Generation}
\label{ssec:Training_Set_Generation}
The achieved BER performances of model-based and NN-based equalizers are generally evaluated in a specified $E_{\text{b}}/N_0$ interval, where $E_{\text{b}}$ is the mean bit energy and $N_0$ is the noise power spectral density on receiver side, i.e., $E_{\text{b}}/N_0$ is a measure for the SNR. To generate the training set for NN-based equalizers, typically sample data transmissions over channel realizations\footnote{These channel realizations are not used for evaluation, but the same statistical channel model is used to generate channels used for evaluation.} drawn from a statistical channel model are conducted, where $E_{\text{b}}/N_0$ for the data transmission is selected randomly with a uniform distribution within a specified range, short training SNR range. The upper and lower limits of the training SNR range are typically hyperparameters, which are to be selected carefully, since they have a significant influence on the performance of trained NN-based equalizers~\cite{Hao23, Baumgartner23_J1, Wiesmayr23}. Despite a careful selection of the training SNR range, the issue of ``flattening out'' BER curves can occur. That is, although NN-based equalizers perform well for a wide $E_{\text{b}}/N_0$ range, at higher $E_{\text{b}}/N_0$ values, and thus low BERs (of, e.g., $10^{-5}$ or $10^{-6}$), their BER curves do not fall as steeply as those of many model-based equalizers. Although this issue occurs for most NN-based equalizers (shown, e.g., in~\cite{Goutay20, Baumgartner23_C1}), interestingly, there are only very few works like~\cite{Wiesmayr23}, where proper training of NN-based equalizers is addressed. In this work, we propose a novel approach for the generation of training sets for NN-based equalizers. By training NN-based data estimators with these specifically generated training sets, the issue of flattening out BER curves at high SNRs can be mitigated significantly. 

Typically, even low complex equalizers like the LMMSE equalizer achieve low BERs at high SNRs (as it can be seen, e.g., in Fig.~\ref{fig:BER_performance_equalizers}, where the LMMSE estimator achieves a BER of $5\cdot 10^{-5}$ at $E_{\text{b}}/N_0 = 14\,\si{dB}$). In other words, the decision boundaries of the baseline LMMSE equalizer and the optimal bit-wise MAP equalizer differ only slightly. That is, by randomly generating data transmissions for the training set at high SNRs, for those data transmissions even with the baseline LMMSE estimator only very few data symbol estimation errors occur. However, the NN-based equalizers are expected to approximate the optimal bit-wise MAP estimator. Hence, for NN training, exactly those few received data symbols are of interest, where with the baseline LMMSE estimator a wrong estimate is obtained for the corresponding transmitted data symbol, while the optimal estimator still achieves a correct data symbol estimate. Since only a few of those important received data symbols are contained in the training set when generating the training data randomly, their influence on the training loss of the NN is small, leading to the aforementioned issue of flattening out BER curves. With the aforementioned observations in mind, we suggest the following method for generating the training set of NN-based equalizers for SC-FDE systems: instead of randomly selecting an SNR value within the SNR training range for the transmission of every data burst that is contained in the training set, we define an evenly spread grid on the SNR training range (on a linear scale). The number of grid points coincides with the number of channels over which data transmissions are to be conducted to generate the training set. For every SNR grid point, a channel realization is drawn from the assumed statistical channel model, and a burst of $N_{\text{burst}}$ data vectors $\ve{d}$ is transmitted over this channel. The corresponding received vectors $\ve{y}$ are equalized using a baseline LMMSE estimator. Instead of including all data vectors of the transmitted burst in the training set, only those are retained where the baseline equalizer produces at least $N_{\text{epd}}$ errors per data vector. Since particularly at higher SNRs the number of retained data vectors is generally far lower than $N_{\text{burst}}$, another burst of data vectors is generated and transmitted over the same channel, again followed by keeping only the data vectors where at least $N_{\text{epd}}$ errors per data vector are produced by the baseline estimator. This procedure is repeated until $N_{\text{burst}}$ data vectors are found for the specific channel, which are then included in the training set. However, for flat channels, even with the baseline equalizer no or too few errors occur such that no data vectors are found which could be included in the training set. Therefore, a stopping criterion has to be introduced, where after $N_{\text{check}}$ burst generations the number of retained data vectors is checked. If the the number of retained data vectors is smaller than, e.g., ${0.1 N_{\text{burst}}}$, the current channel realization is discarded. While keeping the SNR value corresponding to the specified SNR grid point, a new channel realization is drawn from the statistical channel model, and the same data vector selection process as described above is carried out. The parameters for the training set generation depend on the communication system setup for which the NN-based equalizers are trained, and thus they are specified in Sec.~\ref{ssec:Simulation_Setup_and_NN_Training} individually for every setup.

\subsection{Data Normalization}
Normalizing the input data of NNs is generally considered to be very important for training convergence when optimizing their learnable parameters via backpropagation~\cite{LeCun12, Bishop95}. While in many current publications on NN-based data estimation in MIMO systems over uncorrelated Rayleigh fading channels no normalization of the NN input data is applied (cf., e.g.,~\cite{Samuel19, He20, Pratik21}), we showed in~\cite{Baumgartner21_C1} and~\cite{Baumgartner23_J1} that for a so-called UW-OFDM communication system a proper data normalization is of   major importance for the performance of NN-based equalizers (for a visualization of the influence of data normalization on the performance of NN-based equalizers for UW-OFDM, we refer to~\cite{Baumgartner21_C1}).  With the same idea as for UW-OFDM in mind, namely to apply a normalization scheme leading to variances of the elements of the noise vector that are independent of the multipath channel, we implement a data normalization scheme for SC-FDE systems. This data normalization scheme for SC-FDE is elucidated in~\cite{Baumgartner23_C1}, and thus we repeat here only the result.  
To obtain channel-independent noise variances $\text{var}(w_i)$, the system model~\eqref{eq:system_model_SC-FDE} has to be multiplied by $\m{K} = \kappa\widetilde{\m{H}}^{-\nicefrac{1}{2}}$, where
\begin{gather}
\kappa = \sqrt{\text{tr}\{\widetilde{\m{H}}\}/\text{tr}\{\widetilde{\m{H}}\m{M}\m{M}^H\widetilde{\m{H}}\}}\,.
\end{gather}
The normalization of the input data of the NN-based equalizers is implemented by multiplying both $\ve{y}$ and $\widetilde{\m{H}}$ by $\m{K}$ as part of pre-processing, and is neglected in the remainder of this paper for the sake of readability. 

\section{Results}
\label{sec:Results}
In this section, we investigate the proposed versions of SICNN thoroughly by means of simulations of data transmission in an indoor frequency selective environment. To demonstrate the wide applicability the proposed NN-based approaches, we evaluate them for a number of different SC-FDE communication system setups. We show simulation results for SC-FDE systems with both UW and CP guard intervals. Besides simulations with a QPSK modulation alphabet, also results for a 16-QAM alphabet are provided. Most of the simulations are conducted assuming perfect channel knowledge on receiver side. The robustness of the proposed NN-based data estimators in case of imperfect channel knowledge, is proven by simulating their performance for estimated channel impulse responses. Further, we highlight the performance improvements of NN-based equalizers when being trained on a training set generated with our proposed approach, presented in Sec.~\ref{ssec:Training_Set_Generation}.
We compare SICNNv1 and SICNNv2, and their corresponding parameter-reduced versions SICNNv1Red and SICNNv2Red, with state-of-the-art model-based and NN-based equalizers in terms of both their achieved BER performance over a specified SNR range and their computational complexity. More specifically, for comparison with model-based equalizers, we use the LMMSE estimator~\cite{Huemer10}, the iterative DFE (implemented in the same way as described in~\cite{Baumgartner21_C1} for UW-OFDM systems), and the iterative SIC method, where the approximation~\eqref{eq:C_vv_SIC_SC-FDE_approx} is employed. We compare the proposed NNs with the state-of-the-art NN-based data estimators OAMP-Net2~\cite{He20} and DetNet~\cite{Samuel19}, whereby we do not use DetNet as proposed in~\cite{Samuel19} for MIMO systems, but a better performing version that is adapted for SC-FDE systems~\cite{Baumgartner23_C1}. Moreover, we show the BER performance and computational complexity of KAFCNN from~\cite{Baumgartner23_C1}, which is an FCNN that is designed for equalization in SC-FDE systems by using a layer conducting an inverse DFT as a last layer, i.e., the knowledge that the data symbols being defined in time domain are to be estimated given a received vector in frequency domain is incorporated.

Moreover, for an SC-FDE system with a UW guard interval we present the influence of a limited training set size on the BER performance of selected NN-based equalizers to investigate the ``data hunger'' of an NN depending on its number of learnable parameters. 

Finally, we also present performance results for SICNNv2 being employed as an equalizer in communication system utilizing the so-called UW-OFDM signaling scheme. With these results we want the highlight the wide applicability and the versatility of the proposed NN-based equalizers.

\subsection{Simulation Setup and Neural Network Training}
\label{ssec:Simulation_Setup_and_NN_Training}

The shown simulation results are obtained by simulating data transmission without channel coding. Apart from Sec.~\ref{ssec:BER_performance_SICNNv2_UW-OFDM}, all simulation settings, results, interpretations, and conclusions in this work are given for SC-FDE systems. The simulation setup and the results for the simulation of SICNNv2 as an equalizer in a UW-OFDM system are detailed in Sec.~\ref{ssec:BER_performance_SICNNv2_UW-OFDM}. For SC-FDE communications with a UW guard interval, simulations are conducted with the SC-FDE system parameters $N_{\text{d}}=20$, $N_{\text{g}}=12$ (i.e., $N=32$), RRC roll-off factor $\alpha = 0.25$, a baseband sampling time $T_{\text{s}} = 52\,\si{ns}$. Further, unless noted otherwise, a QPSK modulation alphabet is employed, and perfect channel knowledge is assumed on receiver side. The parameters of the simulations of SC-FDE systems with a CP guard interval differ from those with a UW guard interval by the data vector length $N_{\text{d}}=32$, all other parameters are maintained.

The achieved BER performances of the different equalizers are plotted in a specified $E_{\text{b}}/N_0$ interval. The presented BER performances for SC-FDE systems are averaged results over $7000$ different multipath channel realizations, which are modeled as described in~\cite{Fakatselis97} in form of tapped delay lines with uniformly distributed phase, Rayleigh distributed magnitude, and an exponentially decaying power profile with a root mean square delay spread of $\tau_{\text{RMS}}=100\,\si{ns}$. The data transmission is conducted in form of data bursts containing $1000$ blocks of payload data per burst, where the channel is assumed to be stationary for one burst and changes independently of its previous realizations for every burst.

For training the NN-based equalizers, we generate training sets with the proposed approach described in Sec.~\ref{ssec:Training_Set_Generation}. The selected parameters of the training set generation method $N_{\text{epd}}$ and $N_{\text{burst}}$, the training $E_{\text{b}}/N_0$ range, as well as the employed baseline equalizer for selecting the data vectors for the training set are summarized for all SC-FDE system setups in Tab.~\ref{tab:Trainingset_generation_parameters}. Unless stated otherwise, every training set consists of data transmissions over $30000$ different channels. For training of all NNs early stopping is used. That is, the BER performance on a validation set is evaluated after every epoch, and the set of learnable NN parameters achieving the best validation performance is chosen after training for a pre-defined maximum number of epochs.

{
\renewcommand{\aboverulesep}{0pt}
\renewcommand{\belowrulesep}{0pt}
\renewcommand{\arraystretch}{1.3}
\begin{table}[t]
\caption{Parameters of the training set generation approach for different SC-FDE system settings.}
\label{tab:Trainingset_generation_parameters}
\begin{center}
\vspace{-0.2cm}
\begin{tabularx}{11cm}{>{\raggedright}m{3cm}  |>{\raggedright}m{2.5cm} |>{\raggedleft}X |>{\raggedleft}m{0.9cm} |  >{\raggedleft\arraybackslash}m{2cm}}
\toprule
\centering SC-FDE parameters & \centering Baseline equalizer & \centering $N_{\text{epd}}$ & \centering $N_{\text{burst}}$ & \centering $E_{\text{b}}/N_0$ train. range in dB\tabularnewline
\hline
UW guard, QPSK & LMMSE, eq.~\eqref{eq:LMMSE_v1} & $3$ & $100$ & $[2, 12.5]$\\
\hline
CP guard, QPSK & LMMSE, eq.~\eqref{eq:LMMSE_diag} & $2$ & $100$ & $[5, 18]$\\
\hline
UW guard, 16-QAM & LMMSE, eq.~\eqref{eq:LMMSE_v1} & $3$ & $100$ & $[6, 19]$ \\
\hline
UW guard, QPSK, channel est. & approx. LMMSE, eq.~\eqref{eq:LMMSE_diag} & $3$ & $100$ & $[3, 16]$ \\
\bottomrule
\end{tabularx}
\end{center}
\vspace{-0.3cm}
\end{table}
}

The hyperparameters of the NN-based equalizers are found using the hyperparameter optimization framework Optuna~\cite{Akiba19}. For SICNNv1, the best hyperparameter settings found are given in Tab.~\ref{tab:Hyperparameter_settings_SICNN}, and we train it for $25$ epochs at most. For SICNNv1Red, the same hyperparameters as for SICNNv1 are used apart from a learning rate $\eta_{\text{SICNNv1Red}}$. The hyperparameters of SICNNv2 are also given in Tab.~\ref{tab:Hyperparameter_settings_SICNN}, and we train it for a maximum of $25$ epochs. The hyerparameters of SICNNv2Red differ from those of SICNNv2 only by $\eta_{\text{SICNNv2Red}}$. For DetNet, OAMP-Net2, and KAFCNN the best hyperparameter found are shown in Tab.~\ref{tab:Hyperparameter_settings_SOTA_NNs}. Moreover, DetNet and KAFCNN are trained for $60$ epochs at most, and OAMP-Net2 for a maximum of $15$ epochs.

{
\renewcommand{\aboverulesep}{0pt}
\renewcommand{\belowrulesep}{0pt}
\renewcommand{\arraystretch}{1.3}
\begin{table*}[t]
\caption{Hyperparameter settings of the proposed NN equalizers. The hyperparameters of the parameter-reduced NNs SICNNv1Red and SICNNv2Red differ from the hyperparameters of SICNNv1 and SICNNv2 only by the learning rate $\eta_{\text{SICNNv1Red}}$ and $\eta_{\text{SICNNv2Red}}$, respectively.}
\label{tab:Hyperparameter_settings_SICNN}
\begin{center}
\vspace{-0.2cm}
\begin{tabularx}{\textwidth}{>{\raggedright}m{2.4cm}  |>{\raggedleft}m{1.15cm} >{\raggedleft}m{1.5cm} >{\raggedleft}X >{\raggedleft}X >{\raggedleft}X  >{\raggedleft}X >{\raggedleft}X |>{\raggedleft}m{1.15cm} >{\raggedleft}m{1.5cm} >{\raggedleft}X >{\raggedleft}X >{\raggedleft\arraybackslash}X}
\toprule
SC-FDE parameters & \multicolumn{7}{c|}{SICNNv1 / SICNNv1Red} &\multicolumn{5}{c}{SICNNv2 / SICNNv2Red}\\
& \centering $\eta_{\text{SICNNv1}}$ & \centering $\eta_{\text{SICNNv1Red}}$ & \centering $Q$ & \centering $n_{\text{L,C}}$ & \centering $n_{\text{H,C}}$ & \centering $n_{\text{L,pr}}$ & \centering $n_{\text{H,pr}}$ &  \centering $\eta_{\text{SICNNv2}}$ & \centering $\eta_{\text{SICNNv2Red}}$ & \centering $Q$ & \centering $n_{\text{L}}$ & \centering $n_{\text{H}}$ \tabularnewline
%\cmidrule{2-6}\cmidrule(l){7-9}\cmidrule(l){10-15}
%\cline{2-6}\cline{7-9}\cline{10-15}
\hline
UW guard, QPSK & $6\cdot 10^{-4}$ & $3\cdot 10^{-5}$ & $7$ & $3$ & $70$ & $2$ & $10$ & $5\cdot 10^{-4}$ & $1\cdot 10^{-4}$ & $7$ & $4$ & $200$\\
\hline
CP guard, QPSK & $1\cdot 10^{-3}$ & $7\cdot 10^{-5}$ & $7$ & $3$ & $100$ & $2$ & $10$ & $9\cdot 10^{-4}$ & $1\cdot 10^{-4}$ & $7$ & $4$ & $250$ \\
\hline
UW guard, 16-QAM & $9\cdot 10^{-4}$ & $4\cdot 10^{-5}$ & $7$ & $3$ & $70$ & $3$ & $20$ & $1\cdot 10^{-3}$ & $2.5\cdot 10^{-4}$ & $7$ & $4$ & $230$ \\
\bottomrule
\end{tabularx}
\end{center}
\end{table*}
}

{
\renewcommand{\aboverulesep}{0pt}
\renewcommand{\belowrulesep}{0pt}
\renewcommand{\arraystretch}{1.3}
\begin{table*}[t]
\caption{Hyperparameter settings of the state-of-the-art NN equalizers used for comparison. Similar to the original publication~\cite{Samuel19}, for DetNet $\eta$ denotes  the learning rate, $L$ the number of layers, $d_{\text{h}}$ the number of hidden neurons in the single-hidden-layer FCNN, $d_{\text{v}}$ the dimension of the auxiliary variable passing unconstrained information thourgh the network, and $\beta$ the residual weighting factor. For OAMP-Net2~\cite{He20}, $\eta$ is the learning rate, and $T$ is the number of layers. For KAFCNN~\cite{Baumgartner23_C1}, $\eta$ is the learning rate, $L$ the number of layers, $d_{\text{h}}$ the number of neurons per hidden layer, and $\beta$ the residual weighting factor.}
\label{tab:Hyperparameter_settings_SOTA_NNs}
\begin{center}
\vspace{-0.2cm}
\begin{tabularx}{\textwidth}{>{\raggedright}m{2.4cm}  |>{\raggedleft}m{1.15cm} >{\raggedleft}X >{\raggedleft}X >{\raggedleft}X  >{\raggedleft}X | >{\raggedleft}m{1.15cm} >{\raggedleft}X | >{\raggedleft}m{1.15cm} >{\raggedleft}X >{\raggedleft}X >{\raggedleft\arraybackslash}X}
\toprule
SC-FDE parameters & \multicolumn{5}{c|}{DetNet} &\multicolumn{2}{c|}{OAMP-Net2} &\multicolumn{4}{c}{KAFCNN}\\
& \centering $\eta$ & \centering $L$ & \centering $d_{\text{h}}$ & \centering $d_{\text{v}}$ & \centering $\beta$ & \centering $\eta$ & \centering $T$ &  \centering $\eta$ & \centering $L$ & \centering $n_{\text{h}}$ & \centering $\beta$\tabularnewline
%\cmidrule{2-6}\cmidrule(l){7-9}\cmidrule(l){10-15}
%\cline{2-6}\cline{7-9}\cline{10-15}
\hline
UW guard, QPSK & $6\cdot 10^{-4}$ & $15$ & $200$  & $20$ & $0.5$ & $1\cdot 10^{-3}$ & $8$ & $1\cdot 10^{-3}$ & $12$ & $250$ & $0.8$\\
\hline
CP guard, QPSK & $6\cdot 10^{-4}$ & $15$ & $250$ & $30$ & $0.5$ & $1 \cdot 10^{-3}$ & $10$ & $1\cdot 10^{-3}$ & $12$ & $300$ & $0.8$\\
\hline
UW guard, 16-QAM & $6\cdot 10^{-4}$ & $15$  & $220$ & $25$ & $0.5$ & $1\cdot 10^{-3}$ & $8$ & $1\cdot 10^{-3}$ & $12$ & $280$ & $0.8$\\
\bottomrule
\end{tabularx}
\end{center}
\end{table*}
}

\subsection{Bit Error Ratio Performance for SC-FDE}
\label{ssec:BER_performance}
\subsubsection{Unique Word Guard Interval, QPSK}

\begin{figure}[t]
\begin{center}
\begin{tikzpicture}
\begin{semilogyaxis}[compat=newest, width=8cm, height=8cm, grid=both, ylabel={\small BER}, ymax = 1e-1, ymin = 1e-6, xmax=14, xmin=2, xlabel={\small $E_{\text{b}}/N_0$ (dB)}, scaled ticks = false,  x tick label style={/pgf/number format/.cd,fixed,precision=2,/tikz/.cd}, ticklabel style={font=\scriptsize}, legend columns = {2}, legend cell align=left, legend style={font=\scriptsize, at={(0.5,1.01)}, anchor=south}, every axis plot/.append style={thick}]
\addplot[color=myred, mark=x, smooth, mark size=0.08cm,mark options=solid] table[col sep=semicolon, x=EbN0_dB, y=ber_SICNNDiagCvv]{./fig/20231029224651_Nr_Stages_SICNNv1_SICNNv1cpxRed/BER_estimators_SICNNv1_Q2.csv};
\addlegendentry{\scriptsize SICNNv1 $Q=2$}
\addplot[color=myred, smooth, mark size=0.035cm,dashed,mark options=solid] table[col sep=semicolon, x=EbN0_dB, y=ber_SICNNDiagCvv]{./fig/20231029224651_Nr_Stages_SICNNv1_SICNNv1cpxRed/BER_estimators_SICNNv1_Q3.csv};
\addlegendentry{\scriptsize SICNNv1 $Q=3$}
%\addplot[color=myred,  smooth, mark size=0.08cm,densely dotted,mark options=solid] table[col sep=semicolon, x=EbN0_dB, y=ber_SICNNDiagCvv]{./fig/20230710153511_Nr_Stages_SICNNv1/BER_estimators_Q=5.csv};
%\addlegendentry{\scriptsize SICNNv1 $Q=5$}
\addplot[color=myred,  smooth, mark size=0.05cm,mark options=solid] table[col sep=semicolon, x=EbN0_dB, y=ber_SICNNDiagCvv]{./fig/20231029224651_SC-FDE_UW_QPSK/BER_estimators_SICNNDiagCvv.csv};
\addlegendentry{\scriptsize SICNNv1 $Q=7$}
\addplot[color=myblue, mark=*, smooth, mark size=0.05cm] table[col sep=semicolon, x=EbN0dB, y=SIC_MMSE]{./fig/20230104122811_Iterative_SIC_BER_Nr_Iterations/Matlab_results_Q=1.csv};
\addlegendentry{\scriptsize iter. SIC $Q=1$}
\addplot[color=myblue, mark=x, smooth, mark size=0.08cm] table[col sep=semicolon, x=EbN0dB, y=SIC_MMSE]{./fig/20230104122811_Iterative_SIC_BER_Nr_Iterations/Matlab_results_Q=2.csv};
\addlegendentry{\scriptsize iter. SIC $Q=2$}
\addplot[color=myblue, mark=square, smooth, mark size=0.04cm] table[col sep=semicolon, x=EbN0dB, y=SIC_MMSE]{./fig/20230104122811_Iterative_SIC_BER_Nr_Iterations/Matlab_results_Q=3.csv};
\addlegendentry{\scriptsize iter. SIC $Q=3$}
%\addplot[color=myblue, smooth, mark size=0.06cm,dashed,mark options=solid] table[col sep=semicolon, x=EbN0dB, y=SIC_MMSE]{./fig/20230104122811_Iterative_SIC_BER_Nr_Iterations/Matlab_results_Q=4.csv};
%\addlegendentry{\scriptsize iter. SIC $Q=4$}
%\addplot[color=myblue, smooth, mark size=0.05cm,densely dotted,mark options=solid] table[col sep=semicolon, x=EbN0dB, y=SIC_MMSE]{./fig/20230104122811_Iterative_SIC_BER_Nr_Iterations/Matlab_results_Q=5.csv};
%\addlegendentry{\scriptsize iter. SIC $Q=5$}
\addplot[color=myblue, smooth, mark size=0.08cm,dash dot,mark options=solid] table[col sep=semicolon, x=EbN0dB, y=SIC_MMSE]{./fig/20230104122811_Iterative_SIC_BER_Nr_Iterations/Matlab_results_Q=6.csv};
\addlegendentry{\scriptsize iter. SIC $Q=6$}
\addplot[color=mygreen, mark=x, smooth, mark size=0.08cm,dashed,mark options=solid] table[col sep=semicolon, x=EbN0_dB, y=ber_DF]{./fig/20230710153511_Nr_Stages_SICNNv1_old/BER_estimators_Q=2.csv};
\addlegendentry{\scriptsize DFE}
\addplot[color=myorange, mark=x, smooth, mark size=0.05cm,dashed,mark options=solid] table[col sep=semicolon, x=EbN0_dB, y=ber_LMMSE]{./fig/20230710153511_Nr_Stages_SICNNv1_old/BER_estimators_Q=2.csv};
\addlegendentry{\scriptsize LMMSE}
\end{semilogyaxis}
\end{tikzpicture}
\caption{BER performance of the iterative SIC method and SICNNv1 for different numbers of iterations / stages $Q$ (SC-FDE with UW guard, QPSK).}
\label{fig:BER_performance_vs_iterations}
\end{center}
\vspace{-0.3cm}
\end{figure}
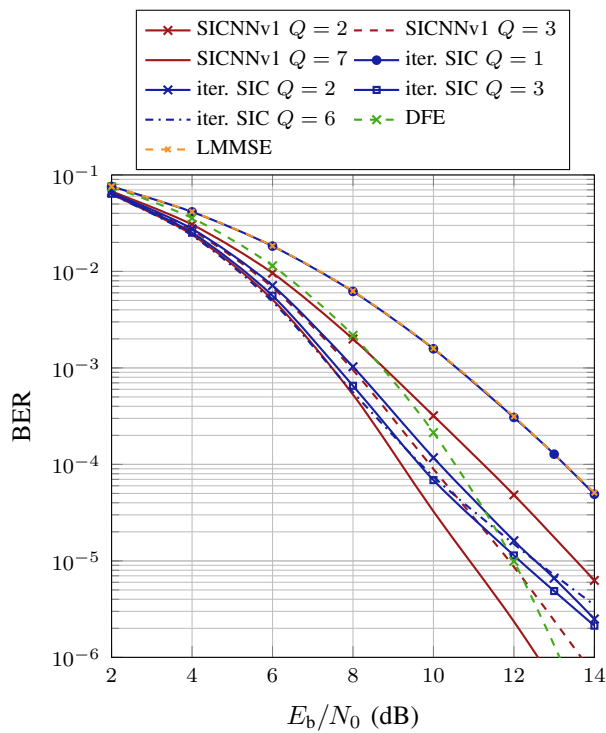

We start with investigations on an SC-FDE system with UW guard interval and QPSK modulation alphabet. First, we investigate the influence of the number of iterations $Q$ of the model-based iterative SIC method as well as the number of stages $Q$ of SICNNv1 on the achieved BER performance to highlight similarities and differences between the model-based and the NN-based approach. As shown in Fig.~\ref{fig:BER_performance_vs_iterations}, our simulation result validates the proof of the equivalence of the bit error probabilities of the LMMSE hard decision estimates and the estimates of the iterative SIC method after one iteration. Moreover, the BER performance of the iterative SIC method considerably improves when conducting a second iteration ($Q=2$), outperforming the DFE over a wide $E_{\text{b}}/N_0$ range. For $Q \geq 3$, two interesting effects are visible. Firstly, the BER performance flattens out at higher $E_{\text{b}}/N_0$ values. Secondly, although more iterations lead to an improvement of the BER performance at lower $E_{\text{b}}/N_0$ values (which is, however, rather small), at higher $E_{\text{b}}/N_0$ values the performance even slightly degrades the more iterations are conducted,  which can be explained by the error caused by approximating the covariance matrix $\m{C}_{\ve{v}\ve{v},k}^{(q)}$. For SICNNv1, more stages than iterations of the model-based method are required to obtain good BER performance, however, for $Q=7$ stages, the iterative SIC method is considerably outperformed by SICNNv1.

\begin{figure}[t]
\begin{center}
\begin{tikzpicture}[
% Style for the spy nodes and the connection line
    spy/.style={%
        draw,black,
        line width=0.4pt,
        circle,inner sep=0pt,
    },
]%[spy using outlines={rectangle, magnification=10, connect spies}]
%% size of the spy-in nodes
    \def\spyviewersize{3cm}

    %% (line width of the spy nodes) / 2
    %% we need this for clipping later
    \def\spyonclipreduce{0.4pt}

    %% first zoom
    %%% factor
    \def\spyfactorI{6}
    %%% spy in point
    \coordinate (spy-in 1) at (1.6,1.65);
    %%% spy on point
    \coordinate (spy-on 1) at (3.75,2.95);% sould be on the curve

\def\pik{
\begin{semilogyaxis}[compat=newest, width=8cm, height=8cm, grid=both, ylabel={\small BER}, ymax = 1e-1, ymin = 1e-6, xmax=14, xmin=2, xlabel={\small $E_{\text{b}}/N_0$ (dB)}, scaled ticks = false,  x tick label style={/pgf/number format/.cd,fixed,precision=2,/tikz/.cd}, ticklabel style={font=\scriptsize}, legend columns = {2}, legend cell align=left, legend style={font=\scriptsize, inner sep = 0.3pt, at={(0.5,1.01)}, anchor=south}, every axis plot/.append style={thick}]
\addplot[color=mygreen, mark=o, smooth, mark size=0.05cm] table[col sep=semicolon, x=EbN0dB, y=DF]{./fig/20231029224651_SC-FDE_UW_QPSK/Matlab_results.csv};
\addlegendentry{\scriptsize DFE}
\addplot[color=myorange, mark=x, smooth, mark size=0.05cm] table[col sep=semicolon, x=EbN0dB, y=LMMSE low rate]{./fig/20231029224651_SC-FDE_UW_QPSK/Matlab_results.csv};
\addlegendentry{\scriptsize LMMSE}
\addplot[color=myblue, mark=square, smooth, mark size=0.05cm] table[col sep=semicolon, x=EbN0dB, y=SIC_MMSE]{./fig/20230104122811_Iterative_SIC_BER_Nr_Iterations/Matlab_results_Q=3.csv};
\addlegendentry{\scriptsize iter. SIC $Q=3$}
\addplot[color=myviolet, mark=*, smooth, mark size=0.05cm,mark options=solid] table[col sep=semicolon, x=EbN0_dB, y=ber_OAMPNet2]{./fig/20231029224651_SC-FDE_UW_QPSK/BER_estimators_OAMP-Net2.csv};
\addlegendentry{\scriptsize OAMP-Net2}
\addplot[color=red, mark=+, smooth, mark size=0.05cm,mark options=solid] table[col sep=semicolon, x=EbN0_dB, y=ber_DetNet]{./fig/20231029224651_SC-FDE_UW_QPSK/BER_estimators_remaining.csv};
\addlegendentry{\scriptsize DetNet}
\addplot[color=blue, mark=*, smooth, mark size=0.05cm,mark options=solid] table[col sep=semicolon, x=EbN0_dB, y=ber_FCNN_IFFT_incorp]{./fig/20231029224651_SC-FDE_UW_QPSK/BER_estimators_remaining.csv};
\addlegendentry{\scriptsize KAFCNN}
\addplot[color=myred, mark=o, smooth, mark size=0.05cm,mark options=solid] table[col sep=semicolon, x=EbN0_dB, y=ber_SICNNDiagCvv]{./fig/20231029224651_SC-FDE_UW_QPSK/BER_estimators_SICNNDiagCvv.csv};
\addlegendentry{\scriptsize SICNNv1}
\addplot[color=myred, mark=+, smooth, mark size=0.07cm,dashed,mark options=solid] table[col sep=semicolon, x=EbN0_dB, y=ber_SICNNcpxRedDiagCvv]{./fig/20231029224651_SC-FDE_UW_QPSK/BER_estimators_SICNNCpxRed.csv};
\addlegendentry{\scriptsize SICNNv1Red}
\addplot[color=mykaki, mark=x, smooth, mark size=0.07cm,mark options=solid] table[col sep=semicolon, x=EbN0_dB, y=ber_SICNNDirectProbEst]{./fig/20231029224651_SC-FDE_UW_QPSK/BER_estimators_remaining.csv};
\addlegendentry{\scriptsize SICNNv2}
\addplot[color=mykaki, mark=square, smooth, mark size=0.05cm,dashed,mark options=solid] table[col sep=semicolon, x=EbN0_dB, y=ber_SICNNcpxRedDirectProbEst]{./fig/20231029224651_SC-FDE_UW_QPSK/BER_estimators_SICNNv2Red.csv};
\addlegendentry{\scriptsize SICNNv2Red}
\end{semilogyaxis}}
\pik

\def\pik2{
\begin{semilogyaxis}[compat=newest, width=8cm, height=8cm, grid=both, minor grid style={line width=0.05pt}, major grid style={line width=0.05pt}, ylabel={\small BER}, ymax = 1e-1, ymin = 1e-6, xmax=14, xmin=2, xlabel={\small $E_{\text{b}}/N_0$ (dB)}, scaled ticks = false,  x tick label style={/pgf/number format/.cd,fixed,precision=2,/tikz/.cd}, ticklabel style={font=\scriptsize}, legend columns = {2}, legend cell align=left, legend style={font=\scriptsize, inner sep = 0.3pt, at={(0.5,1.01)}, anchor=south}, every axis plot/.append style={line width=0.2pt}]
\addplot[color=mygreen, mark=o, smooth, mark size=0.05cm] table[col sep=semicolon, x=EbN0dB, y=DF]{./fig/20231029224651_SC-FDE_UW_QPSK/Matlab_results.csv};

\addplot[color=myorange, mark=x, smooth, mark size=0.02cm] table[col sep=semicolon, x=EbN0dB, y=LMMSE low rate]{./fig/20231029224651_SC-FDE_UW_QPSK/Matlab_results.csv};

\addplot[color=myblue, mark=square, smooth, mark size=0.02cm] table[col sep=semicolon, x=EbN0dB, y=SIC_MMSE]{./fig/20230104122811_Iterative_SIC_BER_Nr_Iterations/Matlab_results_Q=3.csv};

\addplot[color=myviolet, mark=*, smooth, mark size=0.02cm,mark options=solid] table[col sep=semicolon, x=EbN0_dB, y=ber_OAMPNet2]{./fig/20231029224651_SC-FDE_UW_QPSK/BER_estimators_OAMP-Net2.csv};

\addplot[color=red, mark=+, smooth, mark size=0.02cm,mark options=solid] table[col sep=semicolon, x=EbN0_dB, y=ber_DetNet]{./fig/20231029224651_SC-FDE_UW_QPSK/BER_estimators_remaining.csv};

\addplot[color=blue, mark=*, smooth, mark size=0.02cm,mark options=solid] table[col sep=semicolon, x=EbN0_dB, y=ber_FCNN_IFFT_incorp]{./fig/20231029224651_SC-FDE_UW_QPSK/BER_estimators_remaining.csv};

\addplot[color=myred, mark=o, smooth, mark size=0.02cm,mark options=solid] table[col sep=semicolon, x=EbN0_dB, y=ber_SICNNDiagCvv]{./fig/20231029224651_SC-FDE_UW_QPSK/BER_estimators_SICNNDiagCvv.csv};

\addplot[color=myred, mark=+, smooth, mark size=0.022cm,dashed,mark options=solid] table[col sep=semicolon, x=EbN0_dB, y=ber_SICNNcpxRedDiagCvv]{./fig/20231029224651_SC-FDE_UW_QPSK/BER_estimators_SICNNCpxRed.csv};

\addplot[color=mykaki, mark=x, smooth, mark size=0.022cm,mark options=solid] table[col sep=semicolon, x=EbN0_dB, y=ber_SICNNDirectProbEst]{./fig/20231029224651_SC-FDE_UW_QPSK/BER_estimators_remaining.csv};

\addplot[color=mykaki, mark=square, smooth, mark size=0.02cm,dashed,mark options=solid] table[col sep=semicolon, x=EbN0_dB, y=ber_SICNNcpxRedDirectProbEst]{./fig/20231029224651_SC-FDE_UW_QPSK/BER_estimators_SICNNv2Red.csv};

\end{semilogyaxis}}

% first zoom
    %% spy on node
    \node[spy,minimum size={\spyviewersize/\spyfactorI}] (spy-on node 1) at (spy-on 1) {};
    %% spy in node
    \node[spy,minimum size=\spyviewersize] (spy-in node 1) at (spy-in 1) {};
    \begin{scope}
        \clip (spy-in 1) circle (0.5*\spyviewersize-\spyonclipreduce);
        \node[circle, fill=white, draw=white, minimum size=\spyviewersize] (background_circ) at (spy-in 1) {};
        \pgfmathsetmacro\sI{1/\spyfactorI}
        \begin{scope}[
            shift={($\sI*(spy-in 1)-\sI*(spy-on 1)$)},
            scale around={\spyfactorI:(spy-on 1)}
        ]
            \pik2
        \end{scope}
    \end{scope}
    %% connect the nodes
    \draw [spy] (spy-on node 1) -- (spy-in node 1);

\end{tikzpicture}
\vspace{-0.3cm}
\caption{BER performance of NN-based and model-based equalizers for SC-FDE with a UW guard interval and QPSK alphabet.}
\label{fig:BER_performance_equalizers}
\end{center}
\vspace{-0.3cm}
\end{figure}
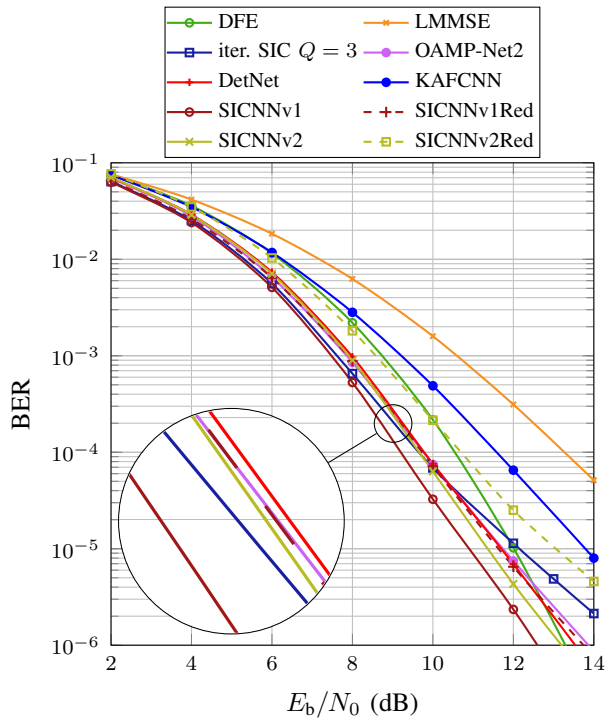

Next, we compare the proposed NN-based data estimators \mbox{SICNNv1}, SICNNv2, SICNNv1Red and SICNNv2Red with the aforementioned state-of-the-art model-based and NN-based equalizers in terms of achieved BER performance. Their training and hyperparameter optimization is conducted with the same training set as used for the proposed NN-based equalizers. As shown in Fig.~\ref{fig:BER_performance_equalizers}, SICNNv1 is the best performing equalizer for a wide $E_{\text{b}}/N_0$ range, followed by SICNNv2 and OAMP-Net2. The parameter-reduced variant SICNNv1Red exhibits approximately the same performance as DetNet. All of the aforementioned NN-based equalizers can outperform or perform similarly as the model-based equalizers considered for comparison. SICNNv2Red, in turn, is the worst performing among the proposed NN-based equalizers, but still has a far better BER performance than KAFCNN. From this simulation result we can conclude that using the same sub-NNs in all stages of the proposed NN-based equalizers leads to a reduction of the number of learnable parameters at the cost of a performance decrease. However, since fewer parameters have to be optimized, the reduction of learnable parameters decreases the computational effort for training the NNs.

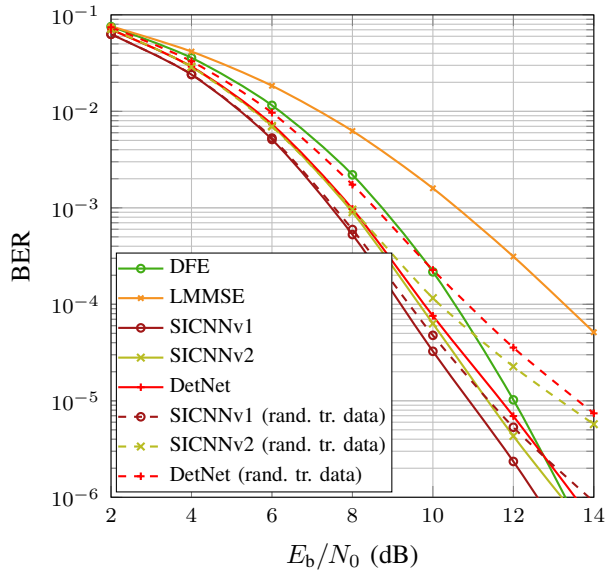
\begin{figure}[t]
\begin{center}
\begin{tikzpicture}
\begin{semilogyaxis}[compat=newest, width=8cm, height=8cm, grid=both, ylabel={\small BER}, ymax = 1e-1, ymin = 1e-6, xmax=14, xmin=2, xlabel={\small $E_{\text{b}}/N_0$ (dB)}, scaled ticks = false,  x tick label style={/pgf/number format/.cd,fixed,precision=2,/tikz/.cd}, ticklabel style={font=\scriptsize}, legend columns = {1}, legend cell align=left, legend style={font=\scriptsize, inner sep = 0.3pt, at={(0.01,0.01)}, anchor=south west}, every axis plot/.append style={thick}]
\addplot[color=mygreen, mark=o, smooth, mark size=0.05cm] table[col sep=semicolon, x=EbN0dB, y=DF]{./fig/20231029224651_SC-FDE_UW_QPSK/Matlab_results.csv};
\addlegendentry{\scriptsize DFE}
\addplot[color=myorange, mark=x, smooth, mark size=0.05cm] table[col sep=semicolon, x=EbN0dB, y=LMMSE low rate]{./fig/20231029224651_SC-FDE_UW_QPSK/Matlab_results.csv};
\addlegendentry{\scriptsize LMMSE}
%\addplot[color=myviolet, mark=+, smooth, mark size=0.05cm,mark options=solid] table[col sep=semicolon, x=EbN0_dB, y=ber_OAMPNet2]{./fig/20231029224651_SC-FDE_UW_QPSK/BER_estimators_SICNNCpxRed_OAMPNet.csv};
%\addlegendentry{\scriptsize OAMP-Net2}
%\addplot[color=blue, mark=*, smooth, mark size=0.05cm,mark options=solid] table[col sep=semicolon, x=EbN0_dB, y=ber_FCNN_IFFT_incorp]{./fig/20231029224651_SC-FDE_UW_QPSK/BER_estimators_remaining.csv};
%\addlegendentry{\scriptsize KAFCNN}
\addplot[color=myred, mark=o, smooth, mark size=0.05cm,mark options=solid] table[col sep=semicolon, x=EbN0_dB, y=ber_SICNNDiagCvv]{./fig/20231029224651_SC-FDE_UW_QPSK/BER_estimators_SICNNDiagCvv.csv};
\addlegendentry{\scriptsize SICNNv1}
%\addplot[color=myred, mark=+, smooth, mark size=0.07cm,dashed,mark options=solid] table[col sep=semicolon, x=EbN0_dB, y=ber_SICNNcpxRedDiagCvv]{./fig/20231029224651_SC-FDE_UW_QPSK/BER_estimators_SICNNCpxRed_OAMPNet.csv};
%\addlegendentry{\scriptsize SICNNv1Red}
\addplot[color=mykaki, mark=x, smooth, mark size=0.07cm,mark options=solid] table[col sep=semicolon, x=EbN0_dB, y=ber_SICNNDirectProbEst]{./fig/20231029224651_SC-FDE_UW_QPSK/BER_estimators_remaining.csv};
\addlegendentry{\scriptsize SICNNv2}
%\addplot[color=mykaki, mark=square, smooth, mark size=0.05cm,dashed,mark options=solid] table[col sep=semicolon, x=EbN0_dB, y=ber_SICNNcpxRedDirectProbEst]{./fig/20231029224651_SC-FDE_UW_QPSK/BER_estimators_SICNNCpxRed_OAMPNet.csv};
%\addlegendentry{\scriptsize SICNNv2Red}
\addplot[color=red, mark=+, smooth, mark size=0.05cm,mark options=solid] table[col sep=semicolon, x=EbN0_dB, y=ber_DetNet]{./fig/20231029224651_SC-FDE_UW_QPSK/BER_estimators_remaining.csv};
\addlegendentry{\scriptsize DetNet}
%%%%%% old results %%%%%%%%%
\addplot[color=myred, mark=o, smooth, mark size=0.05cm,mark options=solid, dashed] table[col sep=semicolon, x=EbN0_dB, y=ber_SICNNDiagCvv]{./fig/20230710225640_BER_performance_equalizers/BER_estimators_SICNNv1_E15.csv};
\addlegendentry{\scriptsize SICNNv1 (rand. tr. data)}
%\addplot[color=myred, mark=+, smooth, mark size=0.07cm,dashed,mark options=solid] table[col sep=semicolon, x=EbN0_dB, y=ber_SICNNcpxRedDiagCvv]{./fig/20230710225640_BER_performance_equalizers/BER_estimators_SICNNv1Red.csv};
%\addlegendentry{\scriptsize SICNNv1Red}
\addplot[color=mykaki, mark=x, smooth, mark size=0.07cm,mark options=solid, dashed] table[col sep=semicolon, x=EbN0_dB, y=ber_SICNNDirectProbEst]{./fig/20230710225640_BER_performance_equalizers/BER_estimators_SICNNv2_SICNNv2Red.csv};
\addlegendentry{\scriptsize SICNNv2 (rand. tr. data)}
%\addplot[color=mykaki, mark=square, smooth, mark size=0.05cm,dashed,mark options=solid] table[col sep=semicolon, x=EbN0_dB, y=ber_SICNNcpxRedDirectProbEst]{./fig/20230710225640_BER_performance_equalizers/BER_estimators_SICNNv2_SICNNv2Red.csv};
%\addlegendentry{\scriptsize SICNNv2Red}
%\addplot[color=myviolet, mark=+, smooth, mark size=0.05cm,mark options=solid] table[col sep=semicolon, x=EbN0_dB, y=ber_OAMPNet2]{./fig/20230710225640_BER_performance_equalizers/BER_estimators_OAMPNet2.csv};
%\addlegendentry{\scriptsize OAMP-Net2}
\addplot[color=red, mark=+, smooth, mark size=0.05cm,mark options=solid, dashed] table[col sep=semicolon, x=EbN0_dB, y=ber_DetNet]{./fig/20230710225640_BER_performance_equalizers/BER_estimators_all.csv};
\addlegendentry{\scriptsize DetNet (rand. tr. data)}
%\addplot[color=blue, mark=*, smooth, mark size=0.05cm,mark options=solid] table[col sep=semicolon, x=EbN0_dB, y=ber_FCNN_IFFT_incorp]{./fig/20230710225640_BER_performance_equalizers/BER_estimators_all.csv};
%\addlegendentry{\scriptsize KAFCNN}
\end{semilogyaxis}
\end{tikzpicture}
\vspace{-0.3cm}
\caption{BER performance comparison of NN-based equalizers for SC-FDE with a UW guard interval and QPSK alphabet, when being trained with randomly generated training data (rand. tr. data), or on a training set generated by the proposed approach.}
\label{fig:BER_performance_equalizers_SC-FDE_UW_QPSK_influence_train_set_generation}
\end{center}
\vspace{-0.3cm}
\end{figure}

For an SC-FDE system with a UW guard interval and QPSK modulation alphabet, we also show the influence of our proposed approach for training set generation, described in Sec.~\ref{ssec:Training_Set_Generation}, on the BER performance of trained NN-based equalizers. Exemplary for SICNNv1, SICNNv2, and DetNet, we compare their performance when being trained on a dataset generated with our approach with the case that they are trained with randomly generated training data. For the randomly generated training data, the $E_{\text{b}}/N_0$ values of the sample transmissions contained in the training set are chosen randomly (with uniform distribution on the linear $E_{\text{b}}/N_0$ scale) in the range $[3\,\si{dB}, 14\,\si{dB}]$, which is a state-of-the-art approach for the training of NN-based equalizers. As shown in Fig.~\ref{fig:BER_performance_equalizers_SC-FDE_UW_QPSK_influence_train_set_generation}, particularly at high $E_{\text{b}}/N_0$ values, the performance of the aforementioned NN-based equalizers can be significantly improved by training them on a training set generated by our proposed method. As elucidated in Sec.~\ref{ssec:Training_Set_Generation}, we assume that the main reason for this performance improvement is, that at high SNRs distinctly more training samples lying close to the decision boundaries of the optimal equalizer are available, allowing the NNs to approximate the optimal decision boundaries. For all following results, the NNs are trained with datasets that are generated with our proposed method.

\subsubsection{Unique Word Guard Interval, 16-QAM}
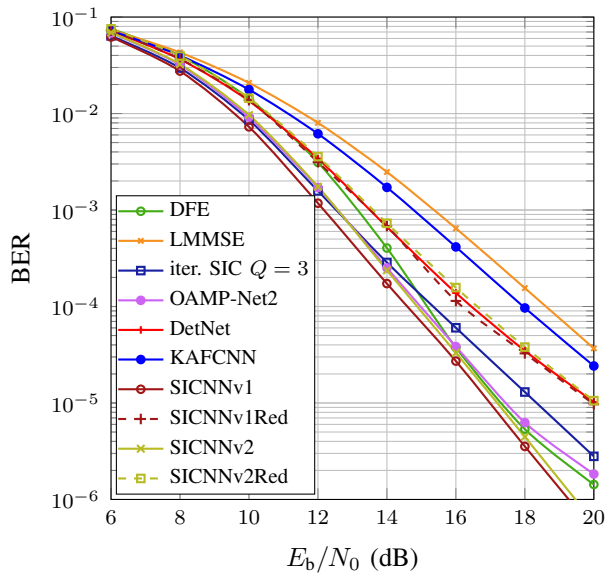
\begin{figure}[t]
\begin{center}
\begin{tikzpicture}
\begin{semilogyaxis}[compat=newest, width=8cm, height=8cm, grid=both, ylabel={\small BER}, ymax = 1e-1, ymin = 1e-6, xmax=20, xmin=6, xlabel={\small $E_{\text{b}}/N_0$ (dB)}, scaled ticks = false,  x tick label style={/pgf/number format/.cd,fixed,precision=2,/tikz/.cd}, ticklabel style={font=\scriptsize}, xtick={6,8,10,12,14,16,18,20}, legend columns = {1}, legend cell align=left, legend style={font=\scriptsize, inner sep = 0.3pt, at={(0.01,0.01)}, anchor=south west}, every axis plot/.append style={thick}]
\addplot[color=mygreen, mark=o, smooth, mark size=0.05cm] table[col sep=semicolon, x=EbN0dB, y=DF]{./fig/20231025155255_SC-FDE_UW_16QAM/Matlab_results.csv};
\addlegendentry{\scriptsize DFE}
\addplot[color=myorange, mark=x, smooth, mark size=0.05cm] table[col sep=semicolon, x=EbN0dB, y=LMMSE low rate]{./fig/20231025155255_SC-FDE_UW_16QAM/Matlab_results.csv};
\addlegendentry{\scriptsize LMMSE}
\addplot[color=myblue, mark=square, smooth, mark size=0.05cm] table[col sep=semicolon, x=EbN0dB, y=SIC_MMSE]{./fig/20231227105916_SC-FDE_UW_16QAM_ISIC_MMSE/Matlab_results_Q=3.csv};
\addlegendentry{\scriptsize iter. SIC $Q=3$}
\addplot[color=myviolet, mark=*, smooth, mark size=0.05cm,mark options=solid] table[col sep=semicolon, x=EbN0_dB, y=ber_OAMPNet2]{./fig/20231123080758_SC-FDE_UW_16QAM/BER_estimators_all.csv};
\addlegendentry{\scriptsize OAMP-Net2}
\addplot[color=red, mark=+, smooth, mark size=0.05cm,mark options=solid] table[col sep=semicolon, x=EbN0_dB, y=ber_DetNet]{./fig/20231123080758_SC-FDE_UW_16QAM/BER_estimators_all.csv};
\addlegendentry{\scriptsize DetNet}
\addplot[color=blue, mark=*, smooth, mark size=0.05cm,mark options=solid] table[col sep=semicolon, x=EbN0_dB, y=ber_FCNN_IFFT_incorp]{./fig/20231123080758_SC-FDE_UW_16QAM/BER_estimators_all.csv};
\addlegendentry{\scriptsize KAFCNN}
\addplot[color=myred, mark=o, smooth, mark size=0.05cm,mark options=solid] table[col sep=semicolon, x=EbN0_dB, y=ber_SICNNDiagCvv]{./fig/20231123080758_SC-FDE_UW_16QAM/BER_estimators_all.csv};
\addlegendentry{\scriptsize SICNNv1}
\addplot[color=myred, mark=+, smooth, mark size=0.07cm,dashed,mark options=solid] table[col sep=semicolon, x=EbN0_dB, y=ber_SICNNcpxRedDiagCvv]{./fig/20231025155255_SC-FDE_UW_16QAM/BER_estimators_SICNNv1Red.csv};
\addlegendentry{\scriptsize SICNNv1Red}
\addplot[color=mykaki, mark=x, smooth, mark size=0.07cm,mark options=solid] table[col sep=semicolon, x=EbN0_dB, y=ber_SICNNDirectProbEst]{./fig/20231123080758_SC-FDE_UW_16QAM/BER_estimators_all.csv};
\addlegendentry{\scriptsize SICNNv2}
\addplot[color=mykaki, mark=square, smooth, mark size=0.05cm,dashed,mark options=solid] table[col sep=semicolon, x=EbN0_dB, y=ber_SICNNcpxRedDirectProbEst]{./fig/20231025155255_SC-FDE_UW_16QAM/BER_estimators_SICNNv2Red.csv};
\addlegendentry{\scriptsize SICNNv2Red}
\end{semilogyaxis}
\end{tikzpicture}
\vspace{-0.3cm}
\caption{BER performance of NN-based and model-based equalizers for SC-FDE with a UW guard interval and 16-QAM alphabet. }
\label{fig:BER_performance_equalizers_SC-FDE_UW_16QAM}
\end{center}
\vspace{-0.3cm}
\end{figure}

To show that the proposed NN-based equalizers can also cope with higher order modulation alphabets, we present BER performance results for an SC-FDE system with a UW guard interval and 16-QAM modulation alphabet. As shown in Fig.~\ref{fig:BER_performance_equalizers_SC-FDE_UW_16QAM}, SICNNv1 is also the best performing among all considered equalizers for this system setup. Similar as for a QPSK modulation alphabet, the best performing equalizer behind SICNNv1 are SICNNv2 and OAMP-Net, outperforming the model-based DFE in lower $E_{\text{b}}/N_0$ regions and performing similarly in higher $E_{\text{b}}/N_0$ regions. SICNNv1Red, SICNNv2Red, DetNet, and KAFCNN, in turn, exhibit a significantly worse performance, where KAFCNN is the worst performing among all considered NN-based equalizers.

\subsubsection{Imperfect Channel Knowledge}

\begin{figure}[t]
\begin{center}
\begin{tikzpicture}
\begin{semilogyaxis}[compat=newest, width=8cm, height=8cm, grid=both, ylabel={\small BER}, ymax = 1e-1, ymin = 1e-6, xmax=14, xmin=2, xlabel={\small $E_{\text{b}}/N_0$ (dB)}, scaled ticks = false,  x tick label style={/pgf/number format/.cd,fixed,precision=2,/tikz/.cd}, ticklabel style={font=\scriptsize}, xtick={2,4,6,8,10,12,14},, legend columns = {1}, legend cell align=left, legend style={font=\scriptsize, inner sep = 0.3pt, at={(0.01,0.01)}, anchor=south west}, every axis plot/.append style={thick}]
\addplot[color=mygreen, mark=o, smooth, mark size=0.05cm, dashed, mark options=solid] table[col sep=semicolon, x=EbN0dB, y=DF]{./fig/20231203213315_SC-FDE_UW_QPSK_ch_est/Matlab_results.csv};
\addlegendentry{\scriptsize DFE, true ch.}
\addplot[color=mygreen, mark=o, smooth, mark size=0.05cm] table[col sep=semicolon, x=EbN0dB, y=DF]{./fig/20231203213315_SC-FDE_UW_QPSK_ch_est/Matlab_results_h_est.csv};
\addlegendentry{\scriptsize DFE, ch. est.}
\addplot[color=myorange, mark=x, smooth, mark size=0.05cm, dashed, mark options=solid] table[col sep=semicolon, x=EbN0dB, y=LMMSE low rate]{./fig/20231203213315_SC-FDE_UW_QPSK_ch_est/Matlab_results.csv};
\addlegendentry{\scriptsize LMMSE, true ch.}
\addplot[color=myorange, mark=x, smooth, mark size=0.05cm] table[col sep=semicolon, x=EbN0dB, y=LMMSE low rate]{./fig/20231203213315_SC-FDE_UW_QPSK_ch_est/Matlab_results_h_est.csv};
\addlegendentry{\scriptsize LMMSE, ch. est.}
\addplot[color=myviolet, mark=*, dashed, smooth, mark size=0.05cm,mark options=solid] table[col sep=semicolon, x=EbN0_dB, y=ber_OAMPNet2]{./fig/20231029224651_SC-FDE_UW_QPSK/BER_estimators_OAMP-Net2.csv};
\addlegendentry{\scriptsize OAMP-Net2, true ch.}
\addplot[color=myviolet, mark=*, smooth, mark size=0.05cm,mark options=solid] table[col sep=semicolon, x=EbN0_dB, y=ber_OAMPNet2_h_est]{./fig/20231203213315_SC-FDE_UW_QPSK_ch_est/BER_estimators_h_est_OAMPNet.csv};
\addlegendentry{\scriptsize OAMP-Net2, ch. est}
%\addplot[color=red, mark=+, smooth, mark size=0.05cm,mark options=solid] table[col sep=semicolon, x=EbN0_dB, y=ber_DetNet_h_est]{./fig/20231203213315_SC-FDE_UW_QPSK_ch_est/BER_estimators_h_est_all.csv};
%\addlegendentry{\scriptsize DetNet}
%\addplot[color=blue, mark=*, smooth, mark size=0.05cm,mark options=solid] table[col sep=semicolon, x=EbN0_dB, y=ber_FCNN_IFFT_incorp_h_est]{./fig/20231203213315_SC-FDE_UW_QPSK_ch_est/BER_estimators_h_est_all.csv};
%\addlegendentry{\scriptsize KAFCNN}
\addplot[color=myred, mark=o, dashed, smooth, mark size=0.05cm,mark options=solid] table[col sep=semicolon, x=EbN0_dB, y=ber_SICNNDiagCvv]{./fig/20231029224651_SC-FDE_UW_QPSK/BER_estimators_SICNNDiagCvv.csv};
\addlegendentry{\scriptsize SICNNv1, true ch.}
\addplot[color=myred, mark=o, smooth, mark size=0.05cm,mark options=solid] table[col sep=semicolon, x=EbN0_dB, y=ber_SICNNDiagCvv_h_est]{./fig/20231203213315_SC-FDE_UW_QPSK_ch_est/BER_estimators_h_est_SICNNv1.csv};
\addlegendentry{\scriptsize SICNNv1, ch.est}
%\addplot[color=myred, mark=+, smooth, mark size=0.07cm,dashed,mark options=solid] table[col sep=semicolon, x=EbN0_dB, y=ber_SICNNcpxRedDiagCvv_h_est]{./fig/20231203213315_SC-FDE_UW_QPSK_ch_est/BER_estimators_h_est_all.csv};
%\addlegendentry{\scriptsize SICNNv1Red}
\addplot[color=mykaki, mark=x, dashed, smooth, mark size=0.07cm,mark options=solid] table[col sep=semicolon, x=EbN0_dB, y=ber_SICNNDirectProbEst]{./fig/20231029224651_SC-FDE_UW_QPSK/BER_estimators_remaining.csv};
\addlegendentry{\scriptsize SICNNv2, true ch.}
\addplot[color=mykaki, mark=x, smooth, mark size=0.07cm,mark options=solid] table[col sep=semicolon, x=EbN0_dB, y=ber_SICNNDirectProbEst_h_est]{./fig/20231203213315_SC-FDE_UW_QPSK_ch_est/BER_estimators_h_est_all.csv};
\addlegendentry{\scriptsize SICNNv2, ch. est}
%\addplot[color=mykaki, mark=square, smooth, mark size=0.05cm,dashed,mark options=solid] table[col sep=semicolon, x=EbN0_dB, y=ber_SICNNcpxRedDirectProbEst_h_est]{./fig/20231203213315_SC-/BER_estimators_h_est_all.csv};
%\addlegendentry{\scriptsize SICNNv2Red}
\end{semilogyaxis}
\end{tikzpicture}
\vspace{-0.3cm}
\caption{BER performance of NN-based and model-based equalizers for SC-FDE with a UW guard interval, QPSK alphabet, and with perfect and imperfect channel knowledge.}
\label{fig:BER_performance_equalizers_ch_est}
\end{center}
\vspace{-0.3cm}
\end{figure}
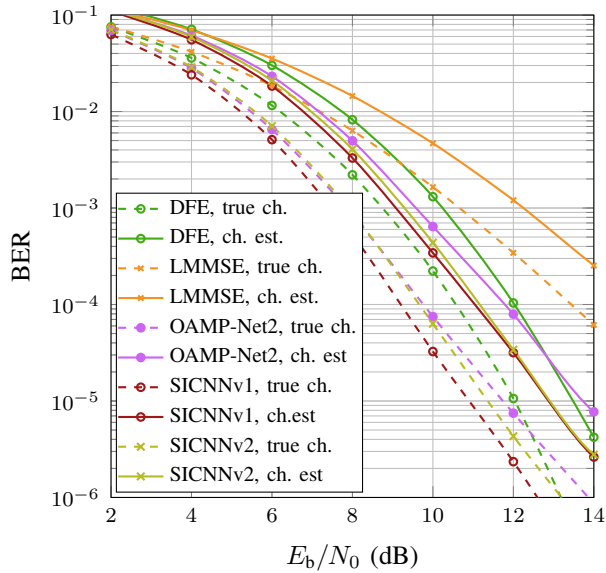

We investigate the influence of imperfect channel knowledge on the performance of NN-based and model-based equalizers. To this end, the channel, which is assumed to be stationary for one transmitted data burst, is estimated as described in~\cite{Huemer14} using a known preamble. This preamble is transmitted prior to the data burst and contains two identical pilot vectors $\ve{x}_{\text{p}}$. Based on the two corresponding received pilot vectors $\ve{y}_{\text{p}}$, the channel frequency response is estimated with the best linear unbiased estimator (BLUE)~\cite{Kay93}. For further details on the estimation of the channel frequency response, we refer to~\cite{Huemer14}.

For this evaluation, the NN-based equalizers are trained in the same manner as for perfect channel knowledge, however, as an input a channel matrix is employed which is computed using the estimated channel frequency response. As shown in Fig.~\ref{fig:BER_performance_equalizers_ch_est} for SICNNv1, SICNNv2, and OAMP-Net2, an imperfect channel knowledge has a similar influence on the BER performance as for model-based equalizers, demonstrating their robustness in terms of imperfect channel knowledge at the receiver.

\subsubsection{Cyclic Prefix Guard Interval, QPSK}

\begin{figure}[t]
\begin{center}
\begin{tikzpicture}
\begin{semilogyaxis}[compat=newest, width=8cm, height=8cm, grid=both, ylabel={\small BER}, ymax = 1e-1, ymin = 1e-6, xmax=18, xmin=4, xlabel={\small $E_{\text{b}}/N_0$ (dB)}, scaled ticks = false,  x tick label style={/pgf/number format/.cd,fixed,precision=2,/tikz/.cd}, ticklabel style={font=\scriptsize}, xtick={4,6,8,10,12,14,16,18}, legend columns = {2}, legend cell align=left, legend style={font=\scriptsize, inner sep = 0.3pt, at={(0.5,1.01)}, anchor=south}, every axis plot/.append style={thick}]
\addplot[color=mygreen, mark=o, smooth, mark size=0.05cm] table[col sep=semicolon, x=EbN0dB, y=DF]{./fig/20231030173856_SC-FDE_CP_QPSK/Matlab_results.csv};
\addlegendentry{\scriptsize DFE}
\addplot[color=myorange, mark=x, smooth, mark size=0.05cm] table[col sep=semicolon, x=EbN0dB, y=LMMSE low rate]{./fig/20231030173856_SC-FDE_CP_QPSK/Matlab_results.csv};
\addlegendentry{\scriptsize LMMSE}
\addplot[color=myblue, mark=square, smooth, mark size=0.05cm] table[col sep=semicolon, x=EbN0dB, y=SIC_MMSE]{./fig/20231223140433_SC-FDE_CP_QPSK_ISIC_MMSE/Matlab_results_Q=3.csv};
\addlegendentry{\scriptsize iter. SIC $Q=3$}
\addplot[color=myviolet, mark=*, smooth, mark size=0.05cm,mark options=solid] table[col sep=semicolon, x=EbN0_dB, y=ber_OAMPNet2]{./fig/20231030173856_SC-FDE_CP_QPSK/BER_estimators_all.csv};
\addlegendentry{\scriptsize OAMP-Net2}
\addplot[color=red, mark=+, smooth, mark size=0.05cm,mark options=solid] table[col sep=semicolon, x=EbN0_dB, y=ber_DetNet]{./fig/20231030173856_SC-FDE_CP_QPSK/BER_estimators_all.csv};
\addlegendentry{\scriptsize DetNet}
\addplot[color=blue, mark=*, smooth, mark size=0.05cm,mark options=solid] table[col sep=semicolon, x=EbN0_dB, y=ber_FCNN_IFFT_incorp]{./fig/20231030173856_SC-FDE_CP_QPSK/BER_estimators_all.csv};
\addlegendentry{\scriptsize KAFCNN}
\addplot[color=myred, mark=o, smooth, mark size=0.05cm,mark options=solid] table[col sep=semicolon, x=EbN0_dB, y=ber_SICNNDiagCvv]{./fig/20231030173856_SC-FDE_CP_QPSK/BER_estimators_SICNNDiagCvv.csv};
\addlegendentry{\scriptsize SICNNv1}
\addplot[color=myred, mark=+, smooth, mark size=0.07cm,dashed,mark options=solid] table[col sep=semicolon, x=EbN0_dB, y=ber_SICNNcpxRedDiagCvv]{./fig/20231030173856_SC-FDE_CP_QPSK/BER_estimators_SICNNcpxRed.csv};
\addlegendentry{\scriptsize SICNNv1Red}
\addplot[color=mykaki, mark=x, smooth, mark size=0.07cm,mark options=solid] table[col sep=semicolon, x=EbN0_dB, y=ber_SICNNDirectProbEst]{./fig/20231030173856_SC-FDE_CP_QPSK/BER_estimators_all.csv};
\addlegendentry{\scriptsize SICNNv2}
\addplot[color=mykaki, mark=square, smooth, mark size=0.05cm,dashed,mark options=solid] table[col sep=semicolon, x=EbN0_dB, y=ber_SICNNcpxRedDirectProbEst]{./fig/20231030173856_SC-FDE_CP_QPSK/BER_estimators_SICNNcpxRed.csv};
\addlegendentry{\scriptsize SICNNv2Red}
\end{semilogyaxis}
\end{tikzpicture}
\vspace{-0.3cm}
\caption{BER performance of NN-based and model-based equalizers for SC-FDE with a CP guard interval and QPSK alphabet. }
\label{fig:BER_performance_equalizers_SC-FDE_CP}
\end{center}
\vspace{-0.3cm}
\end{figure}
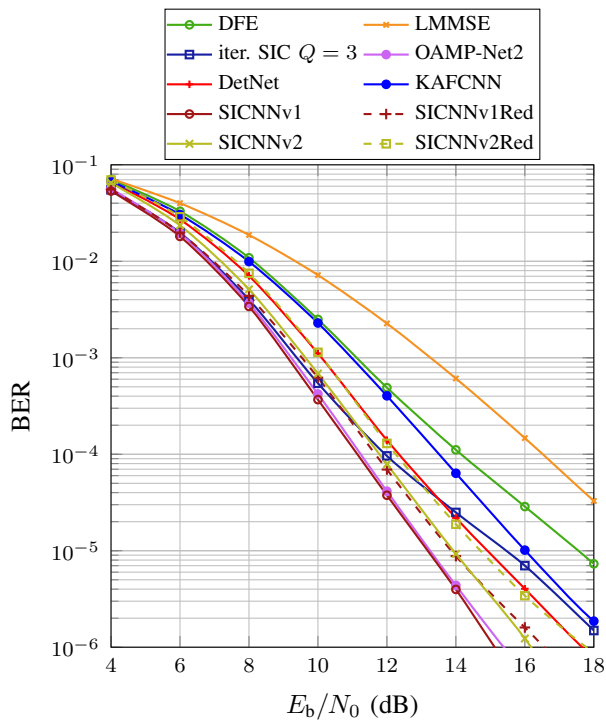

For all the previously shown results we have employed a UW as a guard interval. In this section, we investigate the influence of using a CP as a guard interval on the performance of the regarded NN-based and model-based equalizers. As presented in Fig.~\ref{fig:BER_performance_equalizers_SC-FDE_CP}, SICNNv1 is also the best performing equalizer for this system setup, where its performance is very similar to that of OAMP-Net2. SICNNv2 and SICNNv1Red exhibit a very similar BER performance and can clearly outperform DetNet, which achieves similar BER results as SICNNv2Red. The worst performing NN-based equalizer is KAFCNN, still outperforming the model-based DFE. The LMMSE equalizer performs worst, however, as mentioned in Sec.~\ref{ssec:Single_Carrier_Frequency_Domain_Equalization}, stands out due to its very low complexity for SC-FDE communications with CP guard intervals.

\subsection{Influence of a Reduced Training Set Size}

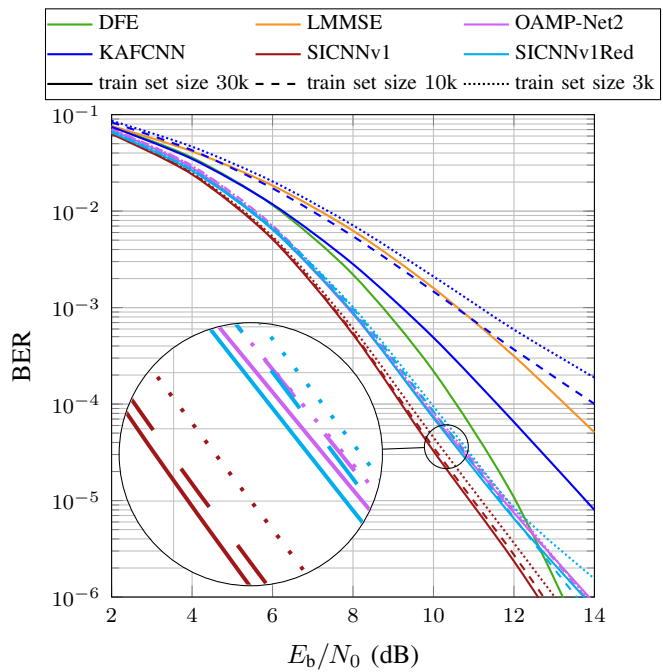
\begin{figure}[t]
\begin{center}
\begin{tikzpicture}[
% Style for the spy nodes and the connection line
    spy/.style={%
        draw,black,
        line width=0.4pt,
        circle,inner sep=0pt,
    },
]%[spy using outlines={rectangle, magnification=10, connect spies}]
%% size of the spy-in nodes
    \def\spyviewersize{3.5cm}

    %% (line width of the spy nodes) / 2
    %% we need this for clipping later
    \def\spyonclipreduce{0.4pt}

    %% first zoom
    %%% factor
    \def\spyfactorI{6}
    %%% spy in point
    \coordinate (spy-in 1) at (1.85,1.9);
    %%% spy on point
    \coordinate (spy-on 1) at (4.45,2);% sould be on the curve

\def\pik{
\begin{semilogyaxis}[compat=newest, width=8cm, height=8cm, grid=both, ylabel={\small BER}, ymax = 1e-1, ymin = 1e-6, xmax=14, xmin=2, xlabel={\small $E_{\text{b}}/N_0$ (dB)}, scaled ticks = false,  x tick label style={/pgf/number format/.cd,fixed,precision=2,/tikz/.cd}, ticklabel style={font=\scriptsize}, legend columns = {3}, legend cell align=left, legend style={font=\scriptsize, inner sep = 0.3pt, at={(0.5,1.03)}, anchor=south}, every axis plot/.append style={thick}, legend entries={DFE, LMMSE, OAMP-Net2, KAFCNN, SICNNv1, SICNNv1Red, train set size $30$k, train set size $10$k, train set size $3$k}]
\addlegendimage{no markers, mygreen}
\addlegendimage{no markers, myorange}
\addlegendimage{no markers, myviolet}
\addlegendimage{no markers, blue}
\addlegendimage{no markers, myred}
\addlegendimage{no markers, cyan}
\addlegendimage{no markers, black}
\addlegendimage{no markers, black, dashed}
\addlegendimage{no markers, black, densely dotted}

\addplot[color=mygreen, smooth, mark size=0.05cm] table[col sep=semicolon, x=EbN0dB, y=DF]{./fig/20230710225640_BER_performance_equalizers/Matlab_results.csv};
%\addlegendentry{\scriptsize DFE}
\addplot[color=myorange,  smooth, mark size=0.05cm] table[col sep=semicolon, x=EbN0dB, y=LMMSE low rate]{./fig/20230710225640_BER_performance_equalizers/Matlab_results.csv};
%\addlegendentry{\scriptsize LMMSE}

\addplot[color=myviolet, smooth, mark size=0.05cm,mark options=solid] table[col sep=semicolon, x=EbN0_dB, y=ber_OAMPNet2]{./fig/20231029224651_SC-FDE_UW_QPSK/BER_estimators_OAMP-Net2.csv};
%\addlegendentry{\scriptsize OAMP-Net2}
%\addplot[color=myviolet,  smooth, dash dot, mark size=0.05cm,mark options=solid] table[col sep=semicolon, x=EbN0_dB, y=ber_OAMPNet2]{./fig/20230716104841_20k_train_channels/BER_estimators_all.csv};
%\addlegendentry{\scriptsize OAMP-Net2 $20$k}
\addplot[color=myviolet, smooth, dashed, mark size=0.05cm,mark options=solid] table[col sep=semicolon, x=EbN0_dB, y=ber_OAMPNet2]{./fig/20231216140616_10k_train_channels/BER_estimators_all.csv};
%\addlegendentry{\scriptsize OAMP-Net2 $10$k}
\addplot[color=myviolet, smooth,densely dotted, mark size=0.05cm,mark options=solid] table[col sep=semicolon, x=EbN0_dB, y=ber_OAMPNet2]{./fig/20231216140701_3k_train_channels/BER_estimators_all.csv};
%\addlegendentry{\scriptsize OAMP-Net2 $3$k}

%\addplot[color=red, mark=+, smooth, mark size=0.05cm,mark options=solid] table[col sep=semicolon, x=EbN0_dB, y=ber_DetNet]{./fig/20230710225640_BER_performance_equalizers/BER_estimators_all.csv};
%\addlegendentry{\scriptsize DetNet}
\addplot[color=blue,  smooth, mark size=0.05cm,mark options=solid] table[col sep=semicolon, x=EbN0_dB, y=ber_FCNN_IFFT_incorp]{./fig/20231029224651_SC-FDE_UW_QPSK/BER_estimators_remaining.csv};
%\addlegendentry{\scriptsize KAFCNN}
%\addplot[color=blue,  smooth,dash dot, mark size=0.05cm,mark options=solid] table[col sep=semicolon, x=EbN0_dB, y=ber_FCNN_IFFT_incorp]{./fig/20230716104841_20k_train_channels/BER_estimators_all.csv};
%\addlegendentry{\scriptsize KAFCNN $20$k}
\addplot[color=blue,  smooth, dashed, mark size=0.05cm,mark options=solid] table[col sep=semicolon, x=EbN0_dB, y=ber_FCNN_IFFT_incorp]{./fig/20231216140616_10k_train_channels/BER_estimators_all.csv};
%\addlegendentry{\scriptsize KAFCNN $10$k}
\addplot[color=blue,  smooth, densely dotted, mark size=0.05cm,mark options=solid] table[col sep=semicolon, x=EbN0_dB, y=ber_FCNN_IFFT_incorp]{./fig/20231216140701_3k_train_channels/BER_estimators_all.csv};
%\addlegendentry{\scriptsize KAFCNN $3$k}

\addplot[color=myred,  smooth, mark size=0.05cm,mark options=solid] table[col sep=semicolon, x=EbN0_dB, y=ber_SICNNDiagCvv]{./fig/20231029224651_SC-FDE_UW_QPSK/BER_estimators_SICNNDiagCvv.csv};
%\addlegendentry{\scriptsize SICNNv1}
%\addplot[color=myred,  smooth, dash dot, mark size=0.05cm,mark options=solid] table[col sep=semicolon, x=EbN0_dB, y=ber_SICNNDiagCvv]{./fig/20230716104841_20k_train_channels/BER_estimators_SICNNv1_SICNNv1Red.csv};
%\addlegendentry{\scriptsize SICNNv1 $20$k}
\addplot[color=myred,  smooth, dashed, mark size=0.05cm,mark options=solid] table[col sep=semicolon, x=EbN0_dB, y=ber_SICNNDiagCvv]{./fig/20231216140616_10k_train_channels/BER_estimators_all.csv};
%\addlegendentry{\scriptsize SICNNv1 $10$k}
\addplot[color=myred,  smooth, densely dotted, mark size=0.05cm,mark options=solid] table[col sep=semicolon, x=EbN0_dB, y=ber_SICNNDiagCvv]{./fig/20231216140701_3k_train_channels/BER_estimators_all.csv};
%\addlegendentry{\scriptsize SICNNv1 $3$k}

\addplot[color=cyan,  smooth, mark size=0.07cm,solid,mark options=solid] table[col sep=semicolon, x=EbN0_dB, y=ber_SICNNcpxRedDiagCvv]{./fig/20231029224651_SC-FDE_UW_QPSK/BER_estimators_SICNNCpxRed.csv};
%\addlegendentry{\scriptsize SICNNv1Red}
%\addplot[color=cyan,  smooth, mark size=0.07cm,dash dot,mark options=solid] table[col sep=semicolon, x=EbN0_dB, y=ber_SICNNcpxRedDiagCvv]{./fig/20230716104841_20k_train_channels/BER_estimators_SICNNv1_SICNNv1Red.csv};
%\addlegendentry{\scriptsize SICNNv1Red $20$k}
\addplot[color=cyan,  smooth, mark size=0.07cm,dashed,mark options=solid] table[col sep=semicolon, x=EbN0_dB, y=ber_SICNNcpxRedDiagCvv]{./fig/20231216140616_10k_train_channels/BER_estimators_SICNNv1Red.csv};
%\addlegendentry{\scriptsize SICNNv1Red $10$k}
\addplot[color=cyan,  smooth, mark size=0.07cm,densely dotted,mark options=solid] table[col sep=semicolon, x=EbN0_dB, y=ber_SICNNcpxRedDiagCvv]{./fig/20231216140701_3k_train_channels/BER_estimators_all.csv};
%\addlegendentry{\scriptsize SICNNv1Red $3$k}

\end{semilogyaxis}}
\pik

\def\pik2{
\begin{semilogyaxis}[compat=newest, width=8cm, height=8cm, grid=both, minor grid style={line width=0.05pt}, major grid style={line width=0.05pt}, ylabel={\small BER}, ymax = 1e-1, ymin = 1e-6, xmax=14, xmin=2, xlabel={\small $E_{\text{b}}/N_0$ (dB)}, scaled ticks = false,  x tick label style={/pgf/number format/.cd,fixed,precision=2,/tikz/.cd}, ticklabel style={font=\scriptsize}, legend columns = {3}, legend cell align=left, legend style={font=\scriptsize, inner sep = 0.3pt, at={(0.5,1.01)}, anchor=south west}, every axis plot/.append style={line width=0.25pt}]

\addplot[color=mygreen, smooth, mark size=0.05cm] table[col sep=semicolon, x=EbN0dB, y=DF]{./fig/20230710225640_BER_performance_equalizers/Matlab_results.csv};
%\addlegendentry{\scriptsize DFE}
\addplot[color=myorange,  smooth, mark size=0.05cm] table[col sep=semicolon, x=EbN0dB, y=LMMSE low rate]{./fig/20230710225640_BER_performance_equalizers/Matlab_results.csv};
%\addlegendentry{\scriptsize LMMSE}

\addplot[color=myviolet, smooth, mark size=0.05cm,mark options=solid] table[col sep=semicolon, x=EbN0_dB, y=ber_OAMPNet2]{./fig/20231029224651_SC-FDE_UW_QPSK/BER_estimators_OAMP-Net2.csv};
%\addlegendentry{\scriptsize OAMP-Net2}
%\addplot[color=myviolet,  smooth, dash dot, mark size=0.05cm,mark options=solid] table[col sep=semicolon, x=EbN0_dB, y=ber_OAMPNet2]{./fig/20230716104841_20k_train_channels/BER_estimators_all.csv};
%\addlegendentry{\scriptsize OAMP-Net2 $20$k}
\addplot[color=myviolet, smooth, dashed, mark size=0.05cm,mark options=solid] table[col sep=semicolon, x=EbN0_dB, y=ber_OAMPNet2]{./fig/20231216140616_10k_train_channels/BER_estimators_all.csv};
%\addlegendentry{\scriptsize OAMP-Net2 $10$k}
\addplot[color=myviolet, smooth,densely dotted, mark size=0.05cm,mark options=solid] table[col sep=semicolon, x=EbN0_dB, y=ber_OAMPNet2]{./fig/20231216140701_3k_train_channels/BER_estimators_all.csv};
%\addlegendentry{\scriptsize OAMP-Net2 $3$k}

%\addplot[color=red, mark=+, smooth, mark size=0.05cm,mark options=solid] table[col sep=semicolon, x=EbN0_dB, y=ber_DetNet]{./fig/20230710225640_BER_performance_equalizers/BER_estimators_all.csv};
%\addlegendentry{\scriptsize DetNet}
\addplot[color=blue,  smooth, mark size=0.05cm,mark options=solid] table[col sep=semicolon, x=EbN0_dB, y=ber_FCNN_IFFT_incorp]{./fig/20231029224651_SC-FDE_UW_QPSK/BER_estimators_remaining.csv};
%\addlegendentry{\scriptsize KAFCNN}
%\addplot[color=blue,  smooth,dash dot, mark size=0.05cm,mark options=solid] table[col sep=semicolon, x=EbN0_dB, y=ber_FCNN_IFFT_incorp]{./fig/20230716104841_20k_train_channels/BER_estimators_all.csv};
%\addlegendentry{\scriptsize KAFCNN $20$k}
\addplot[color=blue,  smooth, dashed, mark size=0.05cm,mark options=solid] table[col sep=semicolon, x=EbN0_dB, y=ber_FCNN_IFFT_incorp]{./fig/20231216140616_10k_train_channels/BER_estimators_all.csv};
%\addlegendentry{\scriptsize KAFCNN $10$k}
\addplot[color=blue,  smooth, densely dotted, mark size=0.05cm,mark options=solid] table[col sep=semicolon, x=EbN0_dB, y=ber_FCNN_IFFT_incorp]{./fig/20231216140701_3k_train_channels/BER_estimators_all.csv};
%\addlegendentry{\scriptsize KAFCNN $3$k}

\addplot[color=myred,  smooth, mark size=0.05cm,mark options=solid] table[col sep=semicolon, x=EbN0_dB, y=ber_SICNNDiagCvv]{./fig/20231029224651_SC-FDE_UW_QPSK/BER_estimators_SICNNDiagCvv.csv};
%\addlegendentry{\scriptsize SICNNv1}
%\addplot[color=myred,  smooth, dash dot, mark size=0.05cm,mark options=solid] table[col sep=semicolon, x=EbN0_dB, y=ber_SICNNDiagCvv]{./fig/20230716104841_20k_train_channels/BER_estimators_SICNNv1_SICNNv1Red.csv};
%\addlegendentry{\scriptsize SICNNv1 $20$k}
\addplot[color=myred,  smooth, dashed, mark size=0.05cm,mark options=solid] table[col sep=semicolon, x=EbN0_dB, y=ber_SICNNDiagCvv]{./fig/20231216140616_10k_train_channels/BER_estimators_all.csv};
%\addlegendentry{\scriptsize SICNNv1 $10$k}
\addplot[color=myred,  smooth, densely dotted, mark size=0.05cm,mark options=solid] table[col sep=semicolon, x=EbN0_dB, y=ber_SICNNDiagCvv]{./fig/20231216140701_3k_train_channels/BER_estimators_all.csv};
%\addlegendentry{\scriptsize SICNNv1 $3$k}

\addplot[color=cyan,  smooth, mark size=0.07cm,solid,mark options=solid] table[col sep=semicolon, x=EbN0_dB, y=ber_SICNNcpxRedDiagCvv]{./fig/20231029224651_SC-FDE_UW_QPSK/BER_estimators_SICNNCpxRed.csv};
%\addlegendentry{\scriptsize SICNNv1Red}
%\addplot[color=cyan,  smooth, mark size=0.07cm,dash dot,mark options=solid] table[col sep=semicolon, x=EbN0_dB, y=ber_SICNNcpxRedDiagCvv]{./fig/20230716104841_20k_train_channels/BER_estimators_SICNNv1_SICNNv1Red.csv};
%\addlegendentry{\scriptsize SICNNv1Red $20$k}
\addplot[color=cyan,  smooth, mark size=0.07cm,dashed,mark options=solid] table[col sep=semicolon, x=EbN0_dB, y=ber_SICNNcpxRedDiagCvv]{./fig/20231216140616_10k_train_channels/BER_estimators_SICNNv1Red.csv};
%\addlegendentry{\scriptsize SICNNv1Red $10$k}
\addplot[color=cyan,  smooth, mark size=0.07cm,densely dotted,mark options=solid] table[col sep=semicolon, x=EbN0_dB, y=ber_SICNNcpxRedDiagCvv]{./fig/20231216140701_3k_train_channels/BER_estimators_all.csv};
%\addlegendentry{\scriptsize SICNNv1Red $3$k}

\end{semilogyaxis}}

% first zoom
    %% spy on node
    \node[spy,minimum size={\spyviewersize/\spyfactorI}] (spy-on node 1) at (spy-on 1) {};
    %% spy in node
    \node[spy,minimum size=\spyviewersize] (spy-in node 1) at (spy-in 1) {};
    \begin{scope}
        \clip (spy-in 1) circle (0.5*\spyviewersize-\spyonclipreduce);
        \node[circle, fill=white, draw=white, minimum size=\spyviewersize] (background_circ) at (spy-in 1) {};
        \pgfmathsetmacro\sI{1/\spyfactorI}
        \begin{scope}[
            shift={($\sI*(spy-in 1)-\sI*(spy-on 1)$)},
            scale around={\spyfactorI:(spy-on 1)}
        ]
            \pik2
        \end{scope}
    \end{scope}
    %% connect the nodes
    \draw [spy] (spy-on node 1) -- (spy-in node 1);
\end{tikzpicture}
\vspace{-0.3cm}
\caption{BER performance comparison for different training set sizes (SC-FDE with UW guard, QPSK alphabet).}
\label{fig:BER_performance_train_set_size}
\end{center}
\vspace{-0.3cm}
\end{figure}

In this section, we investigate the influence of a limited training set size on the BER performance of selected NN-based data estimators. That is, while for the BER performance results shown in Sec.~\ref{ssec:BER_performance} the NN-based equalizers are trained utilizing sample data transmissions over $30\,000$ different multipath channels, the regarded NNs are trained with training sets consisting of $10\,000$ or $3\,000$ channels. We evaluate the influence of a reduced training set size on the BER performance of selected NN-based equalizers for an SC-FDE system with UW-guard interval and QPSK modulation alphabet. The hyperparameters of the NN-based equalizers are the same as stated in Sec.~\ref{ssec:Simulation_Setup_and_NN_Training}, apart from the number of training epochs which is adapted appropriately such that the number of update steps of the learnable NN parameters remains the same for all training set sizes. 

We regard SICNNv1, its parameter-reduced variant SICNNv1Red, the OAMP-Net2, and the KAFCNN for performance comparison, whereby these NNs have for the chosen hyperparameter settings $135\,590$, $19\,370$, $32$, and $746\,628$ learnable parameters, respectively. As shown in Fig.~\ref{fig:BER_performance_train_set_size}, the performance of SICNNv1 and SICNNv1red slightly degrades in case of a reduced training set size of $10\,000$ or $3\,000$ channels.  The BER performance of OAMP-Net2 barely changes when reducing the training set size, while the performance of KAFCNN decreases most and is even worse than that of the LMMSE estimator when being trained with $3\,000$ different channels. That is, the fewer learnable parameters an NN contains, the less it suffers from a limited training set size. This result emphasizes the importance of parameter reduction, e.g., by incorporating model knowledge into the layer architecture of an NN.

\subsection{Bit Error Ratio Performance of SICNNv2 for UW-OFDM}
\label{ssec:BER_performance_SICNNv2_UW-OFDM}
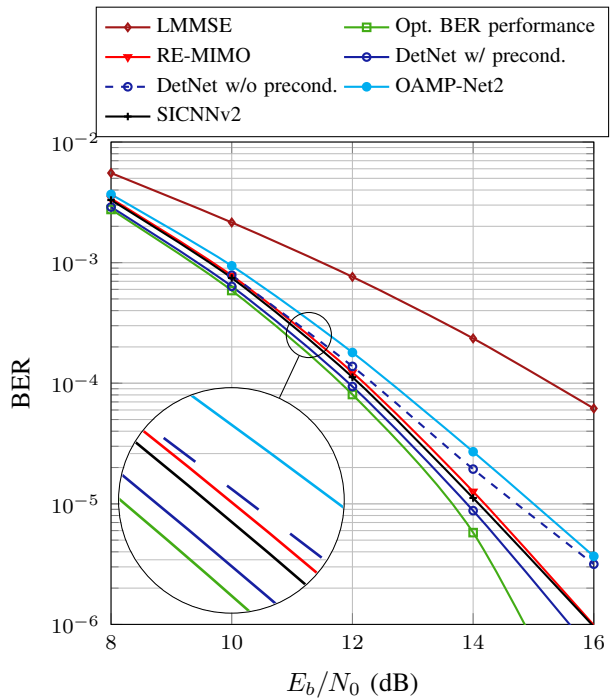
\begin{figure}[t]
\begin{center}
\begin{tikzpicture}[
    % Style for the spy nodes and the connection line
    spy/.style={%
        draw,black,
        line width=0.4pt,
        circle,inner sep=0pt,
    },
]%[spy using outlines={rectangle, magnification=10, connect spies}]
%% size of the spy-in nodes
    \def\spyviewersize{3cm}

    %% (line width of the spy nodes) / 2
    %% we need this for clipping later
    \def\spyonclipreduce{0.4pt}

    %% first zoom
    %%% factor
    \def\spyfactorI{5}
    %%% spy in point
    \coordinate (spy-in 1) at (1.6,1.65);
    %%% spy on point
    \coordinate (spy-on 1) at (2.63,3.85);% sould be on the curve

\def\pik{
\begin{semilogyaxis}[compat=newest, width=8cm, height=8cm, grid=both, ylabel={\small BER}, ymax = 1e-2, ymin = 1e-6, xmax=16, xmin=8, xlabel={\small $E_b/N_0$ (dB)}, scaled ticks = false,  x tick label style={/pgf/number format/.cd,fixed,precision=2,/tikz/.cd}, ticklabel style={font=\scriptsize}, legend style={at={(0.5,1.01)}, anchor=south}, legend columns = {2}, legend cell align=left, legend style={font=\scriptsize}, every axis plot/.append style={thick}]
\addplot[color=myred, mark=diamond, smooth, mark size=0.05cm] table[col sep=semicolon, x=EbN0_dB, y=ber_lin_est]{./fig/20210317113627_small_non_sys_uncoded_QPSK/BER_estimators.csv};
\addlegendentry{\scriptsize LMMSE}
\addplot[color=mygreen, mark=square, smooth, mark size=0.05cm] table[col sep=semicolon, x=Eb_N0_dB, y=ber_mmse]{./fig/20210317113627_small_non_sys_uncoded_QPSK/Matlab_results_MMSE.csv};
\addlegendentry{\scriptsize Opt. BER performance}
\addplot[color=red, mark=triangle, mark options={rotate=180}, smooth, mark size=0.05cm] table[col sep=semicolon, x=EbN0_dB, y=ber_ReMIMO]{./fig/2023056194224_UW-OFDM_small_non_sys_uncoded_QPSK_SICNNv2_RE-MIMO/BER_estimators_RE-MIMO.csv};
\addlegendentry{\scriptsize RE-MIMO}
\addplot[color=myblue, mark=o, smooth, mark size=0.05cm] table[col sep=semicolon, x=EbN0_dB, y=ber_DetNet]{./fig/20210317113627_small_non_sys_uncoded_QPSK/BER_estimators.csv};
\addlegendentry{\scriptsize DetNet w/ precond.}
\addplot[color=myblue, mark=o, dashed, mark options={solid}, smooth, mark size=0.05cm] table[col sep=semicolon, x=EbN0_dB, y=ber_DetNet]{fig/20210112212129_influence_preprocessing/BER_estimators_wo_precond_latest.csv};
\addlegendentry{\scriptsize DetNet w/o precond.}
%\addplot[color=mygreen, mark=triangle, smooth, mark size=0.05cm] table[col sep=semicolon, x=EbN0_dB, y=ber_DF]{./fig/20210317113627_small_non_sys_uncoded_QPSK/BER_estimators.csv};
%\addlegendentry{\scriptsize DFE}
\addplot[color=cyan, mark=*, smooth, mark size=0.05cm] table[col sep=semicolon, x=EbN0_dB, y=ber_OAMPNet2]{./fig/20210317113627_small_non_sys_uncoded_QPSK/BER_estimators_OAMP-Net2.csv};
\addlegendentry{\scriptsize OAMP-Net2}
\addplot[color=black, mark=+, smooth, mark size=0.05cm] table[col sep=semicolon, x=EbN0_dB, y=ber_SICNNDirectProbEst]{./fig/2023056194224_UW-OFDM_small_non_sys_uncoded_QPSK_SICNNv2_RE-MIMO/BER_estimators.csv};
\addlegendentry{\scriptsize SICNNv2}

%\coordinate (spypoint) at (axis cs:9.6,1e-3);
%\coordinate (spyviewer) at (axis cs:10.3,1e-5);
%\spy[width=2.5cm, height=2cm] on (spypoint) in node [fill=white] at (spyviewer);
\end{semilogyaxis}}
\pik

\def\pik2{
\begin{semilogyaxis}[compat=newest, width=8cm, height=8cm, grid=both, minor grid style={line width=0.05pt}, major grid style={line width=0.05pt}, ylabel={\small BER}, ymax = 1e-2, ymin = 1e-6, xmax=16, xmin=8, xlabel={\small $E_b/N_0$ (dB)}, scaled ticks = false,  x tick label style={/pgf/number format/.cd,fixed,precision=2,/tikz/.cd}, ticklabel style={font=\scriptsize}, every axis plot/.append style={line width=0.2pt}]

\addplot[color=myred, mark=diamond, smooth, mark size=0.05cm] table[col sep=semicolon, x=EbN0_dB, y=ber_lin_est]{./fig/20210317113627_small_non_sys_uncoded_QPSK/BER_estimators.csv};

\addplot[color=mygreen, mark=square, smooth, mark size=0.05cm] table[col sep=semicolon, x=Eb_N0_dB, y=ber_mmse]{./fig/20210317113627_small_non_sys_uncoded_QPSK/Matlab_results_MMSE.csv};

\addplot[color=red, mark=triangle, mark options={rotate=180}, smooth, mark size=0.05cm] table[col sep=semicolon, x=EbN0_dB, y=ber_ReMIMO]{./fig/2023056194224_UW-OFDM_small_non_sys_uncoded_QPSK_SICNNv2_RE-MIMO/BER_estimators_RE-MIMO.csv};

\addplot[color=myblue, mark=o, smooth, mark size=0.05cm] table[col sep=semicolon, x=EbN0_dB, y=ber_DetNet]{./fig/20210317113627_small_non_sys_uncoded_QPSK/BER_estimators.csv};

\addplot[color=myblue, mark=o, dashed, mark options={solid}, smooth, mark size=0.05cm] table[col sep=semicolon, x=EbN0_dB, y=ber_DetNet]{fig/20210112212129_influence_preprocessing/BER_estimators_wo_precond_latest.csv};

%\addplot[color=mygreen, mark=triangle, smooth, mark size=0.05cm] table[col sep=semicolon, x=EbN0_dB, y=ber_DF]{./fig/20210317113627_small_non_sys_uncoded_QPSK/BER_estimators.csv};

\addplot[color=cyan, mark=*, smooth, mark size=0.05cm] table[col sep=semicolon, x=EbN0_dB, y=ber_OAMPNet2]{./fig/20210317113627_small_non_sys_uncoded_QPSK/BER_estimators_OAMP-Net2.csv};

\addplot[color=black, mark=+, smooth, mark size=0.05cm] table[col sep=semicolon, x=EbN0_dB, y=ber_SICNNDirectProbEst]{./fig/2023056194224_UW-OFDM_small_non_sys_uncoded_QPSK_SICNNv2_RE-MIMO/BER_estimators.csv};

%\coordinate (spypoint) at (axis cs:9.6,1e-3);
%\coordinate (spyviewer) at (axis cs:10.3,1e-5);
%\spy[width=2.5cm, height=2cm] on (spypoint) in node [fill=white] at (spyviewer);
\end{semilogyaxis}}

% first zoom
    %% spy on node
    \node[spy,minimum size={\spyviewersize/\spyfactorI}] (spy-on node 1) at (spy-on 1) {};
    %% spy in node
    \node[spy,minimum size=\spyviewersize] (spy-in node 1) at (spy-in 1) {};
    \begin{scope}
        \clip (spy-in 1) circle (0.5*\spyviewersize-\spyonclipreduce);
        \node[circle, fill=white, draw=white, minimum size=\spyviewersize] (background_circ) at (spy-in 1) {};
        \pgfmathsetmacro\sI{1/\spyfactorI}
        \begin{scope}[
            shift={($\sI*(spy-in 1)-\sI*(spy-on 1)$)},
            scale around={\spyfactorI:(spy-on 1)}
        ]
            \pik2
        \end{scope}
    \end{scope}
    %% connect the nodes
    \draw [spy] (spy-on node 1) -- (spy-in node 1);

\end{tikzpicture}
\vspace{-0.3cm}
\caption{BER performance comparison for UW-OFDM (system I from~\cite{Baumgartner23_J1}).}
\vspace{-0.3cm}
\label{fig:system_I_non_sys_uncoded_QPSK}
\end{center}
\end{figure}

As mentioned in Sec.~\ref{ssec:SICNNv2}, the layer architecture of SICNNv2 is inferred by deep unfolding iterative SIC, but no properties of any specific communication system are exploited. Hence, we expect SICNNv2 to be universally applicable for any communication system with system models similar to the system model~\eqref{eq:system_model_SC-FDE} of SC-FDE systems. To demonstrate this claim, we apply SICNNv2 as an equalizer in a UW-OFDM system~\cite{Huemer10_2, Huemer12, Lang19}. The data transmission in UW-OFDM systems can be modeled as~\cite{Huemer10_2, Huemer12, Baumgartner23_J1}
\begin{gather}
\ve{y} = \underbrace{\widetilde{\m{H}}\m{G}}_{\m{H}}\ve{d} + \ve{w}\,,
\end{gather}
where $\ve{d}\in\mathbb{S}^{N_{\text{d}}}$ is the transmitted data vector of length $N_{\text{d}}$ to be estimated, $\ve{y}\in\mathbb{C}^{N_{\text{d}}+N_{\text{u}}}$ the received vector at the input of the equalizer, and $N_{\text{u}}$ the length of the UW guard interval. Further, $\widetilde{\m{H}}\in\mathbb{C}^{(N_{\text{d}}+N_{\text{u}})\times (N_{\text{d}}+N_{\text{u}})}$ denotes a diagonal matrix containing the sampled channel frequency response of the channel (excluding at positions of OFDM zero-subcarriers) on the main diagonal, $\m{G}\in\mathbb{C}^{(N_{\text{d}}+N_{\text{u}})\times N_{\text{d}}}$ the so-called generator matrix, which is a full, rectangular matrix, and $\ve{w}\sim\mathcal{NC}(\ve{0}, (N_{\text{d}}+N_{\text{u}})\sigma_{\text{n}}^2\m{I})$, where $\sigma_{\text{n}}^2$ is the variance of AWGN in time domain. For further details on UW-OFDM we refer to~\cite{Huemer10_2, Huemer12}. That is, the models of UW-OFDM and SC-FDE systems are very similar, allowing to employ SICNNv2 unaltered for UW-OFDM, apart from the used data normalization, which is described in~\cite{Baumgartner23_J1}. We train (exactly in the same manner as all other state-of-the-art NNs used for comparison) and evaluate SICNNv2 for a UW-OFDM system referred to as system~I in~\cite{Baumgartner23_J1}, where ${N_{\text{d}} = 8}$, ${N_{\text{u}}=4}$, and the modulation alphabet is QPSK. The best hyperparameter combination found is a learning rate ${\eta = 5\cdot 10^{-4}}$, ${Q=6}$ stages, ${n_{\text{L}} = 2}$ hidden layers of the sub-NNs, and ${n_{\text{H}}=200}$ neurons per hidden layer of the sub-NNs. SICNNv2 is compared to the state-of-the-art NN-based equalizers OAMP-Net2~\cite{He20}, RE-MIMO~\cite{Pratik21}, DetNet~\cite{Samuel19}, and an improved version of DetNet, that employs a preconditioner in its layers~\cite{Baumgartner23_J1}. Due to the small dimension of system~I, even the optimal BER performance can be computed by applying the bit-wise MAP estimator. For all further details on the simulation setup, on the NNs used for comparison, or their training, we refer to~\cite{Baumgartner23_J1}. As shown in Fig.~\ref{fig:system_I_non_sys_uncoded_QPSK}, SICNNv2 can outperform the NN-based equalizers OAMP-Net2, DetNet in its original form~\cite{Samuel19}, and performs similar as RE-MIMO. Further, its performance is very close to that of the improved version of DetNet and to the optimal BER performance achieved by the bit-wise MAP estimator. This result demonstrates the applicability of the proposed SICNN idea for different communication systems.

\subsection{Computational Complexity}
Besides the BER performance of the regarded equalizers, also their computational complexity is an important aspect. We compare the inference complexity of the model-based and the NN-based equalizers regarded in this work in terms of the number of real-valued multiplications required for the equalization of a received vector, where four real-valued multiplications and two real-valued multiplications are accounted for a product of two complex values and a real and a complex value, respectively, and divisions are counted as multiplications. Since NN training can be carried out offline, we do not regard their training complexity. We assume that both $\widetilde{\m{H}}$ and $\m{H} = \widetilde{\m{H}}\m{M}$ are already available and thus the complexity of computing $\m{H}$ is not considered for the following complexity analysis. Unless stated otherwise, we derive the computational complexities for a general matrix $\m{M}$ and a length $N^\prime$ of the received vector, where, as described in Sec.~\ref{ssec:System_Model_for_Singel_Carrier_FDE}, both have to be replaced by the appropriate quantities $\m{M}_{\text{uw}}$ and $N$, or $\m{F}_{N_{\text{d}}}$ and $N_{\text{d}}$, when using a UW or a CP as a guard interval, respectively.

We start by investigating the complexity of the proposed SICNNv1. Let us first consider the operations conducted in a single stage $q$ for estimating the data symbol $d_k$. The computation of $\hat{d}_{k,\text{Re}}^{(q)}$, $\hat{d}_{k,\text{Im}}^{(q)}$, $e_k^{(q)}$, $\ve{y}_{\text{ic},k}^{(q)}$, $\ve{a}_{k}^{(q)}$, takes
\begin{gather*}
\underbrace{|\mathbb{S}^\prime|}_{\hat{d}_{k,\text{Re}}^{(q)}} + \underbrace{|\mathbb{S}^\prime|}_{\hat{d}_{k,\text{Im}}^{(q)}} + \underbrace{4|\mathbb{S}^\prime| + 2}_{e_k^{(q)}} + \underbrace{4N^\prime(N_{\text{d}}-1)}_{\ve{y}_{\text{ic},k}^{(q)}} + \underbrace{2N^\prime(N_{\text{d}}-1)}_{\ve{a}_k^{(q)}}
\end{gather*}
real-valued multiplications, and the inference of FCNN~1 has a complexity of
\begin{gather*}
\underbrace{2(3N^\prime+1)}_{\text{batch norm.}} + (3N^\prime+1)n_{\text{H,C}}+(n_{\text{L,C}}-1)n_{\text{H,C}}^2+n_{\text{H,C}}N^\prime\,.
\end{gather*}
Squaring the outputs of FCNN~1 takes another $N^\prime$ multiplications. For the terms $\ve{h}_k^H\hat{\m{C}}_{\ve{v}\ve{v},k}^{(q)^{-1}}\ve{y}_{\text{ic},k}^{(q)}$ and $\ve{h}_k^H\hat{\m{C}}_{\ve{v}\ve{v},k}^{(q)^{-1}}\ve{h}_{k}$, $\ve{h}_k^H\hat{\m{C}}_{\ve{v}\ve{v},k}^{(q)^{-1}}$ is computed first and multiplied by $\ve{y}_{\text{ic},k}^{(q)}$ and $\ve{h}_k$ subsequently, leading for these two terms in total to a complexity of 
\begin{gather*}
\underbrace{2N^\prime}_{\ve{h}_k^H\hat{\m{C}}_{\ve{v}\ve{v},k}^{(q)^{-1}}} + \underbrace{4N^\prime}_{\cdot \ve{y}_{\text{ic},k}^{(q)}}  + \underbrace{4N^\prime}_{\cdot\ve{h}_k^{\vphantom{(q)}}}\,.
\end{gather*}
The inference of FCNN~2 and the normalization of $\ve{y}_{\text{ic},k}^{(q)}$ and $\ve{h}_k$ require another
\begin{gather*}
6+3n_{\text{H,pr}}+(n_{\text{L,pr}}-1)n_{\text{H,pr}}^2+2n_{\text{H,pr}}|\mathbb{S}^\prime|
\end{gather*}
and $8N^\prime+1$ multiplications, respectively. Consequently, for estimating a single data symbol in a stage, in total 
\begin{gather}
\begin{aligned}
M_{\text{SICNNv1},kq} &= n_{\text{H,pr}}^2 (n_\text{{L,pr}}-1) + n_{\text{H,pr}}(2|\mathbb{S}^\prime|+3)\\&\quad + n_{\text{H,C}}^2(n_{\text{L,C}}-1)+n_{\text{H,C}}(4N^\prime+1)\\&\quad+19N^\prime + 6|\mathbb{S}^\prime|+6N^\prime N_{\text{d}} + 11
\end{aligned}\label{eq:SICNNv1_kq_complexity}
\end{gather}
For data normalization, another 
\begin{gather*}
 \underbrace{4N_{\text{d}}N^\prime+1}_{\kappa} + \underbrace{N^\prime}_{\kappa\widetilde{\m{H}}} + \underbrace{2N^\prime}_{\vphantom{\widetilde{\m{H}}}\m{K}\ve{y}} + \underbrace{N^\prime}_{\m{K}\widetilde{\m{H}}}
\end{gather*}
real-valued multiplications are required, leading to a total complexity of SICNNv1 
\begin{gather}
M_{\text{SICNNv1}} = QN_{\text{d}}M_{\text{SICNNv1},kq} + 4N_{\text{d}}N^\prime + 4N^\prime + 1\,,\label{eq:SICNNv1_complexity}
\end{gather}
with $M_{\text{SICNNv1},kq}$ as specified in~\eqref{eq:SICNNv1_kq_complexity}. 

For SICNNv2, the approach for deriving its complexity is similar. As already stated for SICNNv1, the total complexity for computing $\hat{d}_{k,\text{Re}}^{(q)}$, $\hat{d}_{k,\text{Im}}^{(q)}$, $e_{k,\text{Re}}^{(q)}$, $e_{k,\text{Im}}^{(q)}$, and $\ve{y}_{\text{ic},k}^{(q)}$ is given by ${6|\mathbb{S}^\prime| +4N^\prime(N_{\text{d}}-1)}$.
For computing the scaling factor $\rho_k^{(q)}$, scaling the quantities contained in $\ve{z}_k^{(q)}$ (and squaring $\rho_k^{(q)}$), and the inference of the FCNN, another
\begin{align*}
&\underbrace{4N^\prime}_{\rho_k^{(q)}} + \underbrace{4N^\prime+4}_{\text{scaling}} + 
\big(1+\Big\lfloor \frac{n_{\text{L}}}{3}\Big\rfloor\big)\underbrace{(6N^\prime+2)}_{\text{batch norm.}} + (4N^\prime+3)n_{\text{H}}\\&\quad+(n_{\text{L}}-1)n_{\text{H}}^2+2n_{\text{H}}|\mathbb{S}^\prime|
\end{align*}
multiplications are required. Consequently, the complexity of SICNNv2 follows to
\begin{gather}
M_{\text{SICNNv2}} = QN_{\text{d}}M_{\text{SICNNv2},kq} + 4N_{\text{d}}N^\prime + 4N^\prime + 1\,,\label{eq:SICNNv2_complexity}
\end{gather}
with
\begin{gather}
\begin{aligned}
M_{\text{SICNNv2},kq} &= \Big\lfloor \frac{n_{\text{L}}}{3} \Big\rfloor (6N^\prime+2)+ n_{\text{H}}^2(n_{\text{L}}-1)+ 4N^\prime N_{\text{d}} \\&\quad + 10N^\prime + n_{\text{H}}(4N^\prime+2|\mathbb{S}^\prime|+3) + 6|\mathbb{S}^\prime| + 6\,.
\end{aligned}
\end{gather}

The complexity of DetNet can be derived similarly as for UW-OFDM, which has been conducted in~\cite{Baumgartner23_J1}. For all operations conducted for SC-FDE specific pre-processing and for the inference of DetNet, we refer to~\cite{Baumgartner23_C1}. Here, we only state the final result for the inference complexity of DetNet. In the $q$th DetNet layer inference of a single hidden layer FCNN, one-hot demapping of the data vector estimate in the $q$th layer, and applying weighted residual connections are conducted, which entails
\begin{gather}
\begin{aligned}
M_{\text{DetNet},q} &= 4N_{\text{d}}^2 + 6N_{\text{d}} + 2d_{\text{h}}(N_{\text{d}}(|\mathbb{S}^\prime|+1)+d_{\text{v}})\\&\quad +2N_{\text{d}}|\mathbb{S}^\prime| + d_{\text{v}}
\end{aligned}
\end{gather}
real-valued multiplications, where $d_{\text{h}}$ is the number of neurons in the hidden layer of the FCNN, and $d_{\text{v}}$ is the dimension of an auxiliary variable passing unconstrained information from DetNet layer to DetNet layer~\cite{Samuel19, Baumgartner23_C1, Baumgartner23_J1}.
In total, the inference complexity of DetNet is
\begin{gather}
\begin{aligned}
M_{\text{DetNet}} &= L M_{\text{DetNet},q} - 2N_{\text{d}}|\mathbb{S}^\prime| + \underbrace{4N_{\text{d}}N^\prime + 4N^\prime + 1}_{\text{data normalization}}\\&\quad + \underbrace{6N_{\text{d}}N^\prime + 4N_{\text{d}}^2 N^\prime + 2N^\prime}_{\text{input pre-processing}} \,,
\end{aligned}
\end{gather}
where the subtracted term accounts for no one-hot demapping in the last DetNet layer, $L$ is the number of DetNet layers, and for the input pre-processing the quantities $\m{M}^H\widetilde{\m{H}}^{\nicefrac{1}{2}}\m{M}$ and $\m{M}^H\widetilde{\m{H}}^{-\nicefrac{1}{2}}\ve{y}$ are computed (cf.~\cite{Baumgartner23_C1}).

The KAFCNN~\cite{Baumgartner23_C1} with weighted residual connections and a multiplication by a partial inverse DFT matrix in the last layer requires
\begin{gather}
\begin{aligned}
M_{\text{KAFCNN}} &= \big(3N^\prime + (n_{\text{L}}-1)n_{\text{H}} + 2N^\prime |\mathbb{S}^\prime| + (n_{\text{L}}-2)\big) n_{\text{H}}\\&\quad + \underbrace{4N^\prime N_{\text{d}}|\mathbb{S}^\prime| + 2N_{\text{d}}|\mathbb{S}^\prime|}_{\text{part. IDFT + symb. scaling}} + \underbrace{4N_{\text{d}}N^\prime + 4N^\prime + 1}_{\text{data normalization}}
\end{aligned}
\end{gather}

The OAMP-Net2 layer structure~\cite{He20} also stems from unfolding an iterative model-based algorithm, such that determining its inference complexity is conducted in a similar fashion as for DetNet or SICNNv1/SICNNv2. For the matrix inverse that has to be computed in every layer of OAMP-Net2, we assume that a Cholesky decomposition~\cite{Golub13} is employed for accomplishing this task. 
The computational complexity of OAMP-Net2 can thus be specified as (using the notation from~\cite{He20})
\begin{gather}
\begin{aligned}
M_{\text{ONet2}} &= T\Big(\underbrace{8N_{\text{d}}N^\prime+2N_{\text{d}}}_{\ve{r}_t}+\underbrace{4N_{\text{d}}^2(2N^\prime+1)+8N_{\text{d}}N^\prime+5}_{\tau_t^2}\\&\hspace{-0.6cm} + \underbrace{\frac{14}{3}N^{\prime\,3}+8N^{\prime\,2}(2N_{\text{d}}+1)+8N_{\text{d}}(N^\prime+|\mathbb{S}^\prime|+\frac{1}{2})+2}_{\hat{\ve{x}}_{\text{d},t+1}}\\&\hspace{-0.6cm} +\underbrace{2N_{\text{d}}(2N^\prime+1)+1}_{v_t^2}\Big) + \underbrace{4N_{\text{d}}N^\prime + 4N^\prime + 1}_{\text{data normalization}}
\end{aligned}
\end{gather}

Let us now consider the complexity of the model-based equalizers. For the LMMSE estimator, one has to distinguish between UW and CP guard intervals. We start by regarding its complexity in case of a UW guard interval. Here, we first regard its complexity for determining the LMMSE estimator matrix $\m{E}_{\text{LMMSE}}$, which is independent of the received vector $\ve{y}$ and thus has to be computed only once per data burst (the channel is assumed to be stationary for the whole data burst). By assuming that the inverse in~\eqref{eq:LMMSE_v1} is computed utilizing a Cholesky decomposition~\cite{Golub13},  
\begin{align}
M_{\text{LMMSE,burst}} &= \underbrace{4N_{\text{d}}^2(N_{\text{d}}+N_{\text{g}})}_{\m{M}^H\m{H}}+\underbrace{\frac{14}{3} N_{\text{d}}^3+4N_{\text{d}}^2}_{\text{inverse (Cholesky)}}+\underbrace{4N_{\text{d}}^2(N_{\text{d}}+N_{\text{g}})}_{\cdot\m{M}^H}\nonumber\\
&= \frac{38}{3}N_{\text{d}}^3 + 8N_{\text{d}}^2N_{\text{g}}+4N_{\text{d}}^2
\end{align}
real-valued multiplications are to be carried out for computing $\m{E}_{\text{LMMSE}}$. Given $\m{E}_{\text{LMMSE}}$, the complexity of equalizing one received vector is
\begin{gather}
M_{\text{LMMSE,eq}} = 4N_{\text{d}}(N_{\text{d}} + N_{\text{g}})\,.
\end{gather}

In case of a CP guard interval, equalization becomes for the LMMSE estimator far less complex. As given in~\eqref{eq:LMMSE_diag}, the matrix for which an inverse has to be computed is a diagonal matrix. Hence, obtaining the estimator matrix $\m{E}_{\text{LMMSE,dg}}$ requires $4N_{\text{d}}$ real-valued multiplications, which is already the number of multiplications that have to be carried out per burst
\begin{gather}
M_{\text{LMMSE,burst}} = 4N_{\text{d}}\,.
\end{gather}  
For equalization of a single received data vector, a multiplication with the diagonal estimator matrix $\m{E}_{\text{LMMSE,dg}}$ is required, followed by conducting an inverse DFT, leading to a complexity of
\begin{gather}
M_{\text{LMMSE,eq}} = 4N_{\text{d}} + \underbrace{2N_{\text{d}}\log_2(N_{\text{d}})}_{\text{IDFT}}\,,
\end{gather}
real-valued multiplications.

For the DFE, using a CP guard interval does not reduce the complexity as for the LMMSE, since in every iteration the LMMSE error variances (which are the diagonal elements of the LMMSE error covariance matrix) have to be computed. We distinguish between operations to be carried out only once every data burst and those to be accomplished for every received vector. For a derivation of the complexity of the DFE we refer to~\cite{Baumgartner23_J1}, where an in-depth complexity analysis of the DFE for a similar system, namely a UW-OFDM system is conducted. Here, we only state the final results. The number of real-valued multiplications to be conducted once per data burst is
\begin{gather}
M_{\text{DFE,burst}} = \frac{7}{6}N_{\text{d}}^4 + \frac{11}{3}N_{\text{d}}^3 + \frac{19}{6}N_{\text{d}}^2 + 6N_{\text{d}}^2N^\prime + \frac{2}{3}N_{\text{d}} + 2N_{\text{d}}N^\prime - \frac{14}{3}\,,
\end{gather}
while the inference complexity per received vector is given by
\begin{gather}
M_{\text{DFE,eq}} = 8N_{\text{d}}N^\prime\,.
\end{gather}

For the iterative SIC method, the same steps have to be conducted $Q$ times, where --~as for the LMMSE estimator~-- a Cholesky decomposition is utilized for inverting the (approximated) covariance matrices $\m{C}_{\ve{v}\ve{v},k}^{(q)}$ in every iteration. We assume that the estimated data symbols $\hat{d}_k^{(q-1)}$ are multiplied by the corresponding column of $\m{H}$ only once per iteration, being available for interference cancellation required for estimating any data symbol $d_k$. Hence, the computational complexity of the iterative SIC method with $Q$ iterations is 
\begin{align}
M_{\text{itSIC}} &= Q N_{\text{d}}\Big(\underbrace{\hspace{-0.1cm}6|\mathbb{S}^\prime|+2}_{\hat{d}_k^{(q)}, e_k^{(q)}} + \underbrace{(N_{\text{d}}-1)(2N^\prime+4N^{\prime\,2})+4N^\prime}_{\m{C}_{\ve{v}\ve{v},k}^{(q)}} + \underbrace{\frac{14}{3}N^{\prime\,3}+4N^{\prime\,2}}_{\text{Cholesky}}+\underbrace{\vphantom{\frac{14}{3}}4N^{\prime\,2}+8N^\prime+4}_{f_k^{(q)}(.)} + \hspace{-0.2cm}\underbrace{\vphantom{\frac{14}{3}}4N^\prime}_{\ve{h}_k\hat{d}_k^{(q-1)}}\hspace{-0.2cm}\Big)\nonumber\\
\begin{split}
&= QN_{\text{d}}\Big(\frac{14}{3}N^{\prime\,3} + 4N^{\prime\,2}N_{\text{d}} + 4N^{\prime\,2} + 2N^\prime N_{\text{d}}  + 14N^\prime + 6|\mathbb{S}^\prime|+6\Big)
\end{split}
\end{align}

{
\renewcommand{\aboverulesep}{0pt}
\renewcommand{\belowrulesep}{0pt}
\renewcommand{\arraystretch}{1.1}
\begin{table*}[t]
\caption{Number of required real-valued multiplications (rounded to hundreds) of evaluated equalizers for different SC-FDE system setups.}
\label{tab:Nr_mult_equalizers}
\begin{center}
\vspace{-0.2cm}
\begin{tabularx}{0.93\textwidth}{>{\raggedright}X >{\raggedright}m{1.8cm} >{\raggedleft}m{3cm} >{\raggedleft}m{3cm} >{\raggedleft\arraybackslash}m{3cm}}
\toprule
\multirow{2}{*}{Equalizer} & \multirow{2}{*}{ } & \multicolumn{3}{c}{Nr. multiplications for different setups} \tabularnewline
\cline{3-5}
& & UW guard, QPSK & UW guard, 16-QAM & CP guard, QPSK \\
\hline
SICNNv1 / SICNNv1Red & $M_{\text{SICNNv1}}$ & $3\,288\,600$ & $3\,409\,300$ & $8\,929\,500$\\
\hline
SICNNv2 / SICNNv2Red & $M_{\text{SICNNv2}}$ & $6\,836\,900$ & $31\,552\,200$ & $50\,600\,900$ \\
\hline
DetNet & $M_{\text{DetNet}}$ & $565\,100$ & $911\,400$ & $1\,153\,200$\\
\hline
KAFCNN & $M_{\text{KAFCNN}}$ &  $753\,900$ & $976\,900$ & $1\,072\,700$\\
\hline
OAMP-Net2 & $M_{\text{ONet2}}$ &  $4\,892\,300$ & $4\,894\,800$ & $9\,815\,100$\\
\hline
\multirow{2}{*}{LMMSE} & $M_{\text{LMMSE,burst}}$ & $141\,300$ & $141\,300$ & $100$\\
\cline{2-5}
& $M_{\text{LMMSE,eq}}$ & $2\,600$ & $2\,600$ & $400$\\
\hline
\multirow{2}{*}{DFE} & $M_{\text{DFE,burst}}$ & $295\,400$ & $295\,400$ & $1\,545\,400$\\
\cline{2-5}
& $M_{\text{DFE,eq}}$ & $5\,100$ & $5\,100$ & $8\,200$ \\
\hline
Iterative SIC & $M_{\text{itSIC}}$ & $14\,440\,800$ & $14\,441\,500$ & $27\,897\,500$\\
\bottomrule
\end{tabularx}
\end{center}
\vspace{-0.5cm}
\end{table*}
}

The numerical results of the complexities of the equalizers for the SC-FDE system setup specified in Sec.~\ref{ssec:Simulation_Setup_and_NN_Training} are given in Tab.~\ref{tab:Nr_mult_equalizers}. For all system setups considered, DetNet and KAFCNN are the lowest complex NN-based equalizers. The inference complexity of SICNNv1 is distinctly lower than that of SICNNv2 and OAMP-Net2, and also than the model-based iterative SIC method it is deduced from. However, the LMMSE estimator and the DFE exhibit by far the lowest complexity.

\section{Conclusion}
In this work, we proposed novel NN-based equalizers, called SICNNv1 and SICNNv2, inspired by a model-based soft interference cancellation scheme. SICNNv1 is tailored for an SC-FDE communication system, while SICNNv2 is also applicable for other communication systems with block-based data transmission. In addition, we presented a novel approach for generating training sets for NN-based equalizers, which considerably helps to improve their performance at high SNRs. We evaluated the proposed NN-based equalizers for a number of different SC-FDE system setups, and investigated their robustness with respect to imperfect channel knowledge at the receiver. In particular SICNNv1 exhibits a superior BER performance over all regarded state-of-the-art model-based and NN-based equalization approaches for all SC-FDE system setups considered. To highlight the universal applicability of SICNNv2, we exemplarily presented its state-of-the-art performance for a UW-OFDM system. Further, we investigated the influence of the size of the dataset used to train the NN-based equalizers, and we presented an in-depth complexity analysis.

{\appendices
\section{Noise Statistics in a Soft Interference Cancellation Step}
\label{apx:noise_statistics_in_SIC_step}
Let us consider the system model~\eqref{eq:system_model_interference_cancellation} for estimating the $k$th data symbol $d_k$, $k\in\{0, ..., N_{\text{d}}-1\}$, for any but the first iteration ($q=0$), i.e., $0 < q < Q$,
\begin{gather}
\ve{y}_{\text{ic},k}^{(q)} = \ve{y} - \bar{\m{H}}_k\hat{\bar{\ve{d}}}_k^{(q-1)} = \ve{h}_kd_k \underbrace{- \overbrace{\bar{\m{H}}_k\bar{\bm{\delta}}_k^{(q-1)}}^{\ve{r}_k^{(q)}}+\ve{w}}_{\ve{v}_k^{(q)}}\,,
\end{gather}
which we repeat here for readability. Following central limit theorem arguments, the total noise $\ve{v}_k^{(q)}$ is assumed to feature a multivariate Gaussian distribution, i.e., $p(\ve{v}_k^{(q)})=\mathcal{C N}\big(\bm{\mu}_{\ve{v},k}^{(q)},\m{C}_{\ve{v}\ve{v},k}^{(q)}\big)$, where $\bm{\mu}_{\ve{v},k}^{(q)}$ and $\m{C}_{\ve{v}\ve{v},k}^{(q)}$ are to be specified. We start by computing the statistics of $\bar{\bm{\delta}}_k^{(q-1)}$, followed by those of $\ve{r}_k^{(q)}$ to obtain finally the distribution of $\ve{v}_k^{(q)}$. It is important to note, that for computing noise statistics in iteration $q$, the estimation errors from the previous iteration $(q-1)$ have to be specified, whereby the interference canceled vectors $\ve{y}_{\text{ic},k}^{(q-1)}$ are fixed and available, i.e., the PDFs/PMFs of $\ve{v}_k^{(q)}$, $\ve{r}_k^{(q)}$, and $\bar{\bm{\delta}}_k^{(q-1)}$ are not unconditional, but are conditioned on a given $\ve{y}_{\text{ic},k}^{(q-1)}$. 

A data symbol estimation error $\delta_k^{(q-1)} = \hat{d}_k^{(q-1)} - d_k$, where $\ve{y}_{\text{ic},k}^{(q-1)}$ is given, can only attain a finite number of different values (as many as the cardinality $|\mathbb{S}|$ of the symbol alphabet), and thus its statistics is described by a PMF $p_{d_k|\ve{y}_{\text{ic},k}^{(q)}}\big[\delta_k^{(q-1)}\big|\ve{y}_{\text{ic},k}^{(q-1)}\big]$. 
In order to compute the statistics of an estimation error $\delta_k^{(q-1)} = \hat{d}_k^{(q-1)} - d_k$, we start by reconsidering the MMSE estimate of the preceding iteration 
\begin{align}
\hat{d}_k^{(q-1)} = E_{d_k|\ve{y}_{\text{ic},k}^{(q-1)}}\big[d_k\big|\ve{y}_{\text{ic},k}^{(q-1)}\big]= \sum_{s^\prime\in\mathbb{S}}s^\prime p\big[d_k=s^\prime\big|\ve{y}_{\text{ic},k}^{(q-1)}\big]\,. 
\end{align}
Since 
\begin{gather}
\ve{y}_{\text{ic},k}^{(q-1)} = \ve{y} - \bar{\m{H}}_k\hat{\bar{\ve{d}}}_k^{(q-2)}\,,
\end{gather}
we can reformulate $p\big[d_k\big|\ve{y}_{\text{ic},k}^{(q-1)}\big]$ as $p\big[d_k\big|\ve{y}_{\text{ic},k}^{(q-1)}\big] = p\big[d_k\big|\ve{y}, \hat{\bar{\ve{d}}}_k^{(q-2)}\big]$, where we consider $\hat{\bar{\ve{d}}}_k^{(q-2)}$ to be a fixed vector in iteration $(q-1)$, which does not feature a statistical distribution. It can be observed from the system model for iteration step $(q-1)$
\begin{gather}
\ve{y}_{\text{ic},k}^{(q-1)} = \ve{y} - \bar{\m{H}}_k\hat{\bar{\ve{d}}}_k^{(q-2)} = \ve{h}_kd_k - \bar{\m{H}}_k\bar{\bm{\delta}}_k^{(q-2)}+\ve{w}
\end{gather}
that the PMF $p\big[d_k\big|\ve{y}, \hat{\bar{\ve{d}}}_k^{(q-2)}\big]$ does not depend on the estimate $\hat{d}_k^{(q-2)}$ for a given $\ve{y}$ and $\hat{\bar{\ve{d}}}_k^{(q-2)}$. Hence, $\hat{d}_k^{(q-2)}$ can be included in the condition of $p\big[d_k\big|\ve{y}, \hat{\bar{\ve{d}}}_k^{(q-2)}\big]$ without altering the PMF, i.e., $p\big[d_k\big|\ve{y}, \hat{\bar{\ve{d}}}_k^{(q-2)}\big] = p\big[d_k\big|\ve{y}, \hat{\bar{\ve{d}}}_k^{(q-2)}, \hat{d}_k^{(q-2)}\big] = p\big[d_k\big|\ve{y}, \hat{\ve{d}}^{(q-2)}\big].$
Consequently, the MMSE estimate $\hat{d}_k^{(q-1)}$ and its corresponding conditional MSE can be rewritten as
\begin{align}
\hat{d}_k^{(q-1)} = E_{d_k|\ve{y}_{\text{ic},k}^{(q-1)}}\big[d_k\big|\ve{y}_{\text{ic},k}^{(q-1)}\big] = E_{d_k|\ve{y},\hat{\ve{d}}^{(q-2)}}\big[d_k\big|\ve{y}, \hat{\ve{d}}^{(q-2)}\big] 
\end{align}
and 
\begin{align}
e_k^{(q-1)} = E_{d_k|\ve{y}_{\text{ic},k}^{(q-1)}}\big[\big|d_k-\hat{d}_k^{(q-1)}\big|^2\big|\ve{y}_{\text{ic},k}^{(q-1)}\big] = E_{d_k|\ve{y},\hat{\ve{d}}^{(q-2)}}\big[\big|d_k-\hat{d}_k^{(q-1)}\big|^2\big|\ve{y}, \hat{\ve{d}}^{(q-2)}\big]\,,
\end{align}
respectively. Hence, the conditional PMF of an estimation error $\delta_k$ can be written as
\begin{gather*}
p_{d_k|\ve{y}_{\text{ic},k}^{(q-1)}}\big[\delta_k^{(q-1)}\big|\ve{y}_{\text{ic},k}^{(q-1)}\big] = p_{d_k|\ve{y}, \hat{\ve{d}}^{(q-2)}}\big[\delta_k^{(q-1)}\big|\ve{y},\hat{\ve{d}}^{(q-2)}\big]\,,
\end{gather*}
and the PMF of $\bar{\bm{\delta}}^{(q-2)}$ as $p_{\bar{\ve{d}}_k|\ve{y},\hat{\ve{d}}^{(q-2)}}\big[\bar{\bm{\delta}}_k^{(q-1)}\big|\ve{y},\hat{\ve{d}}^{(q-2)}\big]$. 

Following central limit theorem arguments, we assume the distribution of $\ve{r}_k^{(q)}$ to be multivariate Gaussian. The mean of $\ve{r}_k^{(q)}$ follows to
\begin{align}
\bm{\mu}_{\ve{r},k}^{(q)} &= E_{\bar{\ve{d}}_k|\ve{y},\hat{\ve{d}}^{(q-2)}}\big[\bar{\m{H}}_k\big(\hat{\bar{\ve{d}}}_k^{(q-1)}-\bar{\ve{d}}_k\big)\big|\ve{y},\hat{\ve{d}}^{(q-2)}\big]\nonumber\\ &= \bar{\m{H}}_k\big(\hat{\bar{\ve{d}}}_k^{(q-1)} - \hat{\bar{\ve{d}}}_k^{(q-1)}\big) = \ve{0}\,.
\end{align}
Its covariance matrix, in turn, is given by
\begin{align}
\m{C}_{\ve{r}\ve{r},k}^{(q)} &= E_{\ve{r}_k^{(q)}}\big[\ve{r}_k^{(q)}\ve{r}_k^{(q)\,H}\big]\nonumber\\&= \bar{\m{H}}_k \underbrace{E_{\bar{\ve{d}}_k|\ve{y},\hat{\ve{d}}^{(q-2)}}\big[\bar{\bm{\delta}}_k^{(q-1)}\bar{\bm{\delta}}_k^{(q-1)\,H}\big|\ve{y},\hat{\ve{d}}^{(q-2)}\big]}_{\widetilde{\m{E}}_k^{(q-1)}}\bar{\m{H}}_k^{H}\,.
\end{align}
For computing an element in the $m$th row and the $n$th column of $\widetilde{\m{E}}_k^{(q-1)}$, $m,n\in\{0, ..., N_{\text{d}}-2\}$, let us first consider to which data symbol $d_i$ and $d_j$ in the data vector $\ve{d}$,  $i,j\in\{0, ..., N_{\text{d}}-1\}$, the matrix element $\big[\widetilde{\m{E}}_k^{(q-1)}\big]_{mn}$ belongs. Since $\bar{\bm{\delta}}_k^{(q-1)}$ contains the deviations of the data symbol estimates to the corresponding true data symbols for all but the $k$th data symbols in the data vector, the index mapping follows to
\begin{gather*}
i = \begin{cases}
m & m < k\\
m+1 & m \geq k
\end{cases}\quad\text{and}\quad j = \begin{cases}
n & n < k\\
n+1 & n \geq k
\end{cases}.
\end{gather*}
An off-diagonal element of $\widetilde{\m{E}}_k^{(q-1)}$ can be written as 
\begin{align*}
\big[\widetilde{\m{E}}_{k}^{(q-1)}\big]_{mn} &= E_{\bar{\ve{d}}_k|\ve{y}, \hat{\ve{d}}^{(q-2)}}\big[\delta_i^{(q-1)}\left.\delta_j^{(q-1)}\right.^*\big|\ve{y},\hat{\ve{d}}^{(q-2)}\big]\nonumber\\
&= E_{d_i, d_j|\ve{y}, \hat{\ve{d}}^{(q-2)}}\big[d_id_j^*|\ve{y},\hat{\ve{d}}^{(q-2)}\big] - E_{d_i, d_j|\ve{y}, \hat{\ve{d}}^{(q-2)}}\big[d_i\left.\hat{d}_j^{(q-1)}\right.^*|\ve{y},\hat{\ve{d}}^{(q-2)}\big]\nonumber\\ &\quad - E_{d_i, d_j|\ve{y}, \hat{\ve{d}}^{(q-2)}}\big[\hat{d}_i^{(q-1)}d_j^*|\ve{y},\hat{\ve{d}}^{(q-2)}\big] + E_{d_i, d_j|\ve{y}, \hat{\ve{d}}^{(q-2)}}\big[\hat{d}_i^{(q-1)}\left.\hat{d}_j^{(q-1)}\right.^*|\ve{y},\hat{\ve{d}}^{(q-2)}\big]\nonumber\\
&= E_{d_i, d_j|\ve{y}, \hat{\ve{d}}^{(q-2)}}\big[d_id_j^*|\ve{y},\hat{\ve{d}}^{(q-2)}\big] - E_{d_i, d_j|\ve{y}, \hat{\ve{d}}^{(q-2)}}\big[d_i|\ve{y},\hat{\ve{d}}^{(q-2)}\big]\left.\hat{d}_j^{(q-1)}\right.^*\nonumber\\ &\quad - \hat{d}_i^{(q-1)} E_{d_i, d_j|\ve{y}, \hat{\ve{d}}^{(q-2)}}\big[d_j^*|\ve{y},\hat{\ve{d}}^{(q-2)}\big] + \hat{d}_i^{(q-1)}\left.\hat{d}_j^{(q-1)}\right.^*\nonumber\\
&= E_{d_i, d_j|\ve{y}, \hat{\ve{d}}^{(q-2)}}\big[d_id_j^*|\ve{y},\hat{\ve{d}}^{(q-2)}\big] - \hat{d}_i^{(q-1)}\left.\hat{d}_j^{(q-1)}\right.^*,
\end{align*}
and a diagonal element of $\widetilde{\m{E}}_k^{(q-1)}$ has the value 
\begin{align*}
\big[\widetilde{\m{E}}_k^{(q-1)}\big]_{mm} &= E_{\bar{\ve{d}}_k|\ve{y},\hat{\ve{d}}^{(q-2)}}\big[\delta_i^{(q-1)}\left.\delta_i^{(q-1)}\right.^*\big|\ve{y},\hat{\ve{d}}^{(q-2)}\big]\\&=E_{d_i|\ve{y},\hat{\ve{d}}^{(q-2)}}\big[\delta_i^{(q-1)}\left.\delta_i^{(q-1)}\right.^*\big|\ve{y},\hat{\ve{d}}^{(q-2)}\big]\\
&=e_i^{(q-1)}\,.
\end{align*}

With the above results at hand, $\ve{v}_k^{(q)}$ can be specified to be a zero mean Gaussian distributed vector with a covariance matrix
\begin{align}
\m{C}_{\ve{v}\ve{v},k}^{(q)} &= E_{\ve{v}_k^{(q)}}\big[\ve{v}_k^{(q)}\ve{v}_k^{(q)\,H}\big]\nonumber\\&=E_{\ve{r}_k^{(q)},\ve{w}}\big[\ve{r}_k^{(q)}\ve{r}_k^{(q)\,H}-\ve{r}_k^{(q)}\ve{w}^H-\ve{w}\ve{r}_k^{(q)\,H}+\ve{w}\ve{w}^H\big]\nonumber\\&= \m{C}_{\ve{r}\ve{r},k}^{(q)} - \bar{\m{H}}_k E_{\big(\bar{\ve{d}}_k|\ve{y},\hat{\ve{d}}^{(q-2)}\big),\ve{w}}\big[\bar{\bm{\delta}}_k^{(q-1)}\ve{w}^H\big] - E_{\big(\bar{\ve{d}}_k|\ve{y},\hat{\ve{d}}^{(q-2)}\big),\ve{w}}\big[\ve{w}\bar{\bm{\delta}}_k^{(q-1)\,H}\big]\bar{\m{H}}_k^{H} + \m{C}_{\ve{w}\ve{w}}\nonumber\\
\begin{split}
&= \bar{\m{H}}_{k}\widetilde{\m{E}}_k^{(q-1)}\bar{\m{H}}_{k}^H - \bar{\m{H}}_k E_{\left(\bar{\ve{d}}_k|\ve{y},\hat{\ve{d}}^{(q-2)}\right), \ve{w}}\big[\bar{\bm{\delta}}_{k}^{(q-1)}\ve{w}^H\big] - E_{\left(\bar{\ve{d}}_k|\ve{y},\hat{\ve{d}}^{(q-2)}\right), \ve{w}}\big[\ve{w}\bar{\bm{\delta}}_{k}^{(q-1)\,H}\big]\bar{\m{H}}_k^H + N\sigma_{\text{n}}^2\widetilde{\m{H}}\,,
\end{split}
\end{align}
with $\widetilde{\m{E}}_k^{(q-1)}$ containing
\begin{gather*}
\big[\widetilde{\m{E}}_k^{(q-1)}\big]_{mn} = \begin{cases}
e_i^{(q-1)} & m = n\\
\begin{split}&E_{d_i,d_j|\ve{y},\hat{\ve{d}}^{(q-2)}}\big[d_id_j^*\big|\ve{y},\hat{\ve{d}}^{(q-2)}\big] \\&- \hat{d}_i^{(q-1)}\left.\hat{d}_j^{(q-1)}\right.^*\end{split} & m\neq n
\end{cases}
\end{gather*}
in the $m$th row and $n$th column, and the index mapping as defined above.

\section{Estimates of Iterative Soft Interference Cancellation Method after First Iteration}
\label{apx:estimates_of_soft_interference_cancellation_method_first_iteration}

We show for a QPSK modulation alphabet that the bit error probability of a hard decision estimate produced by the iterative SIC method described in Sec.~\ref{ssec:Iterative_Soft_Interference_Cancellation} after the first iteration is equivalent to the bit error probability of an LMMSE hard decision estimate when initializing all data symbol estimates of the iterative SIC method with $0$, i.e., $\hat{d}_k^{(-1)}=0$, $k\in\{0, ..., N_{\text{d}}-1\}$. To this end, we show that the decision criteria of both estimation methods for the $j$th bit $b_{jk}$ of the $k$th data symbol, $k\in\{0, ..., N_{\text{d}}-1\}$, $j\in\{0, ..., \log_2(|\mathbb{S}|)-1\}$, being $0$ or $1$ are the same, and thus also their bit error probability must coincide. The QPSK bit-to-symbol mapping $(b_{1k} b_{0k}) \mapsto d_k$ is assumed to map $b_{0k}$ to the real part and $b_{1k}$ to the imaginary part of $d_k$. The bit values $0$ and $1$ are mapped to the symbol values $-\rho$ and $\rho$, respectively, with $\rho = 1/\sqrt{2}$ being an energy normalization factor. 

Let us start with the LMMSE data estimator. The SC-FDE system model is given by (cf.~\eqref{eq:system_model_SC-FDE})
\begin{gather}
\ve{y} = \m{H}\ve{d} + \ve{w}\,,\label{eq:system_model_appendix}
\end{gather}
where $\ve{w}\sim\mathcal{C N}(\ve{0}, N\sigma_{\text{n}}^2\widetilde{\m{H}})$, and the corresponding LMMSE data estimator follows to
\begin{gather}
\hat{\ve{d}} = \sigma_{\text{d}}^2\m{H}^H\big(\sigma_{\text{d}}^2\m{H}\m{H}^H + N\sigma_{\text{n}}^2\widetilde{\m{H}}\big)^{-1}\ve{y}\,,\label{eq:LMMSE_v2}
\end{gather}
which is expressed in a different way as in~\eqref{eq:LMMSE_v1}, but can be shown to be mathematically equivalent. Based on the LMMSE estimates $\hat{d}_k$, the hard decision estimate for bit $b_{0k}$ is 
\begin{gather}
\hat{b}_{0k}=\begin{cases}1 & \text{Pr}(b_{0k} = 1|\hat{d}_k) > \text{Pr}(b_{0k} = 0|\hat{d}_k)\\0 & \text{otherwise}\end{cases}\,.\label{eq:LMMSE_dec_criterion_initial}
\end{gather}
It can be shown~\cite{Haselmayr15} that for the LMMSE estimator~\eqref{eq:LMMSE_v2} and the model~\eqref{eq:system_model_appendix} $\text{Pr}(b_{0k}=1|\hat{d}_k) = \kappa \text{Pr}(b_{0k}=1|\ve{y})$ and $\text{Pr}(b_{0k}=0|\hat{d}_k) = \kappa \text{Pr}(b_{0k}=0|\ve{y})$, where $\kappa$ is a proportionality constant. Hence, the condition on $\hat{d}_k$ in criterion~\eqref{eq:LMMSE_dec_criterion_initial} can be replaced by a condition on $\ve{y}$, leading to a decision criterion
\begin{gather}
\hat{b}_{0k}=\begin{cases}1 & \text{Pr}(b_{0k}=1|\ve{y}) > \text{Pr}(b_{0k}=0|\ve{y})\\0 & \text{otherwise}\end{cases}\,.\label{eq:LMMSE_dec_criterion_cond_y}
\end{gather}

By using the Bayesian rule and assuming a uniform prior PMF $p[d_k]$, we can rearrange the criterion in~\eqref{eq:LMMSE_dec_criterion_cond_y} to
\begin{align}
\text{Pr}(b_{0k}=1|\ve{y}) &> \text{Pr}(b_{0k}=0|\ve{y})\nonumber\\
\sum_{s^\prime\in\mathcal{S}_{0}^{(1)}}p[d_k=s^\prime|\ve{y}] &> \sum_{s^\prime\in\mathcal{S}_{0}^{(0)}}p[d_k=s^\prime|\ve{y}] \nonumber\\ \sum_{s^\prime\in\mathcal{S}_{0}^{(1)}} p(\ve{y}|d_k=s^\prime) &> \sum_{s^\prime\in\mathcal{S}_{0}^{(0)}}p(\ve{y}|d_k=s^\prime)\,,\label{eq:LMMSE_dec_criterion_cond_pdf}
\end{align}
where $\mathcal{S}_{0}^{(0)}$ and $\mathcal{S}_{0}^{(1)}$ are the sets of data symbols containing a bit with value $0$ and $1$ at the $0$th position, respectively.
For determining the PDF $p(\ve{y}|d_k=s^\prime)$, we reformulate \eqref{eq:system_model_appendix} as 
\begin{gather}
\ve{y} = \bar{\m{H}}_k\bar{\ve{d}}_k + \ve{h}_k d_k + \ve{w}\,.
\end{gather}
Due to central limit theorem arguments, $\bar{\m{H}}_k\bar{\ve{d}}_k$ can be assumed to be Gaussian distributed, and thus $p(\ve{y}|d_k)$ is approximated to be a multivariate complex Gaussian PDF, i.e., 
\begin{gather}
p(\ve{y}|d_k)=\zeta_{\ve{y}|d_k}\exp\big(-(\ve{y}-\bm{\mu}_{\ve{y}|d_k})^H\m{C}_{\ve{y}\ve{y}|d_k}^{-1}(\ve{y}-\bm{\mu}_{\ve{y}|d_k})\big)\,,\label{eq:LMMSE_cond_pdf}
\end{gather}
with the scaling factor $\zeta_{\ve{y}|d_k}=\frac{1}{\pi^N|\m{C}_{\ve{y}\ve{y}|d_k}|}$, the conditional mean
\begin{gather}
\bm{\mu}_{\ve{y}|d_k} = E[\ve{y}|d_k] = \ve{h}_k d_k\,,
\end{gather}
and the conditional covariance matrix
\begin{align}
\m{C}_{\ve{y}\ve{y}|d_k} = \m{C}_k &= E\big[(\ve{y}-\bm{\mu}_{\ve{y}|d_k})(\ve{y}-\bm{\mu}_{\ve{y}|d_k})^H|d_k\big]\nonumber\\ &= E\big[(\bar{\m{H}}_k\bar{\ve{d}}_k + \ve{w})(\bar{\m{H}}_k\bar{\ve{d}}_k + \ve{w})^H|d_k\big]\nonumber\\ &= \sigma_{\text{d}}^2\bar{\m{H}}_k\bar{\m{H}}_k^H + N\sigma_{\text{n}}^2\widetilde{\m{H}}\,.\label{eq:LMMSE_covar_matrix}
\end{align}
Since $\m{C}_{\ve{y}\ve{y}|d_k} = \m{C}_k$ does not depend on the realization of $d_k$, the scaling factor $\zeta_{\ve{y}|d_k}$ is a constant for any symbol $s^\prime\in\mathbb{S}$. Consequently, by inserting~\eqref{eq:LMMSE_cond_pdf} into~\eqref{eq:LMMSE_dec_criterion_cond_pdf} we arrive at the LMMSE hard decision criterion
\begin{gather}
\hat{b}_{0k} = \begin{cases}
1 & \sum_{s^\prime\in\mathcal{S}_0^{(1)}}g(s^\prime) > \sum_{s^\prime\in\mathcal{S}_0^{(0)}}g(s^\prime)\\
0 & \text{otherwise}
\end{cases},\label{eq:LMMSE_dec_criterion_final}
\end{gather}
with
\begin{gather}
g(s^\prime) = \exp\big(-(\ve{y}-\ve{h}_k s^\prime)^H\m{C}_k^{-1}(\ve{y}-\ve{h}_k s^\prime)\big)\label{eq:g_fun_LMMSE_dec_criterion}
\end{gather}
and $\m{C}_k$ as defined in~\eqref{eq:LMMSE_covar_matrix}.

Let us now consider the iterative SIC method, starting with the system model in its first iteration ($q=0$), which is given by
\begin{gather}
\ve{y} = \ve{h}_k d_k + \ve{v}_k^{(0)}\,,\label{eq:system_model_SIC_first_iter}
\end{gather}
where $\ve{v}_k^{(0)} = \bar{\m{H}}_k\bar{\ve{d}}_k + \ve{w}$ is assumed to be zero mean Gaussian noise with a covariance matrix $\m{C}_{\ve{v}\ve{v},k}^{(0)} = \sigma_{\text{d}}^2\bar{\m{H}}_k\bar{\m{H}}_k^H+N\sigma_{\text{n}}^2\widetilde{\m{H}}$. The MMSE estimate for data symbol $d_k$ is the mean of the posterior PMF $E_{d_k|\ve{y}}[d_k|\ve{y}]$ for the model~\eqref{eq:system_model_SIC_first_iter}, which is given by (cf.~\eqref{eq:MMSE_data_symbol_estimate})
\begin{gather}
\hat{d}_k = \frac{\sum_{s^\prime\in\mathbb{S}}s^\prime p\big(\ve{y}\big|d_k=s^\prime\big)}{\sum_{s^\prime\in\mathbb{S}}p\big(\ve{y}\big|d_k=s^\prime\big)}\label{eq:MMSE_first_iter}
\end{gather}
Considering the QPSK bit-to-symbol mapping defined above, the MMSE hard decision estimate for bit $b_{0k}$ is 
\begin{gather}
\hat{b}_{0k} = \begin{cases}
1 & \text{Re}\{\hat{d}_k\} > 0\\
0 & \text{otherwise}
\end{cases}.\label{eq:initial_dec_criterion_SIC_first_iter}
\end{gather}
Since the denominator of the MMSE estimate given in~\eqref{eq:MMSE_first_iter} is always positive,  the MMSE hard decision estimate $\hat{b}_{0k}$ is estimated to be $1$ if
\begin{gather}
\begin{aligned}
0 &< \text{Re}\big\{\sum_{s^\prime\in\mathbb{S}}s^\prime p\big(\ve{y}\big|d_k=s^\prime\big)\big\}\\&= \sum_{s^\prime\in\mathbb{S}}\text{Re}\{s^\prime\}p\big(\ve{y}\big|d_k=s^\prime\big)\\
&= \sum_{s^\prime\in\mathcal{S}^{(\text{+})}}\text{Re}\{s^\prime\}p\big(\ve{y}\big|d_k=s^\prime\big) + \sum_{s^\prime\in\mathcal{S}^{(\text{-})}}\text{Re}\{s^\prime\}p\big(\ve{y}\big|d_k=s^\prime\big)\\
&=\rho\sum_{s^\prime\in\mathcal{S}^{(\text{+})}}p\big(\ve{y}\big|d_k=s^\prime\big) - \rho\sum_{s^\prime\in\mathcal{S}^{(\text{-})}}p\big(\ve{y}\big|d_k=s^\prime\big)\,,
\end{aligned}
\end{gather}
or, equivalently formulated, if
\begin{gather}
\sum_{s^\prime\in\mathcal{S}^{(\text{+})}}p\big(\ve{y}\big|d_k=s^\prime\big) > \sum_{s^\prime\in\mathcal{S}^{(\text{-})}}p\big(\ve{y}\big|d_k=s^\prime\big)\,,\label{eq:reformulated_dec_criterion_SIC_first_iter}
\end{gather}
with $\mathcal{S}^{(+)}\subset\mathbb{S}$ and $\mathcal{S}^{(-)}\subset\mathbb{S}$ being the sets of symbols containing solely symbols with positive and negative real part, respectively. For the given QPSK bit-to-symbol mapping, these two sets of symbols $\mathcal{S}^{(+)}$ and $\mathcal{S}^{(-)}$ coincide with the sets $\mathcal{S}_{0}^{(1)}$ and $\mathcal{S}_0^{(0)}$ as defined above, respectively. The PDF $p(\ve{y}|d_k)$ for the system model~\eqref{eq:system_model_SIC_first_iter}, in turn, is given by
\begin{gather}
p(\ve{y}|d_k) = \zeta_{\ve{v}_k^{(0)}|d_k}\exp\big(-(\ve{y}-\ve{h}_k d_k)^H\m{C}_{\ve{v}\ve{v},k}^{(0)^{-1}}(\ve{y}-\ve{h}_k d_k)\big)\,,\label{eq:cond_pdf_SIC_first_iter}
\end{gather}
with $\zeta_{\ve{v}_k^{(0)}|d_k} = \frac{1}{\pi^N\big|\m{C}_{\ve{v}\ve{v},k}^{(0)}\big|}$ being constant for every data symbol realization and $\m{C}_{\ve{v}\ve{v},k}^{(0)} \m{C}_k = \sigma_{\text{d}}^2\bar{\m{H}}_k\bar{\m{H}}_k^H+N\sigma_{\text{n}}^2\widetilde{\m{H}}$. Combining~\eqref{eq:initial_dec_criterion_SIC_first_iter}, \eqref{eq:reformulated_dec_criterion_SIC_first_iter}, and~\eqref{eq:cond_pdf_SIC_first_iter} leads to the MMSE hard decision criterion
\begin{gather}
\hat{b}_{0k} = \begin{cases}
1 & \sum_{s^\prime\in\mathcal{S}_0^{(1)}}g(s^\prime) > \sum_{s^\prime\in\mathcal{S}_0^{(0)}}g(s^\prime)\\
0 & \text{otherwise}
\end{cases},
\end{gather}
with $g(s^\prime)$ as defined in~\eqref{eq:g_fun_LMMSE_dec_criterion}. This is the same decision criterion as that of the LMMSE~\eqref{eq:LMMSE_dec_criterion_final}, which concludes the proof. 

}

\end{document}